# DISSERTATION

In order to obtain the

## PhD degree
## Of National Institute of Posts and Telecommunications

Discipline: **Telecommunications and Information Technology**

*Presented by*

## Fatima SALAHDINE

# Compressive Spectrum Sensing for Cognitive Radio Networks

*Directed by: Dr. Naima KAABOUCH and Dr. Hassan El GHAZI*
*January 19th, 2018*

**Dissertation Defense Committee:**

| | | |
|---|---|---|
| Kaabouch Naima | Professor, University of North Dakota, USA | Dissertation Supervisor |
| Hassan El Ghazi | Assoc. Professor, INPT, Morocco | Dissertation Supervisor |
| M'hamed Drissi | Professor, INSA Rennes, France | Chair |
| Jamal El Abbadi | Professor, EMI, Morocco | Reviewer |
| Nawfel Azami | Professor, INPT, Morocco | Reviewer |
| Faissal El Bouanani | Assoc. Professor, ENSIAS, Morocco | Reviewer |
| Badr Abou El Majd | Assoc. Professor, UM5, Morocco | Examiner |
| Esmail Ahouzi | Professor, INPT, Morocco | Examiner |

# ABSTRACT

A cognitive radio system has the ability to observe and learn from the environment, adapt to the environmental conditions, and use the radio spectrum more efficiently. It allows secondary users (SUs) to use the primary users (PUs) channels when they are not being utilized. Cognitive radio involves three main processes: spectrum sensing, deciding, and acting. In the spectrum sensing process, the channel occupancy is measured with spectrum sensing techniques in order to detect unused channels. In the deciding process, sensing results are analyzed and decisions are made based on these results. In the acting process, actions are made by adjusting the transmission parameters to enhance the cognitive radio performance.

Due to multipath fading, shadowing, and varying channel conditions, uncertainty affects all the cognitive radio processes. Measurements taken by the SUs during the sensing process are uncertain. Decisions are taken based on what has already been observed using the SUs knowledge basis, which may have been impacted by uncertainty. This can lead to wrong decisions, and, thus the cognitive radio system can take wrong actions. Thus, uncertainty propagation influences the cognitive radio performance and mitigating it is a necessity.

One of the main challenges of cognitive radio is the wideband spectrum sensing. Existing spectrum sensing techniques are based on a set of observations sampled by an ADC at the Nyquist rate. However, those techniques can sense only one channel at a time because of the hardware limitations on the sampling rate. In addition, in order to sense a wideband spectrum, the wideband is divided into narrow bands or multiple frequency bands. SUs have to sense each band using multiple RF frontends simultaneously, which can result in a very high processing time, hardware cost, and computational complexity. In order to overcome this problem, the signal sampling should be as fast as possible even with high dimensional signals. Compressive sensing has been proposed as a low-cost solution to reduce the processing time and accelerate the scanning process. It allows reducing the number of samples required for high dimensional signal acquisition while keeping the essential information.

This dissertation aims to develop efficient and fast spectrum sensing techniques to improve the spectrum detection over time using compressive sensing solution. It reviews the different spectrum sensing techniques and compares them in terms of the probability of detection and the probability of false alarm. In this dissertation, we propose the use of a dynamic and estimated threshold to perform the matched filter detection technique and improve its efficiency. we give a review of compressive sensing techniques as well as their applications and challenges. we also propose an approach that combines the advantages of a Circulant matrix with Bayesian models to perform compressive sensing under uncertainty. In addition, a Bayesian compressive sensing technique based on the Toeplitz matrix sampling is proposed to enhance the compressive sensing efficiency, reduce the level of randomness, and handle uncertainty. This dissertation also describes a real-time spectrum scanning method using software defined radio units (USRP) to perform the wideband spectrum scanning. It is based on compressive sensing to speed up the scanning process and minimize the computational complexity. Finally, one-bit compressive sensing is reviewed and its performance is compared to the performance of the conventional compressive sensing.

All proposed methods in this dissertation were implemented and extensively tested. Their results were compared to the existing techniques based on a number of metrics, including probability of detection, probability of false alarm, signal to noise ratio (SNR), number of samples, sensing threshold, mean square error, reconstruction error, correlation, recovery time, sampling time, processing time, sparsity, required number of measurements, recovered SNR, hamming distance, and complexity.

# RÉSUMÉ

Avec la prolifération rapide des standards des réseaux sans fils et des services de radio communication, les ressources spectrales radio sont devenues de plus en plus rares et précieuses et la demande en ressources radio devient accrue. En plus, selon des études de l'organise FCC, le spectre fréquentiel n'est pas utilisé tous le temps par les utilisateurs propriétaire de sa licence. Par conséquent, le spectre de fréquence est mal exploité et il nécessite une intervention urgente pour la gestion efficace de cette rare ressource. La radio cognitive est l'une des solutions proposées pour remédier au problème du spectre en permettant un accès opportuniste des autres utilisateurs non licenciés au spectre quand ce dernier est non utilisé.

Afin de réduire le temps de détection, l'acquisition comprimée est proposée. C'est une technique d'acquisition et d'échantillonnage qui permet d'acquérir un signal dense de grande dimension en prenant seulement le minimum de données de ce signal en jetant le reste. A la réception, le signal original peut être reconstitué à partir du signal compressé en résolvant un système linéaire sous-déterminé pour trouver sa solution la plus dense. Cette solution correspond au signal original après avoir estimé ses coefficients perdus.

L'objectif de cette thèse est de contribuer à l'amélioration des techniques de détection de spectre en utilisant l'acquisition comprimée pour accélérer la détection de spectre et améliorer sa performance. Nous avons évalué et comparé les différentes techniques de détection du spectre en se basant sur la probabilité de détection et la probabilité de fausses alarmes. Nous avons proposé d'adopter un seuil de détection dynamique afin d'améliorer l'efficacité de la technique de filtre adapté. Nous avons analysé les différentes techniques d'acquisition comprimée ainsi que leurs applications. Nous avons proposé une technique d'acquisition basée sur la matrice Circulante et les réseaux Bayésien pour réduire l'incertitude dans les mesures. Nous avons proposé également une technique d'acquisition basée sur l'échantillonnage matriciel de Toeplitz et les réseaux Bayésien pour réduire le taux de compression et le temps de détection.

Nous avons aussi présenté une technique de détection en temps réel basée sur la radio logicielle afin d'analyser le spectre large bande en temps réduit. Finalement, nous avons analysé la méthode de l'acquisition comprimée basée sur un seul bit en comparant sa performance avec la méthode de l'acquisition comprimée conventionnelle basée sur plusieurs bits.

Toutes ces méthodes proposées ont été implémentées et testées en se basant sur des métriques de mesure, y compris la probabilité de détection, la probabilité de fausses alarmes, le rapport signal / bruit (SNR), le nombre d'échantillons, le seuil de détection, la corrélation, le temps d'échantillonnage, et le temps de détection.

# ACKNOWLEDGEMENT

I am grateful to my advisors Dr. Hassan El Ghazi and Dr. Naima Kaabouch for their valuable supervision, encouragements, and patience all these years of my Ph.D. program.

I would like to thank Dr. Hassan El Ghazi for his valuable advising during my years of study at the National Institute of Posts and Telecommunications as an engineering student and as a Ph.D. student working under his supervision. Without his intervention, we would never get the opportunity to work and join my American supervisor to perform my research in a very equipped laboratory in the USA.

I would like to express my gratitude to Dr. Naima Kaabouch for her high-level advising, relevant guidance, and support. we would like to thank her for her time and her availability in teaching and training to improve my skills and my research performance. With her welcoming since the first day we went to the USA, as a Fulbright Scholar at UND, she gave me the opportunity to explore the world and join an excellent experience in her laboratory. we cannot find the words to thank her enough for everything she did and she is doing to success my Ph.D. as well as my professional career. Without her welcoming, we would never get the chance to work with a high-level research team and attend my research objectives.

I would like to thank the University of North Dakota for welcoming me as a graduate student to work on my dissertation, allowing me to attend classes, giving me the access to the university resources, and explore the US university system. we would like to thank all the professors, administrators, colleagues, and friends at the electrical engineering department for making my experience there more valuable.

Also, I acknowledge the support of the Fulbright program, the US Department of State, Amideast, and the Moroccan-American Commission for Educational and Cultural Exchange (MACECE) for the Fulbright Scholarship award to follow my research studies at a US university and join my advisor Dr. Kaabouch.

I would like to thank my dissertation reviewers, Dr. Faissal El Bouanani, Dr. Jamal El Abbadi, and Dr. Nawfel Azami, of agreeing to review my dissertation and approving it with valuable comments and significant intervention. I would like to thank them for their interest and availability to read and review my work. I would like to thank Dr. M'hamed Drissi for accepting to be my Ph.D. defense chair. I would like also to thank Dr. Ahouzi Esmail and Dr. Badr Abou El Mjad for being interested to review and examine my work.

I would like to thank all my professors at the National Institute of Posts and Telecommunications and the Doctoral School CEDOC-2TI INPT for near or far support. Also, we thank all my friends and colleagues at INPT. On this occasion, we wish them all the best in their personal and professional life.

Finally, special thanks to all my family members for encouraging me during all these years of pursuing studies and catching dreams.

# TABLE OF CONTENTS

# LIST OF FIGURES

# LIST OF TABLES

# NOTATION

N: number of samples
y(n): SU received signal
s(n): PU signal
w(n): additive white Gaussian noise
$\delta_w^2$: noise variance
h: complex channel gain of the sensing channel
H0 and H1: absence and the presence of the PU signal respectively
T: test statistic of the detector
γ: sensing threshold
$\delta_s^2$: PU signal variance
Ɲ: normal distribution
Q(.): Q-function
$\bar{\lambda}$: average threshold
t: time
s*: complex conjugate of the signal,
$R_{s,s}$: autocorrelation function of the signal s(t)
lag 1 and ag0: first lag and lag zero of the autocorrelation
R: reference line
D: Euclidean distance
ψ(t): continuous wavelet function
u: scaling parameter greater,
v: translating parameter,
m: wavelet edge
xp: signal pilot
$T_{ED}$: test statistic of the energy detector
$T_{MFD}$: test statistic of the matched filter detector
$T_{Aut}$: test statistic of the autocorrelation based detector
$T_{Euc}$: test statistic of the Euclidian distance based detector
E: PU signal energy
Pd: probability of detection,
Pfd: probability of false alarm
$P_{md}$: probability of miss detection
$M_c$: measurements matrix
M: number of measurements
k: sparsity level
Nd: total number of detections
Nt: total number of experiments
Nf: total number of times the signal is not detected
δ: restricted isometry constant
C: Circulant matrix
Mc: Partial Circulant matrix
$\mu_s$: signal mean
$\delta_s$: signal variance
Re: reconstruction error
tr: Recovery time
ts: Sampling time
tp: processing time
Hd: Hamming distance
$M_T$: partial Toeplitz sensing matrix
$t_{sc}$: scanning time

# LIST OF ABBREVIATION

ADC: Analog/Digital Converter
ANRT: National Agency for the Legalization of Communications
AWGN: Additive White Gaussian Noise
BER: Bit Error Rate
BPD: Basis Pursuit Denoising
BR: Bayesian Recovery
CFC: Carrier Frequency Offset
COSAMP: Compressive Sampling Matching Pursuit
CSP: Compressive Signal Processing
FCC: Federal Communications Commission
FFO: Fractional Frequency Offset
FFT: Fast Fourier Transform
FSA: Fixed Spectrum Allocation
GOMP: Generalized Orthogonal Matching Pursuit
GLRT: Generalized Likelihood Ratio Test
GOMP: Generalized Orthogonal Matching Pursuit
HTP: Hard Thresholding Pursuit
IHT: Iterative Hard Thresholding
LMS: Least Mean Square
MIMO: Multiple-Input, Multiple-Output
MLE: Maximum Likelihood Estimator
MMSE: Minimum Mean Squared Error
MP: Matching Pursuit
MSE: Mean Square Error
NHTP: Normalized Hard Thresholding Pursuit
NIHT: Normalized Iterative Hard Thresholding
OFDM: Orthogonal Frequency-Division Multiplexing
OMP: Orthogonal Matching Pursuit
PSO: Particle Swarm Optimization
PU: Primary User
RIP: Restrict Isometry Property
ROC: Receiver Operating Curve
ROMP: Regularized Orthogonal Matching Pursuit
SDR: Software Defined Radio
SINR: Signal to Interference Plus Noise Ratio
SNR: Signal to Noise Ratio
SOMP: Stage Wise Orthogonal Matching Pursuit
SU: Secondary User
USRP: Universal Software Radio Peripheral

# Chapter I

# INTRODUCTION

In this chapter, we present a review of the spectrum sensing under the cognitive radio networks to highlight the different limitations and problems of the spectrum management under the static spectrum allocation. we list our contributions to address some of these problems by adopting compressive sensing framework.

The remaining of the chapter is organized as below: Section I.1 and Section I.2 present the spectrum management problem as well as the cognitive radio cycle; Section I.3 presents the compressive sensing framework; Section I.4 presents the dissertation objectives; Section I.5 and Section I.6 describe the different contributions with the list of all the published papers in international conferences and indexed journals. Finally, the dissertation organization is provided in section I.7.

## I.1 Spectrum Management and Cognitive Radio

Wireless networks and information traffic have grown exponentially over the last decade, which resulted in an excessive demand for the radio spectrum resources [1][2]. The radio spectrum is a limited resource controlled by regulations and the recognized authorities, such as the national agency for the legalization of communications (ANRT) in Morocco and the federal communications commission (FCC) in the US. The current radio spectrum allocation policy consists of assigning the channels to specific users with licenses for specific wireless technologies and services. Those licensed users have access to that spectrum portions to transmit their data, while others are forbidden even when the spectrum is unoccupied [3]. Recent studies reported that the spectrum utilization ranges from 15% to 85% in the US under the fixed spectrum allocation (FSA) policy [4]. FCC measurements also show that some channels are heavily used while others are sparsely used as illustrated in Figure 1[5].

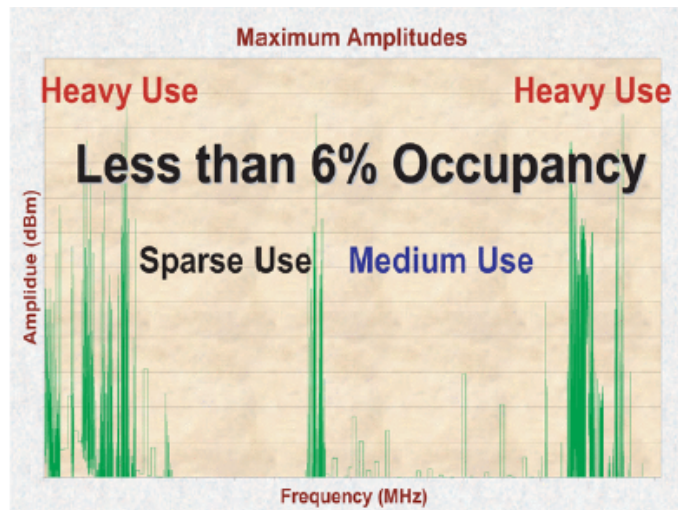

Figure 1: Radio spectrum occupancy [5][6][7].

Allocated spectrum portions are not used all the time by their owners, called primary users (PUs), which creates spectrum holes. A spectrum hole, also called white space, is a frequency band assigned to a PU but it is not being used at a particular time and at a particular location. Therefore, the radio spectrum is inefficiently exploited [8,9]. Thus, the scarcity and inefficiency of the spectrum management require an urgent intervention to enhance the radio spectrum access and achieve high network performance. A better way to overcome the spectrum scarcity issue is to dynamically manage it by sharing unoccupied channels with unlicensed users, called secondary users (SUs), without interfering with the PUs signals. The opportunistic spectrum access (OSA), also called dynamic spectrum access (DSA), has been proposed to address the spectrum allocation problems. In contrast to the FSA, DSA allows the spectrum to be shared between licensed and non-licensed users, in which the spectrum is divided into numerous bandwidths assigned to one or more dedicated users [10,11].

In order to advance the use of the OSA, several solutions have been proposed, including cognitive radio [12,13]. According to Mitola, cognitive radio is an intelligent radio frequency transmitter/receiver designed to detect the available channels in a wireless spectrum and adjust the transmission parameters enabling more communications and improving radio operating behavior [14,15]. It is a new approach to wireless networking in which the radio device is aware of its environment and has the ability to establish and adjust its parameters autonomously. A cognitive radio system can observe and learn from its environment, adapt to the environmental conditions, and make decisions in order to efficiently use the radio spectrum. It allows SUs to use the PU assigned radio spectrum when it is temporally not being utilized as illustrated in Figure 2 [1][3].

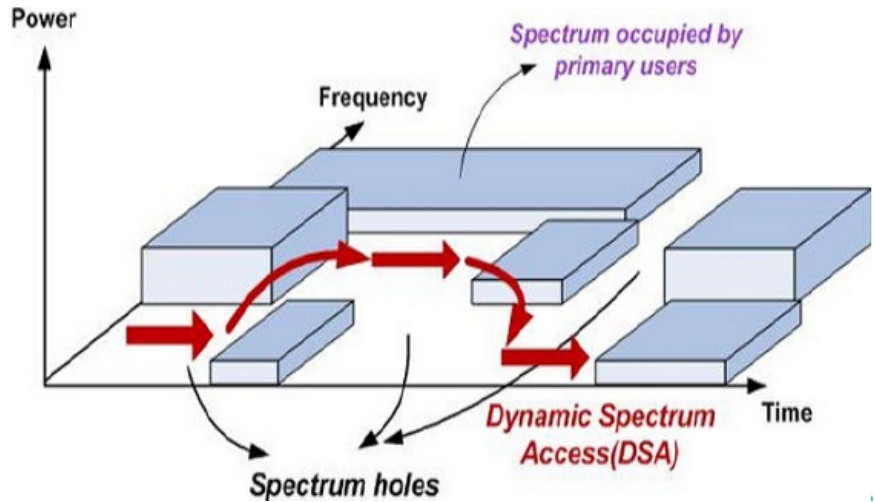

Figure 2: Dynamic spectrum access [3][16][17].

Cognitive radio is considered as the future technology to solve the resource allocation problem that the requirements of the 5th generation of the wireless communication raised. With the 5th generation of the wireless communication, the wide wireless will be interconnected offering high quality of service and data rates. The IEEE 802.22 standard has been defined as the first achievement of the cognitive radio solution to enable SUs to use the TV white spaces in the VHF and UHF bands [2].

## I.2 Cognitive Radio Cycle

As illustrated in Figure 3, a cognitive radio system performs a 3-process cycle: sensing, deciding, and acting [16][17].

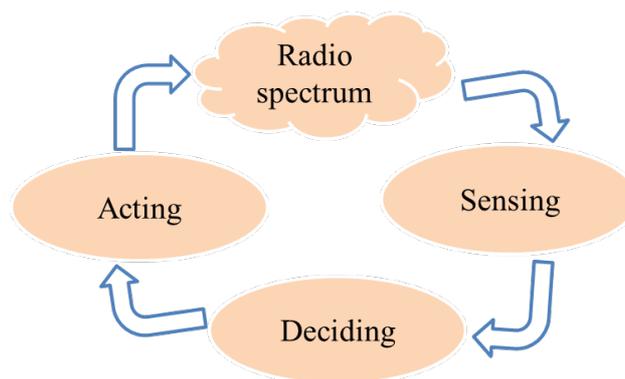

Figure 3: Cognitive radio cycle [17][16].

The first process is critical since it is the stage where the measurements are taken and the spectrum sensing is performed. Due to multipath fading, shadowing, or varying channel conditions [16-19], uncertainty affects this first process. In the observation process, measurements taken by the SUs are also uncertain. In the next process, SUs make a decision based on what has already been observed using their knowledge basis, which may have been impacted by the uncertainty in the detected measurements, leading to the wrong decisions. In the last process, uncertainty spreads over the cognitive radio cycle, and sometimes the wrong actions are taken [1]. Thus, uncertainty propagation impacts all the radio spectrum processes, which degrades the cognitive radio performance [16].

Therefore, it is necessary to address these uncertainty problems in the cognitive radio cycle by sensing the spectrum correctly, making the correct decision, and taking the right action.

## I.3 Compressive Sensing

In order to sense the wideband radio spectrum, communication systems must use multiple RF frontends simultaneously, which can result in long processing time, high hardware cost, and computational complexity. To address these problems, fast and efficient spectrum sensing techniques are needed. Compressive sensing has been proposed as a low-cost solution for dynamic wideband spectrum sensing in cognitive radio networks to speed up the acquisition process and minimize the hardware cost [20][21]. It consists of directly acquiring a sparse signal in its compressed form that includes the maximum information using a minimum number of measurements and then recovering the original signal at the receiver. Over the last decade, a number of compressive sensing techniques have been proposed to enable scanning the wideband radio spectrum at or below the Nyquist rate. However, these techniques suffer from uncertainty due to random measurements, which degrades their performances [22][23]. To enhance the compressive sensing efficiency, reduce the level of randomness, and handle uncertainty, signal sampling requires a fast, structured, and robust sampling matrix; and signal recovery requires an accurate and fast reconstruction algorithm [20,21,26].

Hence, efficient spectrum sensing and compressive sensing techniques are highly required in order to speed up the wideband spectrum scanning, deal with uncertainty, and perform accurate and reliable sensing occupancy measurements.

## I.4 Dissertation Objectives

This dissertation aims to develop efficient spectrum sensing and compressive sensing techniques that deal with uncertainty and enhance the wideband radio spectrum scanning for cognitive radio systems using software defined radio (SDR) units.

To achieve this goal, the following objectives were pursued:

- Develop efficient spectrum sensing techniques

- Develop fast compressive sensing techniques for high dimensional signal sampling

- Implement and extensively test the proposed techniques through simulations and real-time experiments using SDR units, namely Universal Software Radio Peripheral (USRP).

## I.5 Dissertation Contributions

The contributions of this dissertation are:

1) We contributed to the spectrum sensing techniques by proposing an improved matched filter technique with a dynamic and estimated sensing threshold to increase the sensor detection. we evaluated and compared this technique with the existing techniques based on SNR level, number of samples, and sensing threshold. The approach of selecting the sensing threshold allows better assessing the sensing methods performance than with a fixed threshold. This work was published and presented at the International Conference on Wireless Networks and Mobile Communications Conference (WINCOM'15), Marrakech, Morocco.

2) We provided a deep overview of techniques that deal with uncertainty in the cognitive radio cycle. we examined several models and methods to deal with uncertainty. Among these are probabilistic theory, fuzzy set theory, possibility theory, and evidence theory. These techniques were compared and deeply analyzed in the context of cognitive radio networks. This work was published and presented at the $8^{th}$ IEEE Annual Computing and Communication Workshop and Conference (IEEE CCWC'17), Las Vegas, ND, USA.

3) We provided an in depth survey on compressive sensing techniques. we classified these techniques according to which process they target, namely, sparse representation, sensing matrix, or recovery algorithms. we discussed examples of potential applications of these techniques in spectrum sensing, channel estimation, and multiple-input multiple-output (MIMO) based cognitive radio network. This work was published in the Physical Communication Journal, ELSEVIER, May 2016.

4) We developed a compressive sensing technique that combines the strengths of Circulant matrices and Bayesian models for fast and efficient compressive sensing. we evaluated the efficiency of our technique by comparing its results to the results of the basis pursuit technique using several metrics. These metrics are: mean square error, reconstruction error, correlation coefficients, processing time, recovery time, and sampling time. This work was published and presented at the 8th IEEE Annual Ubiquitous Computing, Electronics and Mobile Communication Conference (IEEE UEMCON'16), New York, NY, USA.

5) We developed a method based on the Bayesian model and Toeplitz matrix. The proposed method was implemented and extensively tested. The simulation results were analyzed and compared to those of the two techniques: basis pursuit and orthogonal matching pursuit algorithms with Toeplitz and random matrix. To evaluate the efficiency of the proposed method, several metrics were used, namely sampling time, sparsity, required number of measurements, recovery time, processing time, recovery error, SNR, and mean square error. This work was published in the International Journal of Communication Systems, Wiley Online Library, February 2017.

6) We performed a real time wideband spectrum scanning using software defined radio units. The scanning is based on Bayesian compressive sensing to reduce the processing time and speed up the spectrum scanning. The proposed method as well as the conventional scanning method were implemented over a wideband range of frequencies in real time. The results of the experiments were analyzed and compared in terms of processing time, number of sensed channels, SNR, occupancy, detection rate, and false detection rate. This work was published and presented at the 9th IEEE Annual Ubiquitous Computing, Electronics and Mobile Communication Conference (UENCOM'17), New York, NY, USA.

7) We implemented and investigated the efficiency of one-bit compressive sensing as one of the compressive sensing paradigm. we extensively tested its performance and compared it to the performance of conventional compressive sensing. Metrics for this comparison include speed, robustness to noise, complexity, and reconstruction success rate. This work was submitted to the 18th IEEE International Conference on Industrial Technology (IEEE ICIT'18), Lyon, France.

## I.6 List of Publications

1. <u>Fatima Salahdine</u>, Hassan El Ghazi, Naima Kaabouch, and Wassim Fassi Fihri, "Matched Filter Detection with Dynamic Threshold for Cognitive Radio Networks," International Conference on Wireless Networks and Mobile Communications (WINCOM'15), Marrakech, Morocco, October 2015.

## I.7 Dissertation Organization

The remaining of this dissertation is organized as follows. In Chapter II, we analyze the existing spectrum sensing techniques and we compare their efficiencies. These techniques are energy, autocorrelation, Euclidean distance, wavelet, and matched filter based detection. In Chapter III, we present a deep review of compressive sensing, including techniques classification, methods, applications, and challenges. In Chapter IV, we propose a spectrum sensing technique based on matched filter detection with dynamic threshold estimation. we compare its efficiency to the existing spectrum sensing techniques. In Chapter V, we propose a compressive sensing technique based on Circulant matrix for signal sampling and Bayesian recovery for signal recovery. we compare its efficiency to the existing techniques in terms of sampling and recovery. In Chapter VI, we propose a Bayesian compressive sensing technique based on Toeplitz sensing matrix. we analyze and compare the efficiency of the proposed technique with the existing techniques based on a number of metrics that cover most of the compressive sensing aspects.

In Chapter VII, we present a real-time spectrum occupancy survey with compressive sensing using SDR units. we analyze the efficiency of the spectrum survey with compressive sensing to the efficiency of the spectrum survey without compressive sensing. In Chapter VIII, we describe the one-bit compressive sensing approach and we compare its performance with the conventional compressive sensing using a number of metrics. In Chapter IX, we conclude with the dissertation objectives and contributions and highlight the dissertation future works. In Appendix A, we analyze and compare the different techniques for dealing with uncertainty in cognitive radio networks. In Appendix B, we presented the SNR estimation technique used in the real time experiments. In Appendix C, we presented a general review about how the underdetermined system can be solved using optimization algorithms.

# Chapter II

# SPECTRUM SENSING

In the previous chapter, we have presented the objectives of this dissertation and my proposed contributions to solve the spectrum management problem. In this chapter, we are focusing on the state-of-the-art of the existing spectrum sensing techniques and their comparisons.

The rest of the chapter is organized as below: Section II.1 discusses the spectrum sensing model as well as the performance evaluation metrics; Section II.2 describes existing spectrum sensing techniques, including energy, autocorrelation, Euclidian distance, wavelet, and matched filter based sensing. Finally, a conclusion is given at the end of the chapter.

## II.1 Spectrum Sensing Model

Spectrum sensing is one of the most important processes performed by cognitive radio systems. It allows the SUs to learn about the radio environment by detecting the presence of the PU signals using one or multiple techniques and decide to transmit or not in its frequency band [1]. The spectrum sensing model can be formulated as:

$$y(n) = \begin{cases} w(n) & H_0: \text{PU is absent} \\ h * x(n) + w(n), & H_1: \text{PU is present} \end{cases} \quad (1)$$

where $n=1….N$, $N$ is the number of samples, $y(n)$ is the SU received signal, $x(n)$ is the PU signal, $w(n)$ is the additive white Gaussian noise (AWGN) with zero mean and variance $\delta_w^2$, and $h$ is the complex channel gain of the sensing channel. $H_0$ and $H_1$ denote respectively the absence and the presence of the PU signal. The PU signal detection is performed using one of the spectrum sensing techniques to decide between the two hypotheses $H_0$ and $H_1$. The detector output, also called the test statistic, is then compared to a threshold in order to make the sensing decision about the PU signal presence. The sensing decision is performed as:

$$\begin{cases} \text{if } T \geq \gamma, & H_1 \\ \text{if } T < \gamma, & H_0 \end{cases} \quad (2)$$

where $T$ denotes the test statistic of the detector and $\gamma$ denotes the sensing threshold. If the PU signal is absent, SU can access to the PU channel. Otherwise, it cannot access to that channel at that time. Figure 4 presents the general model of the spectrum sensing [24][25].

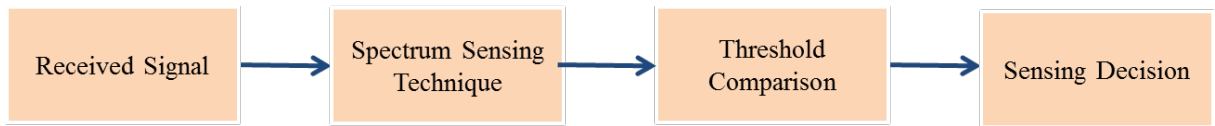

Figure 4: General model of spectrum sensing.

A number of sensing techniques have been proposed in the literature. These techniques are classified into two main categories: cooperative sensing and non-cooperative sensing as illustrated in Figure 5 [1][29].

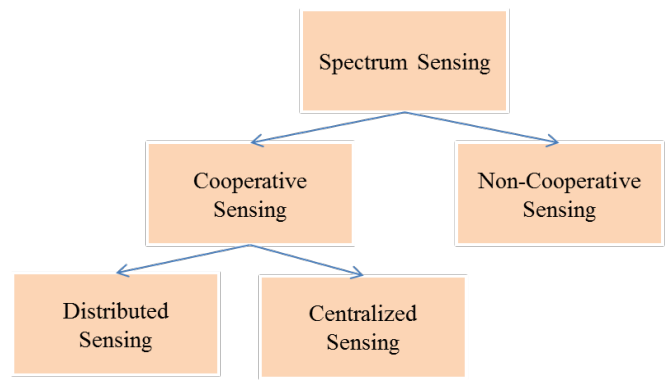

Figure 5: Spectrum sensing classification [29].

Under the non-cooperative sensing category, also called local sensing, each SU seeks for its own objectives and does not take into account other users. As there is no communication or collaboration between the different SUs that sense the same frequency band, the spectrum sensing decision is performed locally. The non-cooperative techniques are simple and do not require high processing time and hardware cost. However, they are subject to errors due to shadowing, fading, interferences, and noise uncertainty. They are mainly adopted when only one sensing terminal is available or when there is no possible communication between SUs.

Under the cooperative spectrum sensing category, the SUs collaborate and coordinate with each other taking into account the objectives of each user to make the final common decision. This cooperation between the different SUs can be divided into two schemes: centralized and distributed schemes. For the distributed scheme, SUs exchange their local observations and sensing results. Each SU takes its own decision taking into account the received results from the other SUs sensing the same frequency band. This approach does not require any common infrastructure for the final decision and the detection is controlled by the SUs. For the centralized scheme, all the SUs send their sensing results to a central unit, called fusion center, as illustrated in Figure 6.

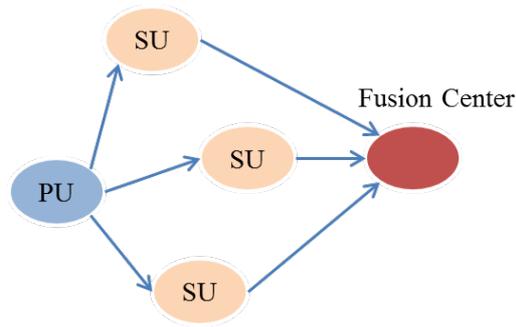

Figure 6: Centralized Cooperative spectrum sensing.

The fusion center decides about the spectrum access based on the received observations. The decision can be soft or hard combining decision with AND/OR rules. Under both spectrum sensing categories, SUs can perform the sensing using a spectrum sensing technique [30][29].

## II.2 Spectrum Sensing Techniques

A number of spectrum sensing techniques have been proposed to identify the presence of the PU signal transmission. These techniques provide more spectrum utilization opportunities to the SUs with no interferences or intrusive to the PUs. Examples of these techniques are presented in Figure 7, namely energy, autocorrelation, Euclidian distance, wavelet, and matched filter based sensing [6][7].

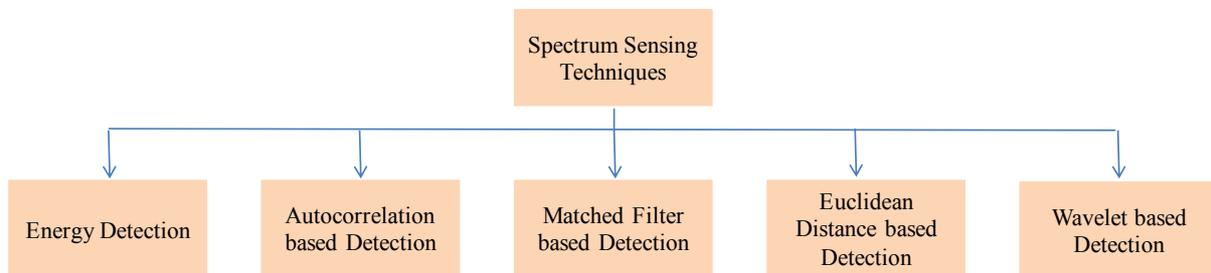

Figure 7: Examples of spectrum sensing techniques.

### II.2.1 Energy detection

Energy detection is the simplest sensing technique, which does not require any information about the PU signal to operate. It performs by comparing the received signal energy with a threshold. The threshold depends only on the noise power. The decision statistic of an energy detector can be calculated from the squared magnitude of the FFT averaged over $N$ samples of the SU received signal as illustrated in Figure 8 [24,25,30].

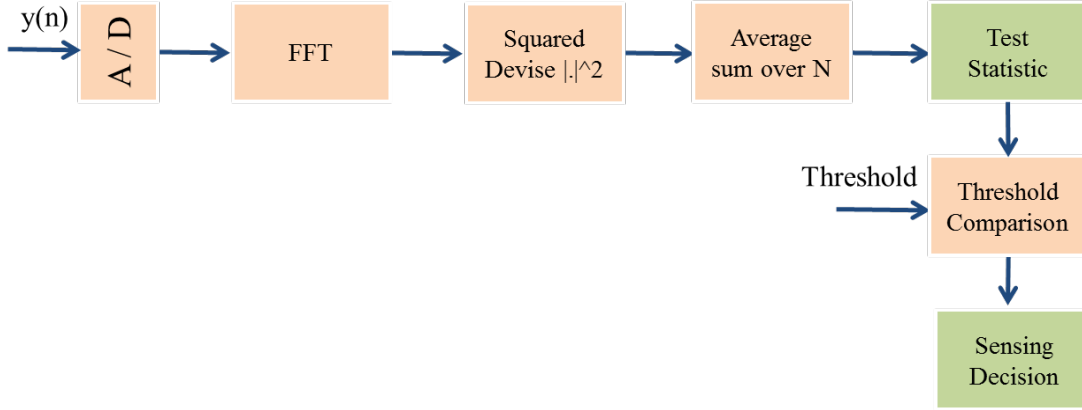

Figure 8: Energy detection model.

The detector output is the received signal energy as given by

$$T_{ED} = \sum_{n=0}^{N} y(n)^2 \qquad n=1\ldots N, \qquad (3)$$

where $N$ is the sample number, and $y(n)$ is the SU received signal, and $T_{ED}$ is the test statistic. Thus, the decision-based energy detection can be expressed as:

$$\begin{cases} \text{If } T_{ED} \geq \lambda, & \text{PU signal is present} \\ \text{If } T_{ED} < \lambda, & \text{PU signal is absent} \end{cases} \qquad (4)$$

where $\lambda$ denotes the sensing threshold. The received signal can be approximated by a Gaussian random signal. Based on the central limit theorem, when the number of samples exceeds 250 ($N>250$), the test statistic has a central chi-square distribution with $N$ degrees of freedom for $H_0$ hypothesis, while it has a non-central chi-square distribution with $N$ degrees of freedom for $H_1$ hypothesis[28]. Thus, the test statistic, $T_{ED}$, is approximated as Gaussian and given by

$$\begin{cases} H_0: T_{ED} \sim \mathbb{N}(N\delta_w^2, 2N\delta_w^4) \\ H_1: T_{ED} \sim \mathbb{N}(N(\delta_w^2 + \delta_s^2), 2N(\delta_w^2 + \delta_s^2)^2) \end{cases} \qquad (5)$$

where $\delta_s^2$ denotes PU signal variance, $\delta_w^2$ denotes the noise variance, and $\mathbb{N}$ denotes the normal distribution. For evaluation metrics, probability of detection, $P_d$, and probability of false alarm, $P_{fd}$, for additive white Gaussian noise (AWGN) channel can be expressed respectively as:

$$Pd = Q\left(\frac{\lambda - N((\delta_w^2 + \delta_s^2))}{\sqrt{2N((\delta_w^2 + \delta_s^2))^2}}\right) \quad , \quad Pfd = Q\left(\frac{\lambda - N\delta_w^2}{\sqrt{2N\delta_w^4}}\right) \qquad (6)$$

where $Q(.)$ denotes Q-function and $\lambda$ denotes the threshold [25][30]. $P_d$ and $P_{fd}$ are formulated as a function of *SNR*

$$P_d = Q\left(\frac{\bar{\lambda}-N(1+\gamma)}{\sqrt{2N(1+\gamma)^2}}\right) \qquad , \qquad P_{fd} = Q\left(\frac{\lambda - N\delta_w^2}{\sqrt{2N\delta_w^4}}\right) \qquad (7)$$

where $\gamma$ denotes *SNR* and $\bar{\lambda}$ denotes the average threshold, $\bar{\lambda} = \lambda/\delta_w^2$. Thus, the sensing threshold depends on the noise power and it is expressed for a target $P_{fd}$ as:

$$\lambda = (Q^{-1}(P_{fd})\sqrt{2N} + N)\delta_w^2 \qquad (8)$$

Each threshold value corresponds to a pair of ($P_d$, $P_{fd}$), representing what called the receiver operating curve (ROC). ROC represents the plotting of the correct detection rate as a function of the false detection rate for several thresholds [9,26,32-36].

The sensing threshold for energy detector is an important parameter. When a detector does not adjust its threshold properly, it suffers from some performance degradation of the spectrum sensing. Various approaches were suggested for energy detection technique [38-44]. As the sensing performance is highly affected by the estimation error of the noise power, a dynamic estimation of the noise power is recommended in [38]. Adaptive threshold control is implemented with linear adaptation on the threshold based on SINR. This approach attains a considerably higher SU throughput than the fixed threshold approach, but maintains unacceptable chances of false alarms [39].

The authors in [40] presented an adaptive threshold in unknown AWGN with noise power estimation, keeping the false alarm rate at a preferred point under any noise level. This technique is based on a concept of dedicated noise estimation channel in which only noise is received by SUs. An improved energy detection is proposed in [41] where misdetection of PU transmission due to a sudden drop in PU transmission power is addressed by keeping an additional updated list of the latest fixed number of sensing events that are used to calculate an average test statistic value. A double-threshold technique is proposed in [42] with the intention of finding and localizing narrowband signals. Another technique is presented in [43] based on wideband spectrum sensing, which senses the signal strength levels within several frequency ranges to improve the opportunistic throughput of SU and decreases the interference to PU. In [44], the authors proposed an improved energy detection with an adaptive threshold to increase the detection rate.

**II.2.2 Autocorrelation based Detection**

This technique is based on the value of the autocorrelation coefficient of the received signal. It exploits the existing autocorrelation features in the transmitted signal and not in the noise [30]. The autocorrelation function is defined as:

$$R_{x,x}(\tau) = \int_{-\infty}^{+\infty} x(t)\, x^*(t-\tau)dt \qquad (9)$$

where $\tau$ is the lag, $t$ is time, $x^*$ is the complex conjugate of the signal, and $R_{x,x}$ is the autocorrelation function of the signal. In spectrum sensing, sensing quality is affected by the noise level and it is difficult to interpret the signals affected by the Gaussian noise [46][47]. In fact, white noise is uncorrelated and its autocorrelation function results in a sharp spike at zero lag while the rest of lags are close to zero as illustrated in Figure 9.

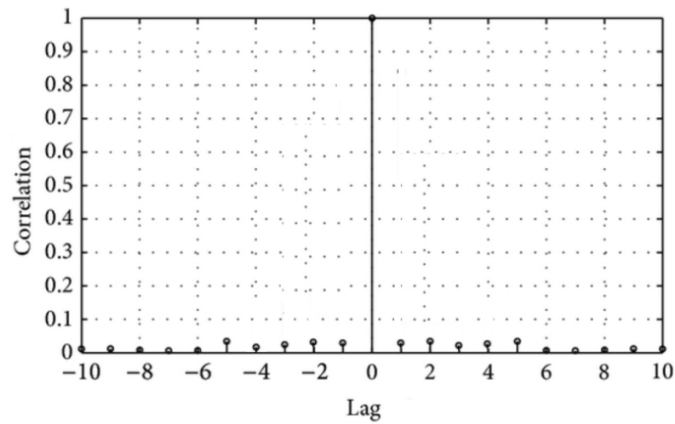

Figure 9: Autocorrelation function of a white noise [7][31].

However, $R_{x,x}(\tau)$ can present some high values depending on the transmitted stream proprieties. The transmitted signal is correlated; the zero lag and the first lag are very close as illustrated in Figure 10.

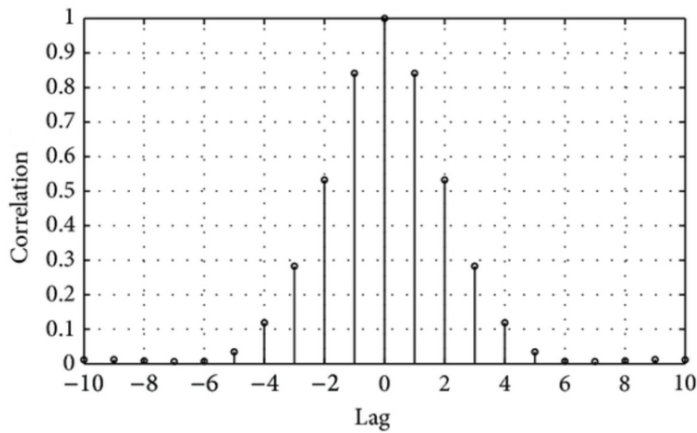

Figure 10: Autocorrelation function of an FM signal [7][31].

Therefore, the autocorrelation of the noise is uncorrelated and the autocorrelation of the signal is correlated as illustrated in Figure 9 and Figure 10. When the degree of correlation is higher, the strength of the signal is higher [31][45]. Thus, the spectrum sensing is performed by exploiting the autocorrelation function to detect the PU signal presence under noise as illustrated in Figure 11.

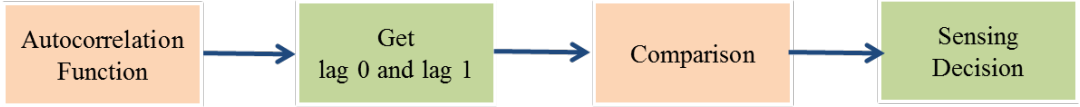

Figure 11: Autocorrelation based sensing model.

The sensing decision is based on the knowledge of the statistical distribution of the autocorrelation function. For random noise, the first lag of the autocorrelation is very small or negative, but when there is a signal the autocorrelation at the first lag represents a significant value. Thus, the sensing method consists on comparing *lag0* and *lag1* of the autocorrelation function of the SU received signal [48]. The sensing decision is expressed as:

$$\begin{cases} \text{if } lag0 \approx lag1 & \text{, PU signal is present} \\ \text{if } lag0 \gg lag1 & \text{, PU signal is absent} \end{cases} \quad (10)$$

The threshold is the margin between the two lag values. For instance, if *lag0* is superior to *lag1* by a value of $\lambda_{Aut}$ %, this value, $\lambda_{Aut}$, is the autocorrelation threshold. Autocorrelation based sensing is able to differentiate between signals and noise, which makes it less sensitive to noise uncertainty. It depends on the autocorrelation features and its performance is limited by the hardware based fractional frequency offset (FFO) for practical implementation; however, it is easy to implement and does not require high computing power [48].

### II.2.3 Euclidian Distance based Detection

Euclidean distance based detection is a new sensing method proposed in [7]. It is mainly based on the autocorrelation of the SU received signal. This detector performs by computing the Euclidean distance between the autocorrelation of the signal and a reference line. The autocorrelation of the received signal, $R_{x,x}(\tau)$, can be presented as the mean of

$$R_{x,x}(\tau) = \sum_{n=1}^{N} x(n)\, x^*(n-\tau) \quad (11)$$

where *N* is the number of samples. The reference line refers to the equation:

$$R = \frac{2}{M} e + 1 \quad (12)$$

where $R$ is the reference line, $M$ denotes the number of lags of the autocorrelation including positive and negative values, and $0 \leq e \leq \frac{M}{2}$. The Euclidean distance, $D$, is the difference between the reference line and the signal autocorrelation [31][45]. It can be expressed as:

$$D = \sqrt{\sum (R_{x,x}(\tau) - R)^2} \qquad (13)$$

The sensing is then performed by comparing this metric with a threshold as illustrated in Figure 12.

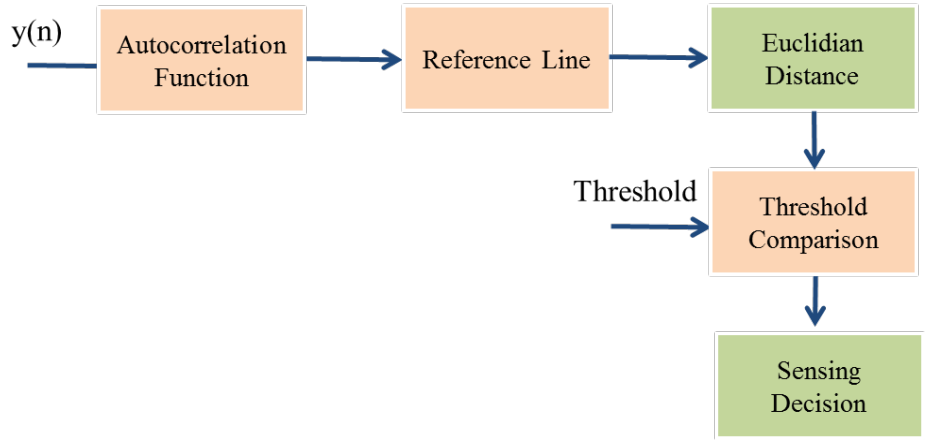

Figure 12: Euclidean distance based sensing model.

The sensing decision is expressed as:

$$\begin{cases} If\ D \geq \lambda, & PU\ signal\ is\ absent \\ If\ D < \lambda, & PU\ signal\ is\ present \end{cases} \qquad (14)$$

where $\lambda$ denotes the sensing threshold. Euclidean distance based sensing is more efficient than the autocorrelation based sensing in terms of the detection success rate [47][48].

### II.2.4 Wavelet based Sensing

Wavelet based sensing, also called edge detection, is based on the continuous wavelet transform, which allows finding the signal decomposed coefficients with the help of a basis [50][51]. For a given signal $s(t)$, the continuous wavelet function, $\psi(t)$, operates in time domain where it includes a certain range and zeros elsewhere. It is given by

$$f(u,v) = <x(t), \psi_{u,v}> = \int_{-\infty}^{+\infty} x(t)\psi_{u,v}^*(t)dt \qquad (15)$$

where $u$ is the scaling parameter greater, $v$ is the translating parameter, and $\psi_{u,v}(t)$ is the basis. The wavelet transform allows mapping from the one dimensional signal to two dimensions coefficients *f(u,v)*. The frequency-time analysis can be performed at frequency corresponds to parameter, *v*, at time instant corresponds to parameter, *u*. The wavelet based sensing is operated by computing the continuous wavelet transform of the signal to perform the power spectral density. The local maximum of the power spectral density corresponds to the edge, which is compared to a threshold to decide about the spectrum occupancy as illustrated in Figure 13.

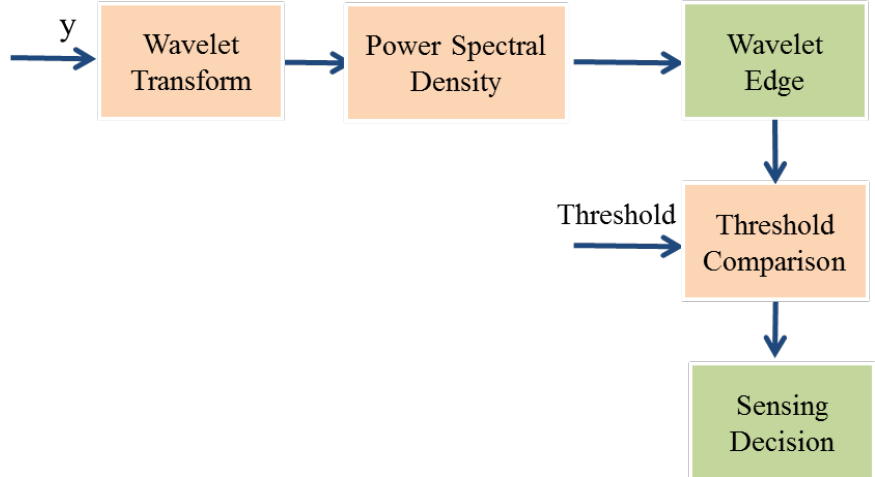

Figure 13: Wavelet based sensing system model.

The sensing decision is expressed as:

$$\begin{cases} \text{If } m \geq \lambda, & PU\ signal\ is\ absent \\ \text{If } m < \lambda, & PU\ signal\ is\ present \end{cases} \quad (16)$$

where *m* is the wavelet edge. The wavelet edge is adopted for sensing decisions because the power density of a given signal corresponds to one spike at the signal frequency while it corresponds to multiple spikes when noise is added as illustrated in Figure 14.

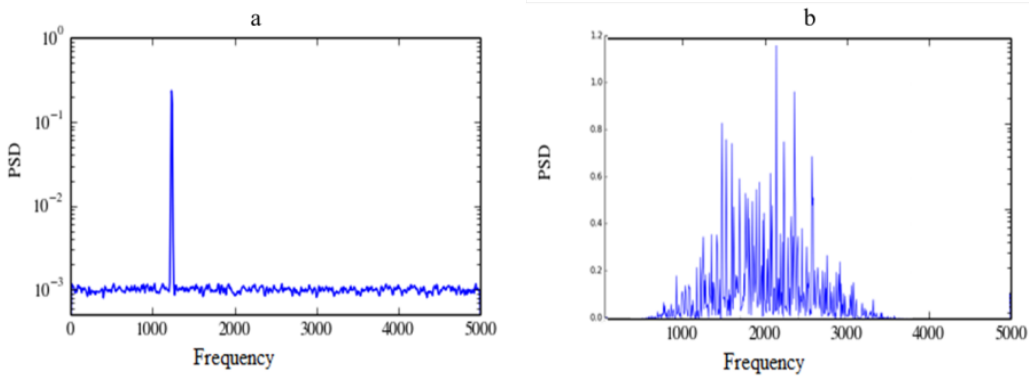

Figure 14: Power density function of (a) noiseless signal and (b) noisy signal (Hz) [50][51].

It has been demonstrated that the wavelet based sensing can be adopted to identify the number of occupied bands over a frequency range, however, it requires high processing time [50]. Thus, wavelet based sensing can easily distinguish between the signal and the noise while deciding about the spectrum occupancy. It can also identify how many frequency bands are utilized. It represents less complexity but requires high processing time.

### II.2.5 Matched Filter Detection

Matched filter detector is a coherent pilot sensor that maximizes the *SNR* at the output of the detector. It is an optimal filter that requires the prior knowledge of the PU signals. This sensing technique is the best choice when some information about the PU signal are available at the SU receiver. Assuming that the PU transmitter sends a pilot stream simultaneously with the data, the SU receives the signal and the pilot stream. Matched filter detection is performed by projecting the received signal in the direction of the pilot, *xp*, as illustrated in Figure 15 [30,48,50].

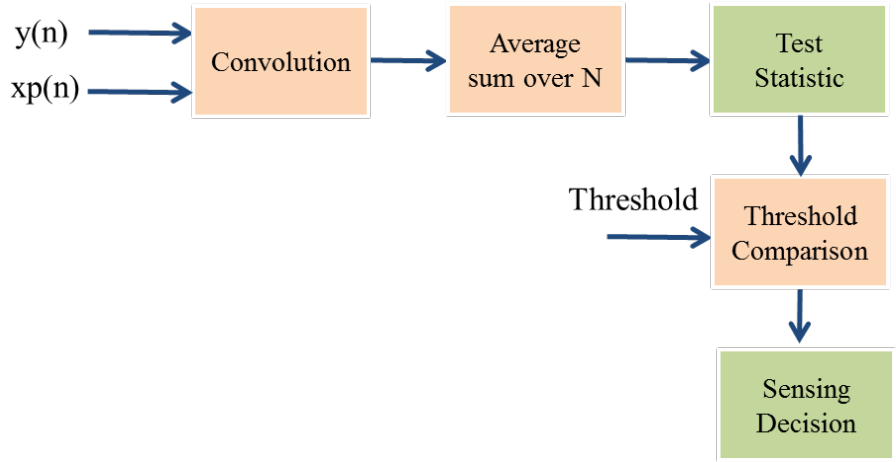

Figure 15: Matched filter detection model [7][31].

The test statistic is expressed as:

$$T_{MFD} = \sum_N y(n)\, x_p(n) \qquad (17)$$

where $x_p$ denotes the PU signal, $y$ denotes the SU received signal, and $T_{MFD}$ denotes the test statistic of the matched filter detector. The test statistics, $T_{MFD}$, is then compared with a threshold in order to decide about the spectrum availability. The SU received signal, as well as the PU signal, are approximated to be random Gaussian variables. As a linear combination of Gaussian random variable, $T_{MFD}$ is also approximated as a Gaussian random variable.

$$\begin{cases} \text{If } T_{MFD} \geq \lambda, & \text{PU signal is present} \\ \text{If } T_{MFD} < \lambda, & \text{PU signal is absent} \end{cases} \quad (18)$$

Based on the Neyman-Pearson criteria, the probability of detection and the probability of false alarm are given by

$$P_d = Q\left(\frac{\lambda - E}{\sqrt{E\delta_w^2}}\right) \quad , \quad P_{fd} = Q\left(\frac{\lambda}{\sqrt{E\delta_w^2}}\right) \quad (19)$$

where $E$ is the PU signal energy, $\lambda$ is the sensing threshold, $Q(.)$ is the $Q$- function, and $\delta_w^2$ is the noise variance. The sensing threshold is expressed as a function of the PU signal energy and noise variance.

$$\lambda = (Q^{-1}(Pfd))\sqrt{E\,\delta_w^2} \quad (20)$$

Assuming that the signal is completely known is unreasonable and impractical. Some communication systems contain pilot stream or synchronization codes for channel estimation and frequency band sensing. A novel hybrid matched filter structure is proposed in [53], based on traditional matched filter by mixing segmented and parallel matched filter to overcome the frequency offset sensitivity. This new structure allows balancing between the sensing time and the hardware complexity. As both carrier frequency offset (CFC) and phase noise (NP) demean the sensing performance of matched filter detection, matched filter detection performance is examined in the presence of CFC and PN in [54]. Robust sensing technique is proposed to overcome the negative impact of CFC and NP on the sensing performance.

On the other hand, the sensing threshold for matched filter detector is an important parameter as for the other sensing techniques in which a number of researchers treated the threshold selection. In [55], the matched filter detection has been used with a static value to decide about the spectrum occupancy. In [56], each pair of ($P_d$, $P_{fd}$) is associated with a particular threshold to make sensing decision. In other research works, the sensing threshold is determined dynamically by multiplying the theoretical threshold by a positive factor [18]. Others do not mention how the threshold was selected. Thus, with a static threshold, the sensing decision is not reliable because of the noise uncertainty. Therefore, the performance of the matched filter based detection depends mainly on the available information about the PU signal, including the bandwidth, central frequency, and modulation scheme. The sensing performance degrades when these data are incorrect or uncertain.

**II.2.6 Evaluation Metrics**

To evaluate the performance of a sensing technique, a number of metrics have been proposed, including $P_d$, $P_{fd}$, and $P_{md}$. $P_d$ is the probability that the SU declares the presence of the PU signal when the spectrum is occupied [3][32]. It is expressed as:

$$P_d = \text{Prob } (H_1/H_1) \tag{21}$$

where $H_0$ and $H_1$ denote respectively the absence and the presence of the PU signal. The higher the $P_d$, the better the PU protection is.

The probability of false alarm, $P_{fd}$, is the probability that the SU declares the presence of the PU signal when the spectrum is actually free (idle). It is expressed as:

$$P_{fd} = \text{Prob } (H_1/H_0) \tag{22}$$

The lower the $P_{fd}$, the more the spectrum access the SUs will obtain. The probability of miss detection, $P_{md}$, is the probability that the SU declares the absence of a PU signal when the spectrum is occupied. It is given by

$$P_{md} = \text{Prob } (H_0/H_1) \tag{23}$$

These three metrics measure the efficiency of the spectrum sensing techniques and can be expressed as:

$$P_d + P_{fd} + P_{md} = 1 \tag{24}$$

There is a tradeoff between the probability of false alarm and the probability of miss detection. False detection of PU activity causes interferences to the PU and missed detection of the PU activity misses spectrum opportunities. This tradeoff can be expressed as conservative with $P_{fd}$ and aggressive with $P_{md}$, and a spectrum sensing technique has to fulfill the constraints on both probabilities [3].

## II.3 Conclusion

In this chapter, we presented the spectrum sensing and its categories. we described several spectrum sensing techniques and their characteristics, namely energy, autocorrelation, Euclidian distance, wavelet, and matched filter based detection. We also presented the different metrics used for performance evaluation.

In the next chapter, we are going to present the compressive sensing solution and its techniques.

# Chapter III

# COMPRESSIVE SENSING

In the previous chapter, we have discussed deeply the state-of-the-art of the spectrum sensing techniques. In this chapter, we focus on the compressive sensing as a low cost solution used to overcome the spectrum sensing problems. A number of papers related to compressive sensing have been published. However, most of them describe compressive sensing techniques corresponding to one process either sparse representation, sensing matrix, or recovery. Other papers focus on one of the compressive sensing categories; and a few focus on the applications. Thus, there is a great need for detailed review papers that compare and analyze the current compressive sensing techniques deeply. Therefore, we provide in this chapter an in depth survey on compressive sensing techniques and classifies these techniques according to which process they target, namely, sparse representation, sensing matrix, or recovery. It also discusses examples of potential applications of these techniques.

The remainder of the chapter is organized as follows: Section III.1 discusses the related works and the motivation behind proposing the compressive sensing solution; Section III.2 presents the compressive sensing theory; Section III.3 discusses the sensing matrix techniques; Section III.4 presents the recovery algorithms, Section III.4 discusses the different applications and challenges of the compressive sensing; Finally, a conclusion is given at the end.

## III.1 Introduction

Existing spectrum sensing techniques, also called non-compressive sensing detectors, are based on a set of observations sampled by an analog/digital converter (ADC) at the Nyquist rate [57-59]. However, these techniques can sense only one band at a time because of the hardware limitations on the sampling rate. In addition, in order to sense a wideband spectrum, the band is divided into narrow bands or multiple frequency channels. The SU has to sense each band using multiple RF frontends simultaneously, which can result in a very high processing time, hardware cost, and computational complexity. To overcome these limitations, compressive sensing has been proposed to reduce the processing time and speed up the spectrum scanning process. It allows reducing the number of samples required for high dimensional signal acquisition while keeping the important information [20].

In cognitive radio, this approach is applicable because of the signal sparsity feature, which is valid in most of spectrum sensing scenarios. It was also applied on channel estimation to overcome the limitations of the existing channel estimation techniques, known as non-compressive sensing channel estimation techniques [20]. These techniques represent high complexity and often require the knowledge of the channel response at the receiver for multipath wireless channels, which is not always possible in practice because of the high number of required antennas. Moreover, multichannel signals are known to be sparse, which can lead to high estimation errors at the receiver. In order to overcome these limitations and exploit the sparsity of the multipath wireless channels, compressive sensing channel estimation techniques have been proposed and investigated, especially for MIMO-OFDM communication systems [20].

A number of papers related to compressive sensing have been published. However, most of them describe compressive sensing techniques corresponding to one process either sparse representation, sensing matrix, or recovery. Other papers focus on one of the compressive sensing categories, and a few focus on the applications. Thus, there is a need for detailed review papers that compare and analyze the current compressive sensing techniques. Therefore, this chapter provides a detailed overview on compressive sensing techniques and classifies these techniques according to which process they target. The chapter also discusses each category, some techniques under each category, and provides a deep comparison of these categories. In addition, some compressive sensing applications are discussed.

## III.2 Compressive Sensing Theory
### III.2.1 General Model

The concept of compressive sensing is firstly introduced by Candes as a new approach to sample signals at or below the Nyquist rate [20]. Traditional approaches were based on the Shannon-Nyquist theorem, in which it is possible to recover a signal at the receiver only if it is sampled at the Nyquist rate. The compressive sensing mechanism is an acquisition process followed by a reconstruction algorithm. It combines sampling and sensing processes in one process. It involves three main processes: sparse representation, encoding, and decoding as illustrated in Figure 16. During the sparse representation, a sparse representation of the signal is found over a basis that permits to reconstruct the signal as accurately as possible. In the next process, the sparse signal is sampled and compressed based on a measurements matrix, $M_c$, of $M$x$N$ elements to extract *M* samples from *N* of the signal, *x*, where *M*<<*N*. In the last process, the compressed signal can be then recovered at the receiver using a recovery algorithm.

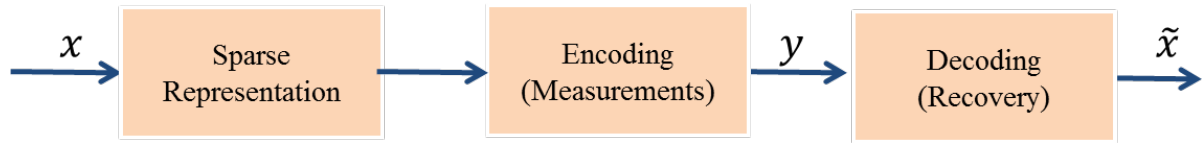

Figure 16: Compressive sensing model [20].

Considering a high dimensional signal, *x*, with a high number of samples *N*, *x* is assumed to be sparse in some domain with *k*-sparsity where *k*<<*N*. Sparse representation consists of representing the signal by a number of projections on a suitable sparse basis, $\varphi$, also known as dictionary or projection. Examples of sparse basis include wavelet transform, Fourier transform, and discrete cosine transform. Non sparse signal can be represented as a sparse signal by some sparse transforms [17][50]. Every sparse signal can be represented in a scarifying basis $\varphi$ as:

$$x = \varphi \, s \qquad (25)$$

where *s* is the signal's projection on the sparse basis $\varphi(NxN)$, $\|s\|_0 = k \ll N$, and $\|x\|_m = \sqrt[m]{\sum_i |x_i|^m}$ is the $\mathcal{L}_m$ norm [60][61]. The sparse signal, *x*, is compressed using the sensing matrix, $\Phi$. The compression consists on multiplying the signal with an *MxN* matrix. *M* is much smaller than *N* and it represents the measurements number that includes the essential information of *x*. For efficient recovery, *M* should respect this condition $M = O(k \log(N))$ for random measurements. The compressed signal is given as:

$$y = \Phi \, x \qquad (26)$$

where *y* (*M*, 1) denotes the signal measurements, which selects only *M* samples from *x* (*N*, 1), *M*<<*N*. $\Phi$ is the sensing matrix (*M*, *N*) as shown in Figure 17.

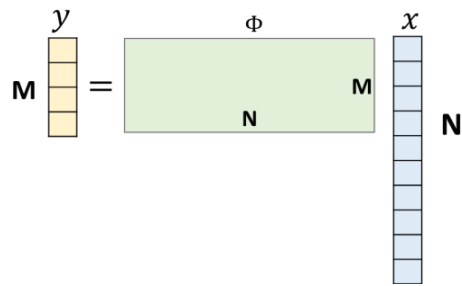

Figure 17: Compressive sensing structure [20].

From equations (25) and (26), compressed signal can be reformulated as:

$$y = \Phi\, x = A\, s \tag{27}$$

where $A = \Phi\, \varphi$ is an *MxN* matrix, known as recovery matrix, it satisfies the restricted isometry property. The last step consists in reconstructing the signal at the receiver, which implies solving the equation (27). It aims to solve an underdetermined system with more unknowns than the number of equations. Because of the sparsity assumption, it is possible to estimate a high dimensional signal *N* from few measurements *M*. To solve this underdetermined system and approximate *x* coefficients, the problem is considered as an optimization problem. Various recovery algorithms have been developed to solve the following optimization problem.

$$\tilde{x} = \underset{y=\Phi x}{\mathrm{argmin}} \|x\|_1 \quad \text{Subject to} \quad y = \Phi\, x \tag{28}$$

where the recovered signal $\tilde{x}$ is the sparsest solution from many possible solutions of the optimization problem. $\|x\|_1 = \sum_i |x_i|$ is the $\mathcal{L}_1$ norm of *x* and represents the sum of the absolute values of *x* coefficients [61][62]. Considering the noise, two cases are raised, noisy case and noiseless case. In practical applications, random noise is added to the acquired measurements during the signal processing. Hence, with the noisy measurements, equation (27) can be reformulated as:

$$y = \Phi\, x + w \tag{29}$$

where *w* is a random noise vector that needs to be estimated during the recovery process. Taking noise into account, the optimization problem can be reformulated as:

$$\tilde{x} = \underset{y=\Phi x+w}{\mathrm{argmin}} \|x\|_1 \quad \text{Subject to} \quad y = \Phi\, x + w \tag{30}$$

A number of recovery algorithms have been proposed to solve the linear programming problem, such as $\mathcal{L}_1$ norm minimization, gradient descent, iterative thresholding, matching pursuit, and orthogonal matching pursuit [22,61-68]. An example of techniques that reconstruct the original signal with noise is described in [70].

### III.2.2 Compressive Sensing Requirements

Determining the sampling matrix to use for compression and which solver to apply depend on some conditions and requirements including sparsity, restrict isometry property, coherence, and measurements number. For sparsity, a sparse signal is a signal with a limited number of non-zero elements and most of its elements are zero or with very low power. A signal with *N* samples is *k*-sparse implies that it has only *k* non-zero coefficients and (*N* - *k*) zero elements, where *N* is greater than *k*. Sparse representation is a major requirement for compressive sensing and ensures the signal compressibility. It consists in presenting the signal in some domain with only the essential information acquired. For the restricted isometry property, it is a characteristic of orthonormal matrices bounded with a restricted isometry constant, which is a positive number between 0 and 1. A matrix that satisfies this property in order *k* implies that

$$\exists \delta \in (0,1) \ / \ (1-\delta)\|x\|_2^2 \leq \|\Phi x\|_2^2 \leq (1+\delta)\|x\|_2^2 \tag{31}$$

where $\delta$ is the restricted isometry constant, $\Phi$ is the sensing matrix, and *x* is the original signal [71]. This property allows guarantying the uniqueness of the recovered solution, $\tilde{x}$, and it has to be considered during the matrix design. When a matrix satisfies the RIP, it guarantees that the underdetermined system solution is unique and robust. Fourier, random Gaussian, and Bernoulli matrices are examples of matrices that satisfy the RIP property [22,71-73].

For the coherence property, it examines the sensing matrix quality and evaluates its efficiency. The mutual coherence of two matrices $\Phi$ and $\varphi$ measures the maximal correlation between any two elements of them. It can be computed as follows:

$$\mu(\Phi, \varphi) = \sqrt{N} \max_{1 \leq i,j \leq N} |\langle \Phi_i | \varphi_j \rangle| \tag{32}$$

where $\mu(\Phi, \varphi)$ denotes the coherence between the two matrices $\Phi$ and $\varphi$ with $1 < \mu(\Phi, \varphi) < \sqrt{N}$. High coherence corresponds to high correlation between $\Phi$ and $\varphi$ elements, which implies that the compressive sensing technique requires more measurements to perform. Compressive sensing requires that $\Phi$ is incoherent with $\varphi$. The small value of coherence means few measurements are required for signal reconstruction. The coherence of the sensing matrix can be defined also as $\mu(\Phi)$, which represents the largest value of correlation between any two normalized columns of $\Phi$. It is given as:

$$\mu(\Phi) = \max_{1 \leq i \neq j \leq N} |\langle \Phi_i | \Phi_j \rangle| \tag{33}$$

where $\mu(\Phi)$ represents the coherence and $\Phi_i$ and $\Phi_j$ represent two columns of $\Phi$ and. The smaller number of measurements necessitates the lower value of coherence [62]. For the number of measurements, it depends on the sensing matrix. Each sensing matrix requires a specific number of measurements to perform well [20].

### III.2.3 Classification of Compressive Sensing Techniques

Compressive sensing techniques can be classified into two main categories: distributed compressive sensing and jointly compressive sensing [75]. Distributed compressive sensing is performed based on two processes by sampling separately and recovering jointly. In other words, the measurements are acquired independently by each node. Each node samples its signal, $x_i$, using its sensing matrix, $\Phi_i$, to get its signal measurements, $y_i$. The same reconstruction algorithm is then applied jointly to recover all the original signals, $x_i$, where we $=1\ldots J$, $J$ is the number of nodes [76] as illustrated in Figure 18. Distributed compressive sensing enables distributed sampling of the received signals to reduce data storage and measurements rates. Examples of distributed techniques include those described in [76][77].

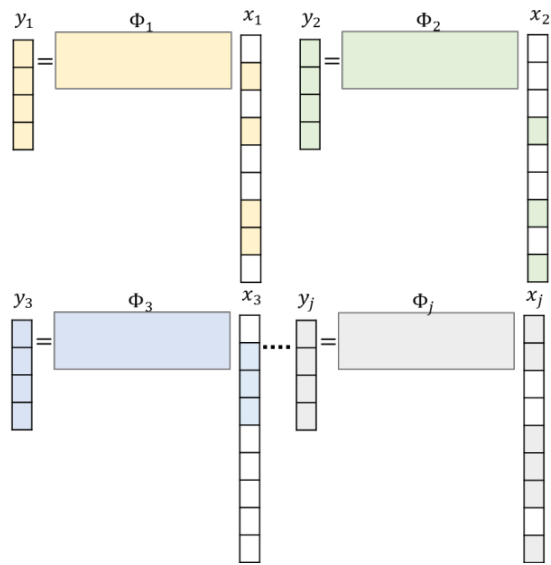

Figure 18: Distributed compressive sensing [20].

Jointly compressive sensing is based on the joint concept, in which measurements and reconstruction processes are performed at the same process using the same sensing matrix. It allows recovering jointly sparse signals as illustrated in Figure 19.

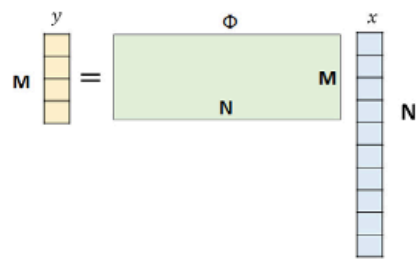

Figure 19: Jointly compressive sensing [20].

As the same compressive sensing techniques are involved for both categories, the classification of techniques is based on which process is involved, namely, sampling matrix process or recovery process. Examples of sampling matrix techniques and recovery techniques are presented in Figure 20.

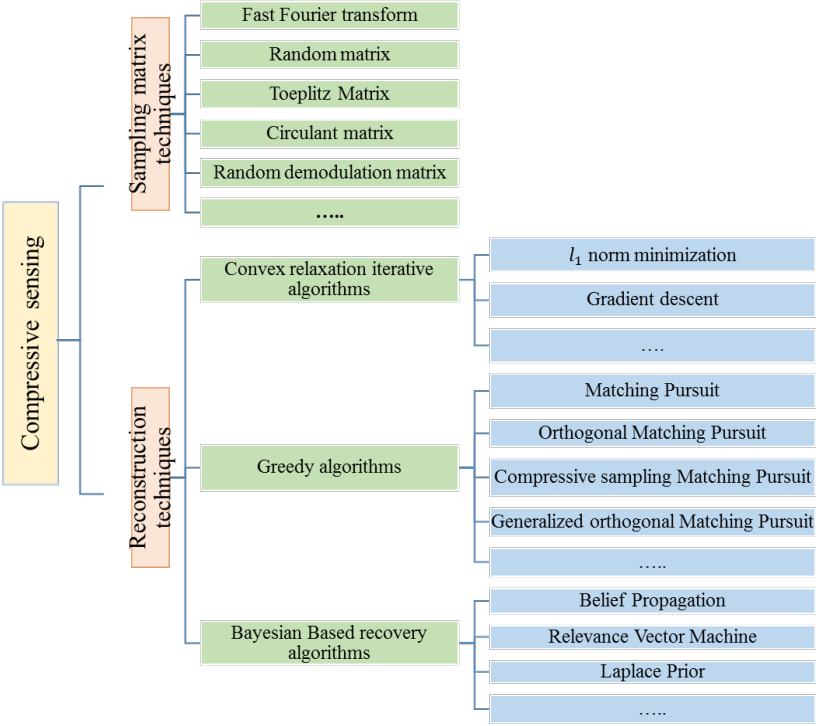

Figure 20: Classification of compressive sensing techniques [20].

## III.3 Sampling Matrix Techniques

Sensing matrix, also called sampling matrix or measurements matrix, is the matrix used to sample a sparse signal below the Nyquist rate. It is designed to reduce the number of measurements as much as possible while preserving the main information included in the signal without losing any important data after recovery process. It must be known, independent of the original signal, stable, incoherent with the sparsity basis matrix, and satisfies the restrict isotropy property. Several matrices are adopted in compressive sensing as sensing matrix that respects the compressive sensing requirements.

Efficient compressive sensing requires suitable sensing matrix as the matrix choice is extremely important. Determining which matrix to use to acquire and reconstruct the signal at the receiver is a fundamental process. It might be based on some criteria including recovery complexity, speed, universality, the number of measurements required, and hardware cost. Universality implies that the signal can be sparse in any domain or any basis. Examples of the most used sensing matrices are matrix randomly sampled from Fourier transform or Walsh-Hadamard transform, random matrix, Toeplitz matrix, Circulant matrix, random demodulation matrix, Gaussian or Bernoulli matrix, deterministic matrix, random convolution matrix, and discrete cosine transform. These sensing matrices are described and compared in Table 1. Other sensing matrices have been proposed as an extension of the listed matrices over [79].

Table 1: Compressive sensing matrices comparison.

| Matrix | Concept | Strengths | Weaknesses |
| --- | --- | --- | --- |
| Fourier transform or Walsh-Hadamard transform | Randomly selected rows of a discrete Fourier transform matrix (DFT) | -Easy to implement<br>-Fast recovery | -Not universal<br>-High uncertainty<br>-Not accurate |
| Random matrix | Random Projection | -Recovery with high efficiency<br>-Easy to implement<br>-Universal | -$M$ knowledge required<br>-More memory required |
| Toeplitz matrix, Circulant matrix | The entries are independently distributed in one row | -Randomness is reduced<br>-Fast acquisition and recovery | Not universal |
| Random demodulation matrix | Pseudorandom binary sequences are used to modulate the input signal | -Easy to implement<br>-Works in noiseless and noisy signals<br>-Universal | -High Uncertainty<br>-Not accurate |
| Gaussian or Bernoulli matrix | Independent and identically distributed (i.i.d) | -Simple and easy to implement<br>- Universal | -Costly hardware implementation |
| Deterministic matrix | Deterministic matrix | -Fast<br>-Reduced memory usage | -Requires high number of measurements |
| Random convolution matrix | Convolution product | -Simple and easy to implement<br>-Universal | -Can only recover noiseless signals<br>-More measurements required |
| Discrete cosine transform (DCT) | Matrix formed from a random subset of $M$ rows | Simple and easy to implement | -More $M$ required<br>-Missing data recovery |

As one can see from Table 1, random matrices and matrices formed from a random subset of Fourier transform matrix or discrete cosine transform are slow but easy to implement. Toeplitz matrices and Circulant matrices are able to reduce the randomness compared to others. Gaussian and Bernoulli matrices are also simple to implement, but their hardware implementation is expensive. Deterministic matrices require the number of measurements to be more than the expected threshold, but random convolution matrices require less measurement.

## III.4 Recovery Algorithms

Recovery algorithms allow reconstructing a compressed signal from few measurements. Adopting these algorithms requires the RIP and coherent properties to be satisfied. A number of recovery algorithms have been proposed to recover the signal by solving the previously described optimization problem. Recovery algorithms can be classified into three classes: convex relaxation iterative [19,21,80], Greedy [68,81-87], and Bayesian recovery [49,77-79].

### III.4.1 Convex Relaxation Iterative Algorithms

The convex relaxation category is the process used to solve the underdetermined system via linear programming system. Examples of convex techniques include $\mathcal{L}_0$ norm minimization, $\mathcal{L}_1$ norm minimization, and $\mathcal{L}_2$ norm minimization. Since $\mathcal{L}_0$ norm represents the number of non-zero elements of the signal, this solution can recover the sparse signal, but according to Donoho, this solution is a non-deterministic polynomial-time hard (NP-hard) problem. The $\mathcal{L}_2$ norm is not also considered because it represents the energy of the signal and it cannot find the sparsest solution. Donoho then suggested $\mathcal{L}1$ norm minimization as the better solution. It represents the sum of absolute values of all signal coefficients and needs to solve the following convex problem.

$$\tilde{x} = \underset{y=\Phi x}{\operatorname{argmin}} \|x\|_1 \quad \text{Subject to} \quad \|y - \Phi x\|_2 \leq \epsilon \tag{34}$$

where $\epsilon$ denotes the amount of noise in the measurements. Taking $\epsilon$ into account, basis pursuit (BP) and basis pursuit de-noisy are two examples of $\mathcal{L}1$ norm minimization algorithms, where $\epsilon = 0$ and $\epsilon \neq 0$ respectively. The $\mathcal{L}1$ norm minimization algorithm requires a sensing matrix that satisfies the restrict isotropy property with a small value of RIC. It can recover the original signal via linear programing and can find the compressible solution with high probability. Considering the basis pursuit solution, equation (10) can be reformulated as:

$$\tilde{x} = \min \|y - \Phi x\|_2 + z\|x\|_1 \tag{35}$$

Basis pursuit minimizes the cost function, $z$ is the regulation parameter. The algorithm operates in an iterative process, step by step. Firstly, the first guest is initialized ($x = x_0 = \Phi' y$), which represents the minimal energy of the signal, then, the cost function ($\min \|y - \Phi x_0\|_2$) is provided. Secondly, for $k$ iterations, the new sensing matrix is computed by selecting the required measurements number and the signal coefficients are updated to attend the minimized form. The algorithm operates iteratively until obtaining enough signal coefficients less than the signal sparsity level. Other convex iterative relaxation algorithms have been proposed such as gradient descent and iterative thresholding [89][90]. This class of recovery algorithms can work with any matrix that satisfies the restrict isometry property.

**III.4.2 Greedy Algorithms**

Greedy algorithms consist on selecting one position of a non-zero element of the signal, *x*, which corresponds to picking one column from the sensing matrix. A number of techniques have been proposed under this category. Examples of these techniques are matching pursuit (MP) [91], orthogonal matching pursuit (OMP) [81], stagewise orthogonal matching pursuit (StOMP) [82], compressive sampling matching pursuit (CoSaMP) [68][89], generalized orthogonal matching pursuit (GOMP) [20][90], and regularized orthogonal matching pursuit (ROMP) [69][83]. Matching pursuit algorithm is the basis solver and the others are inspired from it to handle some of its limitations. Other Greedy algorithms include iterative hard thresholding (IHT) [84], normalized iterative hard thresholding (NIHT) [85], hard thresholding pursuit (HTP) [86], and normalized hard Thresholding pursuit (NHTP) [87].

Matching pursuit algorithm permits to recover the signal by decomposing it into a linear expansion of waveforms selected from a dictionary. It consists in selecting a column from A that maximizes the inner product of the current residual. Firstly, the vector that corresponds to the longest projection of *x* is selected from the dictionary. Secondly, the signal *x* is orthogonalized by removing any element of the selected vector from *x* to get the residual of *x*, that has the lowest energy. Then, the two previous steps are repeated to the remaining of the dictionary in an iterative process until the residual norm is low than a threshold *c*. The residual norm of *x* after t steps denotes $r_t$, satisfied

$$\|r_t\| \leq c/\sqrt{t} \qquad (36)$$

Orthogonal matching pursuit algorithm is inspired from matching pursuit algorithm by removing not only elements of the selected vector from *x*, but also from the basis before repeating the process. It is proved that orthogonal matching pursuit presents best results solution than matching pursuit, but it is more expensive. GOMP algorithm is the generalized version of orthogonal matching pursuit technique. It consists of selecting multiple indices at each iteration instead of one. Greedy techniques are known as the faster techniques to recover a sparse signal. However, the recovered signal has been shown to be not optimal, and cannot be considered as the exact recovered signal. As in practical situations, the original signal is unknown, it is difficult to evaluate the efficiency of this type of algorithms by comparing the original signal to the output of the algorithm.

### III.4.3 Bayesian Recovery

Bayesian compressive sensing algorithms consist of using a Bayesian model to estimate the unknown parameters in order to deal with uncertainty in measurements. Bayesian model is a probabilistic method that handles uncertainty by considering probabilistic distributions instead of deterministic values [49]. It requires prior knowledge of the involved parameters to compute the posterior distributions of the unknown parameters. The Bayesian compressive sensing consists of finding the sparse solution of a regression problem by exploiting the probabilistic distributions. It solves the underdetermined system and finds the accurate solution by estimating efficiently the unknown parameters using the information that we have about the system. It is based on two main elements: the knowledge about the linear relationship between the signal measurements and the original signal, and the knowledge about the fact that the original signal is $k$-sparse.

Examples of techniques classified under this category are Bayesian model using relevance vector machine learning [77], Bayesian model using Laplace priors [78], and Bayesian model via belief propagation [97]. These probabilistic techniques allow estimating the coefficients of the original signal based on hierarchical prior to estimate the full posterior on the signal and its variance. They represent low error rate and less recovery time. All these Bayesian-based algorithms were used only with random matrices.

### III.4.4 Recovery Algorithms Comparison

Bayesian recovery algorithms represent low error rate and less recovery time. Convex algorithms are consistent and Greedy algorithms are fast. They are complementary to each other, but they share some characteristics. Table 2 represents a comparison between some their techniques in terms of complexity and minimum measurements number required for recovery. For most of these techniques, the reconstruction complexity depends on the number of samples, $N$, sparsity level, $k$, and measurements number, $M$. The complexity of convex relaxation algorithms, such as $\mathcal{L}_1$ norm minimization, is very high compared to the Greedy algorithms, which makes them not practical for hardware implementation. However, Greedy algorithms require fewer measurements compared to the convex relaxation algorithms, which makes them faster to process.

Table 2: Convex relaxation and Greedy algorithms comparison [20].

| Reconstruction techniques | Measurements number | Complexity |
|---|---|---|
| $\mathcal{L}_1$ norm minimization | $O(k \log N)$ | $O(N^3)$ |
| Orthogonal matching pursuit | $O(k \log N)$ | $O(kMN)$ |
| Stage wise orthogonal matching pursuit | $O(N \log N)$ | $O(N \log N)$ |
| Regularized orthogonal matching pursuit | $(k \log^2 N)$ | $O(kMN)$ |
| Compressive sampling matching pursuit | $O(k \log N)$ | $O(MN)$ |

Thus, convex relaxation algorithms require a small number of measurements for efficient reconstruction; however, they are time-consuming and complex in terms of computation. Greedy techniques are not efficient, but they are easy to implement. Bayesian recovery is faster and provides minimal errors compared to other techniques [20].

## III.5 Applications and Challenges

### III.5.1 Compressive Sensing Schemes

Compressive sensing schemes illustrate how the compressive sensing operates, and how the signal is sampled and acquired simultaneously. Classical ADC works under Shannon-Nyquist theorem, which limits their utilization and efficiency. Thanks to compressive sensing, which is a revolution of the sampling theorem and digital signal processing field. The compressive measurements concept is used for signal processing as a new concept called compressive signal processing (CSP) [94]. Signal acquisition can be performed in a very short time acquiring only the main information of the sparse signal. Many schemes have been proposed for compressive sensing implementation to replace the classical ADC. Examples of these schemes include random demodulator [79], modulated wideband converter [79], random filtering [99,100], random convolution [101,102], and compressive multiplexer [103,104].

The random demodulator is an acquisition scheme, known also as the analog to information converter [79]. Its architecture is illustrated in Figure 21. This technique assumes that the original signal is sparse in some domain. It is a multiplication block followed by a pass-low filter, in which the signal is modulated by multiplying it to a high-speed sequence generated by a pseudo-random generator in the analog domain. The low-pass filter output is then sampled at a rate lower than Nyquist rate. Finally, a recovery algorithm is applied at the receiver to recover the input signal [100-103]. This scheme aims to process signals in time domain.

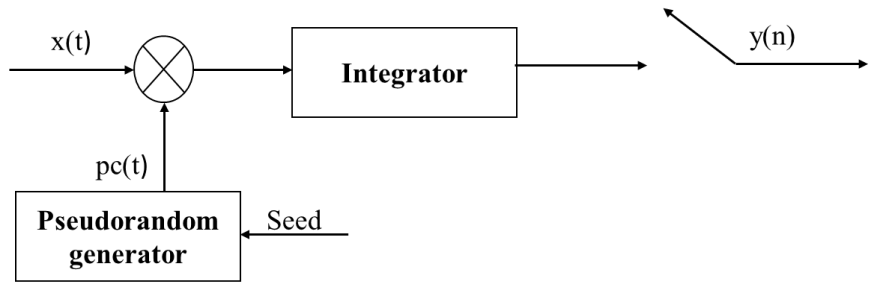

Figure 21: Random demodulator scheme [79].

A modulator wideband converter is a compressive sensing scheme used for multiband sparse signals. In contrast to the random demodulator scheme, the modulator wideband converter can also work in frequency domain [79] as illustrated in Figure 22.

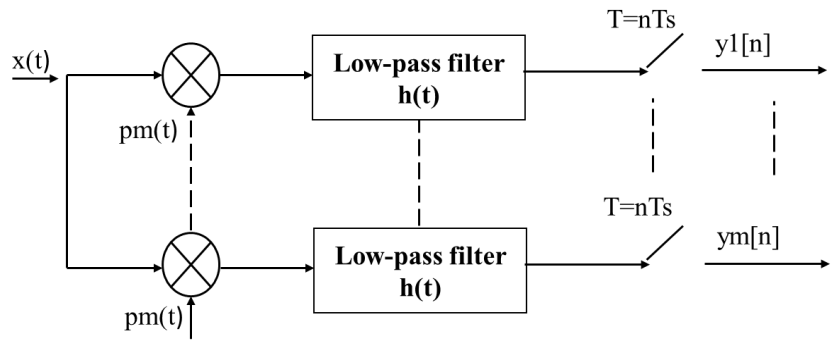

Figure 22: Modulator wideband converter scheme [79].

For *m* channels, the original signal is multiplied by a periodic waveform $p_i(t)$, followed by a low pass filter, $h(t)$. The output sequences, $y_i[n]$, correspond to the input for the recovery algorithm block. Modulated wideband converter is also known as Xampling [100-102]. Random filtering is a generic and structured architecture based on the finite impulse response (FIR) filter. In time domain, the sparse signal is acquired through the convolution with a random tap FIR filter, *h*. This scheme can be used to capture and recover sparse signals when the number of the random taps is known. The random tap can be obtained by using Gaussian or Bernoulli distribution with zero mean and variance 1. This scheme has the potential to be implemented for new ADCs. It is easy to implement in hardware or software and can be used for continuous time signal and streaming [99][100]. Figure 23 presents the random filtering scheme.

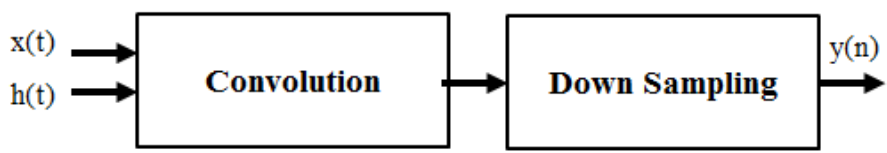

Figure 23: Random filtering scheme in time domain [99].

Random filtering in frequency domain is called random convolution and it is based on fast Fourier transform (FFT). The sparse signal is captured through the convolution with a random pulse or random filter, *h*. This process is then followed by a random subsampling. Despite the randomness, this scheme is universal and it represents fast computations by using double FFT [101,102]. Figure 24 represents the random convolution scheme.

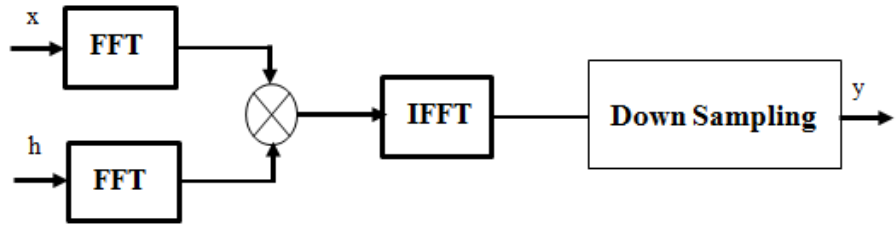

Figure 24: Random convolution scheme in frequency domain [101].

The compressive multiplexer is mostly used for multi-channel compressive sensing and it requires one ADC for all the channels instead of one ADC by channel, as shown in Figure 25. It consists of coding each channel with an orthogonal code pi where $i=1...l$, $l$ is the number of channels. All the coded channels are summed and sampled through the ADC. The code is a pseudo-random chipping sequence ($\mp 1$).

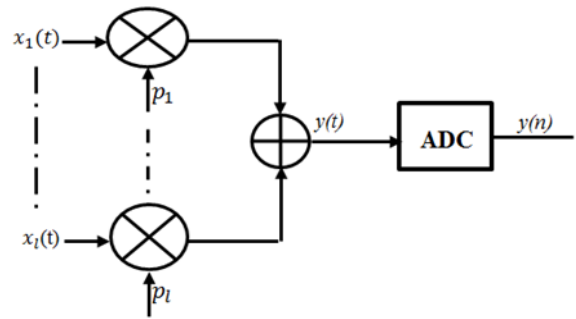

Figure 25: Compressive multiplexer scheme.

The compressive sensing architectures discussed above have been tested and implemented in testbed environments, where it was demonstrated that these schemes can sample sparse signals in practical environments [99]. The implementation of compressive sensing in hardware still represents a great challenge. Thus, schemes need to be developed and should be applicable to real scenarios not only limited to simulations.

### III.5.2 Compressive Sensing Applications

### III.5.2.1 Cognitive Radio Networks

Over the years, compressive sensing has attracted the attention of many researchers who applied it for wideband spectrum sensing and channel estimation in the context of cognitive radio and multiple input, multiple output [104-122]. In cognitive radio, because of the spectrum underutilization, sparsity requirement is approved and transmitted signals are compressible. Several compressive sensing approaches that focus on recovering the signal have been proposed [51,123-128]. In [125], the authors proposed to detect the presence or absence of the primary user (PU) signal from a minimal number of measurements instead of using the entire SU received signal. Thus, the spectrum sensing model can be reformulated in terms of compressive sensing as:

$$y = \begin{cases} \Omega\, n & H_0: \text{PU is absent} \\ \Omega\, (S + n), & H_1: \text{PU is present} \end{cases} \qquad (37)$$

where $y$ denotes the compressive measurements, $n$ denotes the AWGN noise, $\Omega$ denotes a known sensing matrix ($M$x$N$) with ($M \ll N$), $S$ denotes the signal to be detected, $H_0$ denotes the PU signal is absent, and $H_1$ denotes the PU signal is present. The test statistic, $T$, is deduced by applying the likelihood ratio test and it is given by

$$T = y^T \left( \Omega \Omega^T \right)^{-1} \Omega\, S \qquad (38)$$

The test statistic corresponds to the compressive detector and contains sufficient information to decide about the spectrum availability. $T$ is then compared with an appropriate threshold for decision making. In [124], compressive sensing approach is extended to spectrum sensing and PU localization problem in a collaborative compressive sensing approach. The proposed model is implemented with Kalman filter for PU tracking and dynamic sensing for estimating the spectrum utilization. In [125][128], a wideband analog signal is acquired directly in the analogic domain using an analog to information converter. In [129,130], an adaptive sequential compressive sensing approach is proposed to detect white spaces in the spectrum with a minimum $M$. This approach aims to deal with the noise uncertainty in order to allow SU to detect PU presence even at low SNR. In [131], a wideband spectrum sensing is proposed, in which only a portion of the wideband spectrum is estimated by compressive sensing approach to reduce the memory size needed for storage and also reduce computational complexity.

Yet other techniques have been proposed that are based on Bayesian compressive sensing [132]. Bayesian compressive sensing is applied as a recovery algorithm to estimate the unknown parameters including the original signal and noise variance. In [133], a collaborative compressive spectrum sensing is proposed, which operates with multiple nodes for wideband sensing. Sensing is performing by multiple nodes in a distributed environment; each node performs based on its measurements and measurements from other nodes of the same band. In [133], a parallel spectrum sensing scheme is proposed, in which the received signal is compressed by a sensing matrix in a parallel structure. A wavelet edge detector is then applied to combine the outputs from each node. In [104], Kronecker sparse basis is used to perform compressive spectrum sensing. In [127], compressive sensing framework is used as an acquisition system for spectrum sensing in cognitive radio without reconstructing the signal at the receiver, which is known as compressive signal processing. The authors argue that since the main objective of spectrum sensing is to sense the spectrum and decide about its state, it is not necessary to recover it after sensing and analyzing it. According to these authors, this objective is achieved based only on compressed measurements. The spectrum sensing problem is then reformulated based on known sensing matrix and required signal knowledge to determine the channel availability.

In [134], a robust cyclic compressive sensing technique is proposed for spectrum sensing in order to exploit the cyclic sparsity to reconstruct the sparse 2-cyclic spectrum using $\mathcal{L}_1$ norm minimization. This technique aims to overcome the problem of high-dimensional signals sampling by recovering only the cyclic spectrum instead of reconstructing the original signal and its frequency responses. In [135], the authors proposed an adaptive compressive sensing technique that adaptively adjusts the number of measurements independently of the sparsity level. In [104], compressive sensing has been applied for wideband power spectrum sensing, which allows sampling and reconstructing the power spectral density of each channel instead of the signal. This proposed approach uses the multi-coset sampling technique based on the minimal sparse ruler problem. In [111], a cooperative spectrum sensing technique is proposed based on the rank minimization of the measurements signal at the SU receiver. In [137], the wavelet transform is used in order to directly scan the wideband spectrum using a high rate ADC to detect the edges of occupied bands. In [138,139], compressive sensing framework is implemented in each SU receiver to handle the problem of malicious users. This approach detects and discards signals from malicious users based on low-rank matrix completion. In [137], a real-time algorithm is proposed that combines compressive sensing with geo-location database at the SU receiver for spectrum sensing to decide about the band occupancy. In [140], compressive sensing is applied in spectrum sensing to reduce the noise at the SU receiver and reduce the sensing overhead by sensing multiple narrow bands using selective filters.

In order to evaluate the performance of compressive sensing techniques, several metrics have been used, including: recovery success rate [141], failure rate [88], error sparsity, signal sparsity [116], reconstruction error, recovery time [141], mean square error [140], compression ratio, processing time [141], probability of returning the true solution and storage cost of the sensing matrix. Recovery success rate represents the success rate of the recovery algorithm and who much more successful the algorithm is among a number of experiments. It aims to calculate the similarity between the reconstructed signal and the original signal for several values of $k$, $M$ and $N$. Failure rate represents the opposite role of recovery success rate. It calculates the number of time the signal is not reconstructed exactly among a number of experiments. Error sparsity and signal sparsity are two metrics considered to investigate the sparsity level of the estimated signal after reconstruction, compare it with the sparsity level of the original signal before compressive sensing, and calculate the error sparsity. Reconstruction error represents the error level of the reconstruction algorithm. It is calculated as the normalized value of the difference between the reconstructed signal and the original signal. Recovery time evaluates how faster the recovery algorithm is. Processing time evaluates how faster the technique is taking into account the three stages of compressive sensing technique. Compressive ratio ($M/N$) represents the ratio between the sample number of $x$ and the sample number of the measurements. It aims to make sure that a signal with a huge number of samples can be reconstructed from few measurements.

### III.5.2.2 Multiple input, multiple output (MIMO)

In addition, compressive sensing is applied to MIMO based cognitive radio systems to enhance these systems [142-144]. In [147], the authors proposed a new architecture of the receiver for MIMO-OFDM in order to sample multi-channel signals using a single ADC instead of using multiple ones. For different MIMO schemes (2x2, 3x3, and 4x4), this approach allows mixing the transmitted symbols from each antenna, separating the transmitted symbols from each antenna and then recovering them from the mixed signals. It aims to reduce the sampling rate, sampling time, hardware cost, and the number of ADCs by exploring the diversity, the channel occupancy, and the sparsity of MIMO-OFDM channels. In [143], compressive sensing is used to perform the channel estimation with sparse multipath MIMO signals. In [149], the authors proposed a novel architecture of an MIMO-OFDM system that exploits compressive sensing by using a 4x4 MIMO scheme. Under the conventional MIMO-OFDM system, the number of the required ADCs corresponds to the number of antennas at the receiver. With the proposed system, the number of ADCs is reduced to one ADC and the MIMO detector is substituted by the compressive sampling reconstruction as shown in Figure 26.

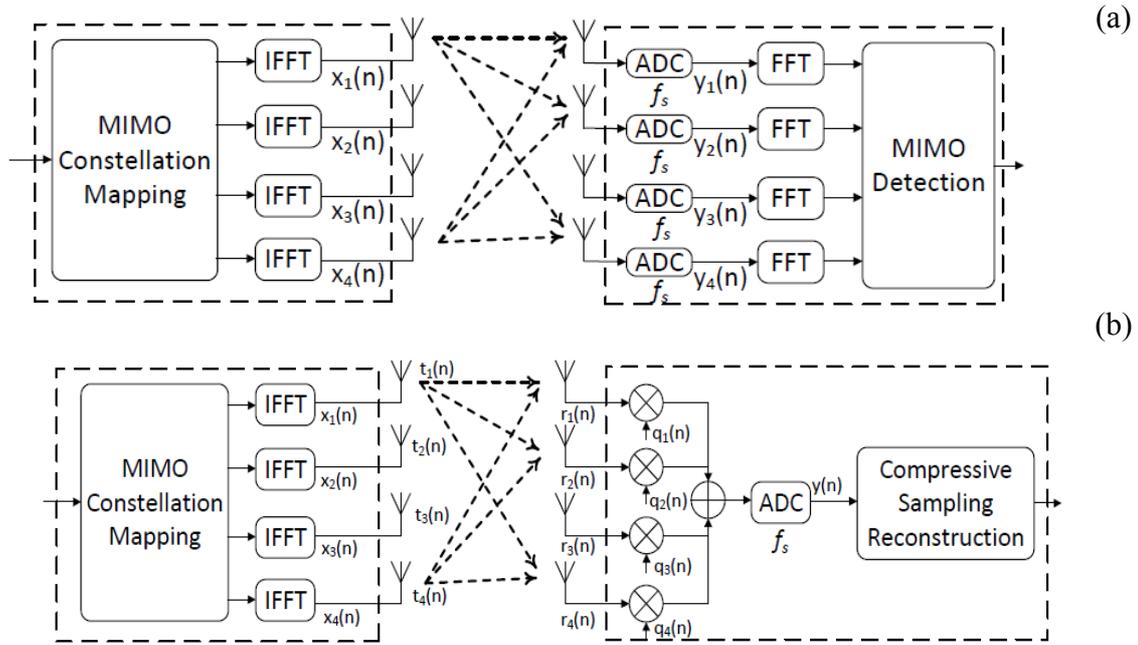

Figure 26: (a) Conventional MIMO-OFDM system; (b) MIMO-PFDM based compressive sensing [149].

The sparse signals from multiple antennas are jointly combined and sampled. The recovery process reconstructs the entire signals directly on frequency domain and separates them [90].

### III.5.2.3 Channel Estimation

Compressive sensing was also applied to channel estimation to overcome the limitations of the existing channel estimation techniques, known as non-compressive sensing channel estimation techniques. A number of these techniques have been proposed in the literature to evaluate and approximate the channel behavior [144]. Examples of these techniques include pilot-aided [145], blind [146], and least mean square (LMS) based channel estimation [20]. The channel estimation pilot aided techniques include a number of schemes such as minimum mean squared error (MMSE) and maximum likelihood estimator (MLE) based technique [147]. In [108], a low-rank estimation technique based on MMSE has been proposed for estimating channel pilots using singular value decomposition. This technique requires the SNR and channel frequency correlation to be known. In [148], the MLE was used for channel estimation in the context of OFDM-MIMO. Low-rank estimation The blind channel estimation technique consists of using the data symbols to estimate the parameters of the channel. This technique is not efficient for fading channels and represents high complexity [148]. LMS based channel estimation represents low complexity compared to the other techniques and it consists on computing the LMS based on the transmitted signal over the channel and the identity matrix [148]. The aforementioned techniques represent high complexity and often require the knowledge of the channel response at the receiver for multipath wireless channels, which is not always possible in practice because of the high number of required antennas [147-152].Moreover, multichannel signals are known to be sparse, which can lead to high estimation errors at the receiver. In order to overcome these limitations and exploit the sparsity of the multipath wireless channels, compressive sensing channel estimation techniques have been proposed and investigated, especially for MIMO-OFDM communication systems [153].

In addition, channel estimation has been investigated in wireless communications to allow the receiver to estimate the impact of the channel on the signal [150]. In [145], compressive channel estimation technique is adopted in MIMO communication systems for wideband spectrum to address the limitations of the conventional techniques. It allows estimating all channels responses, $h_{i,j}$, between the transmitter, $i$, and the receiver, $j$. In [154], compressive sensing is used for pilot allocation in MIMO-OFDM systems to estimate the channel impulse response with reduced number of pilots based on the sparsity of channel impulse response. In [155], the authors proposed a new approach for channel estimation through compressive sensing, in which the sparsity feature of the signal is explored to estimate selective channels in OFDM systems. This approach aims also to estimate and overcome the interferences problems in multi-carrier communications whether inter-symbol or intercarrier interferences. In [156], the authors investigate channel estimation using compressive sensing and evaluate it based on bit error rate and mean square error. To estimate the channel response coefficients, OMP and SAMP algorithms have been applied for non-contiguous OFDM using reduced $M$.

In [157], compressive sensing has been applied to a pilot-aided channel estimation technique in order to overcome its limitations with the assumption that multipath channels are sparse. The authors showed that with compressive sensing, sparse multipath channels can be estimated with high accuracy and low pilot overhead using a fixed number of pilots for the estimation process. The authors suggested adopting banded channel matrix for the sensing process with unknown symbols data at the receiver. The model is tested for different channel types, including time variant and invariant channels.

### III.5.3 Compressive Sensing Limitations and Challenges

A number of issues have been investigated in order to make compressive sensing efficient solution for the next generation of wireless communication systems. These issues are especially related to the implementation aspects for real scenarios. They include designing practical sensing matrix, exact and approximate sparsity, hardware implementation, uncertainty due to the noisy and real environment, recovery uncertainty, and RIP proof [158][159]. The matrix design is one of the main limitations of compressive sensing techniques. Efficient compressive sensing requires designing a practical structured sampling matrix that can sample the signal and take only the main information taking into account the cost and the speed for the encoding process. Implementing a random matrix in practice is challenging because of its unstructured nature and randomness in its elements [159]. Thus, the design is complex, costly, and requires high memory storage. Therefore, these limitations need to be addressed, and, thus represent open research topics to be investigated.

Regarding sparsity, compressive sensing requires signals to be sparse; however, signals can only be approximately sparse in practical scenarios [104,158,159]. In [160], the authors showed that the sparsity of a signal is unknown and needs to be estimated in practice. They proposed a technique to measure the signal sparsity-based on $\mathcal{L}_0$ norm, $\|x\|_0$, that represents the number of non-zero coefficients of the signal. Thus, there is a need for techniques to transform signals into sparse signals in real applications.

Regarding the limitations due to noise, most of the proposed compressive sensing techniques consider the Gaussian noise with known or unknown variance. In addition to the noise, other factors impact compressive sensing techniques' performances such as interference, high level of noise uncertainty, channel uncertainty, and imperfections. Because of these factors, compressive sensing performance degrades in real scenarios [86,158,159].

However, a few papers investigated the impact of uncertainty due to the noise in measurements. Thus, there is a great need for investigating and developing practical compressive sensing techniques that can deal with the imperfections of real networks. In addition, dealing with real signals involves non-linear signals, which is another issue that limits the compressive sensing application for non-linear measurements [158]. Furthermore, the hardware implementation of compressive sensing is also one of the main limiting challenges due to a number of problems, including synchronization, calibration, and the uncertainty in measurements [158,159].

Research shows that it is not easy to select which compressive sensing technique to consider for a specific application since efficient techniques are complex and require a great deal of processing time [158,159]. Indeed, compared to the other alternative acquisition and recovery techniques, compressive sensing approach is characterized by its efficiency in sampling high-dimensional signals below Nyquist rate, which makes the spectrum scanning faster. It can also recover signals from few measurements with high efficiency and with low recovery errors.

Designing fast ADCs is yet another solution for signal sampling in terms of hardware. However, ADCs are costly and they capture the entire signal, which makes them slow and inefficient, compared to the compressive sensing solution.

## III.6 Conclusion

In this chapter, we have provided an in depth survey on compressive sensing techniques and classified these techniques according to which process they target, namely, sparse representation, sensing matrix, or recovery algorithms. we have also presented potential applications of compressive sensing including spectrum sensing, channel estimation, and multiple-input multiple-output in cognitive radio.

Existing compressive sensing techniques suffer from high processing time and low accuracy, improved compressive sensing techniques are proposed through the rest of the dissertation. Starting the next chapter, we are going to present and discuss the different dissertation contributions.

# Chapter IV

# MATCHED FILTER DETECTION BASED DYNAMIC THRESHOLD

In the previous chapter, we have discussed the state-of-the-art of spectrum sensing as well as compressive sensing. In this chapter, we are focusing on one of the dissertation contributions for improving spectrum sensing performance.

Developing new algorithms to perform spectrum sensing is important to enhance the access to the radio spectrum. Researchers are investigating techniques to improve the existing sensing techniques by addressing some of their limitations [44][56]. Here, we propose an estimated and dynamic sensing threshold for matched filter detection to increase the probability of detection and the decision reliability.

The remaining of this chapter is organized as follows. In Section IV.1, we analyze the related works. In section IV.2, we explain the simulation methodologies for each sensing method. The results of this simulation are discussed in section IV.3. Finally, a conclusion is given at the end.

## IV.1 Introduction

A number of sensing techniques have been proposed over the last decade. Examples of these techniques are: energy [7], autocorrelation [28], and matched filter based sensing [6]. Energy detection is based on the received signal energy and does not require any information about the PU signal to make the sensing decision. Autocorrelation based sensing is based on the value of the autocorrelation coefficients of the received signal. Matched filter based sensing detects known PU signals [6,7,48]. These techniques were more detailed in chapter II.2.

Sensing threshold is an important parameter. When a detector does not adjust its threshold properly, it suffers from performance degradation of the sensing results. Various approaches are suggested for energy detection. As the sensing performance is highly affected by the estimation error of noise power, a dynamic estimation style of noise power is recommended in [39]. Adaptive threshold control is implemented in [40] with linear adaption on threshold based on SINR. This approach attains a considerably higher SU throughput than the fixed threshold approach, but maintains unacceptable chances of false alarms. Adaptive threshold in unknown white Gaussian noise is presented in [41] with noise power estimation, keeping the false alarm rate at a preferred point under any noise level; it is based on a concept of dedicated noise estimation channel in which only noise is received by SU. An improved energy detection method is proposed in [42] where misdetection of PU transmission due to sudden drop in PU transmission power is addressed by keeping an additional updated list of latest fixed number of sensing events that are used to calculate an average test statistic value. A double-threshold technique is proposed in [43] with the intention of finding and localizing narrowband signals. Another technique is presented in [18] based on wideband spectrum sensing, which senses the signal strength levels within several frequency ranges to improve the opportunistic throughput of SUs and decreases the interference to the PU.

However, for matched filter detection and autocorrelation based detection, sensing threshold is set to a static value to sense the PU signals. In [161], each pair of ($P_d$, $P_f$) is associated with a particular threshold to make sensing decision. In other works, the sensing threshold is determined dynamically by multiplying the theoretical threshold by a positive factor [53]. Others don't mention how the threshold was selected. However, with a static threshold, sensing decision is not reliable because of the uncertainty of the noise. In this work, we suggest an estimated and dynamic sensing threshold for matched filter detection to increase the probability of detection and the decision reliability.

## IV.2 Methodology

The general system model is presented in Figure 27. The PU signal is generated as a QPSK signal with *N* samples. The AWGN channel adds a Gaussian noise to the input signal with the same number of samples. The noise is generated randomly in range of *SNR* [-20dB, 20dB]. Afterward, the spectrum sensing method is performed on the output. The test statistic is then computed for each technique and compared with the estimated threshold for decision making.

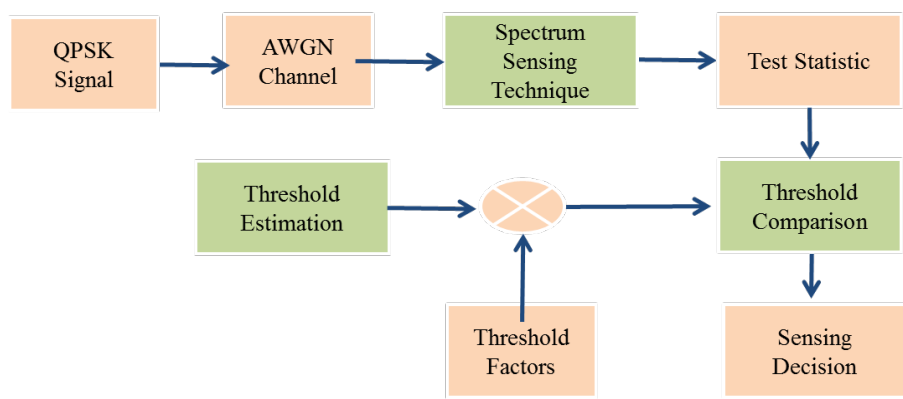

Figure 27: Simulation model of spectrum sensing method [6].

For the threshold estimation block, the threshold is estimated dynamically at each iteration and multiplied by a threshold factor to investigate the impact of the threshold on the detection performance. It implies that the sensing threshold is the estimated threshold multiplied by a predefined factor. For each technique, the threshold is estimated and then generated dynamically to perform the test comparison. The estimated dynamic threshold $\lambda'$ is used to assess the performance of the sensing techniques. It is giving by

$$\lambda' = k\,\lambda \qquad (39)$$

where $\lambda$ is the estimated threshold for each sensing algorithm and $k$ is a positive factor. $P_d$ and $P_{fd}$ are used to evaluate the performance of each technique. The simulation methodology is performed for a number of iteration, called cycle number, which represents the total number of experiments. $P_d$ is simulated as the ratio of the total number of detections, $N_d$, by the total number of experiments, $N_t$. $P_{fd}$ is simulated as the ratio of the total number of times the signal is not detected, $N_f$, by the total number of experiments, $N_t$. These two probabilities are given by

$$P_d = N_d/N_t \quad , \quad P_{fd} = N_f/N_t \qquad (40)$$

For performance evaluation, the spectrum sensing techniques are implemented for $N_t$ experiments to get $P_d$ and $P_{fd}$ as presented in Figure 28. Three variables are created and initialized by zero, namely, *count, n,* and *m*. At each iteration and for each sensing parameter, the variable count is incremented by one at each iteration. If the test statistic is greater than the sensing threshold, a variable, *n*, will be incremented by one. Otherwise, a variable, *m*, will be incremented by one. At the end of the loop, after $N_t$ experiments, the total number of detection is the total number of *n* values. The total number of false detection is the total number of *m* values. Thus, $P_d$ and $P_{fd}$ are the average values over $N_t$ experiments [6].

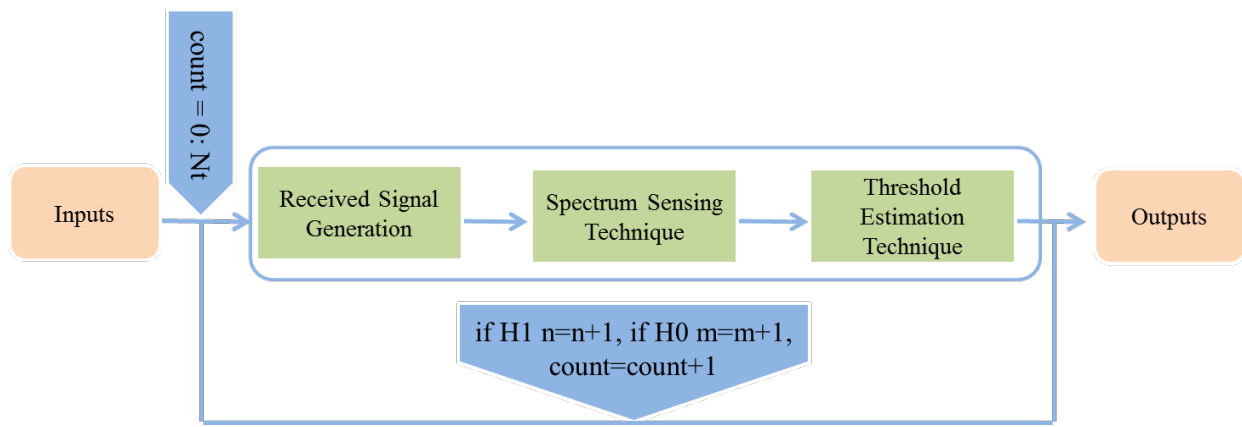

Figure 28: Simulation methodology for spectrum sensing methods [6].

The inputs involved in the simulation are: (1) number of samples corresponds to the maximum value of $P_d$, $N= 1000$; (2) total number of experiments 1000 cycle; (3) *SNR* range -20dB to

+20dB; and (4) threshold factors [1, 2, 3, 4]. For each parameter setting, the outputs, $N_d$ and $N_f$, are obtained by varying one of these parameters: *SNR*, threshold, the number of samples. Afterward, $P_d$ and $P_{fd}$ are computed. In an iterative loop, matched filter detection is performed by convolving the received signal with some PU pilot stream, the output is averaged over N samples to get the test statistics as presented in Figure 29.

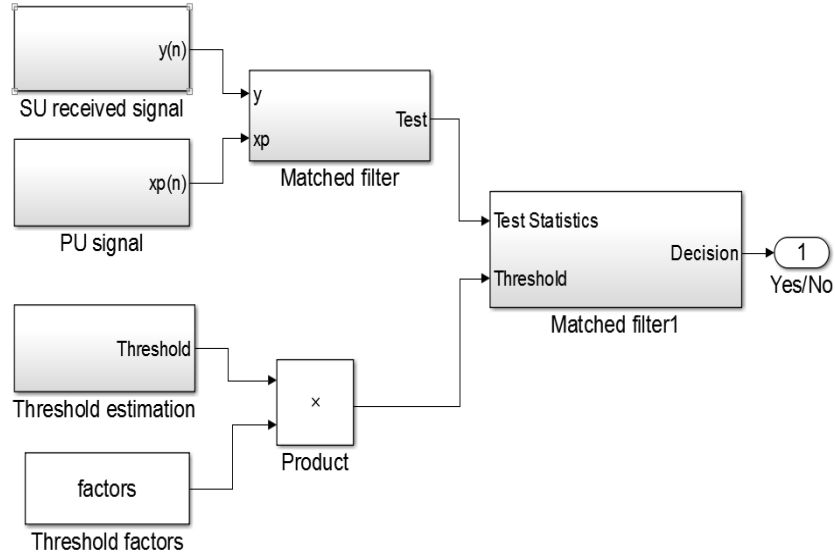

Figure 29: Simulation model for matched filter detection [6].

I proposed to adopt the quite time approach to estimate the sensing threshold. It refers to the time period when it is assumed that PU signal is absent and only noise is transmitted [6]. The threshold is expressed as:

$$\lambda = \sum_N w(n)\, x^*_p(n) \quad n=1...N \tag{41}$$

where $\lambda$ represents the estimated threshold, *N* is the sample number, *w(n)* is the additive white Gaussian noise, and $x^*_p(n)$ is the PU pilot stream. At each iteration, the threshold is generated dynamically using the quit time approach and multiplied by a threshold factor (1 to 4) to investigate the impact of the threshold on the performance of the sensing technique.

## IV.3 Results and Discussion

The matched filter detection based dynamic threshold is implemented and its simulation results are compared to those of energy and autocorrelation based detection. The probability of detection and the probability of false alarm are simulated varying *SNR*, the threshold, and the number of samples. Examples of these results are presented from Figure 30 to Figure 33.

Figure 30 represents $P_d$ as a function of *SNR* for the three sensing techniques for 1000 samples and for a threshold factor equal to 1. Under high *SNR*, in which the signal power is higher than the noise power, matched filter detection achieves its 100% of detection with a small number of samples while energy detection presents a very high detection, which means that this detector is not able to distinguish between the noise and the signal. Autocorrelation based sensing represents a lower $P_d$ comparing to the other techniques. Yet, each technique has a particular behavior under environment conditions. For energy detection, it gives a good performance under high *SNR*, which works only for high power signals, a high threshold, and a high number of samples. However, this method is not reliable because of the non-ability to distinguish between signal and noise while detecting. On the other hand, matched filter detection achieves a high performance for a low number of samples and can work under low *SNR*. For autocorrelation based detection, it represents low $P_{fd}$ compared to the others methods.

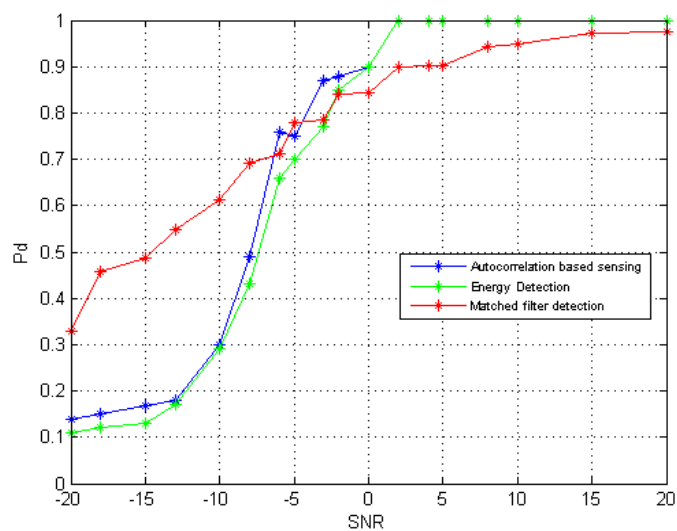

Figure 30: Pd as a function of SNR for the 3 different techniques.

For the matched filter detection, the simulations were performed for 1000 experiments to get an average result of $P_d$ and $P_{fd}$ for each set of parameters. Figure 31 represents the $P_d$ as a function of N for several values of *SNR*. For high *SNR* (- 4dB), $P_d$ is nearly equal to 100%. For low *SNR* (- 20dB), $P_d$ increases for *N* greater than 400 samples. $P_d$ also increases with the increase of N and it achieves its 100% value for *N*=1000. Thus, for a small number of samples, matched filter detection can work even at low SNR and it achieves its 100% for 1000 samples.

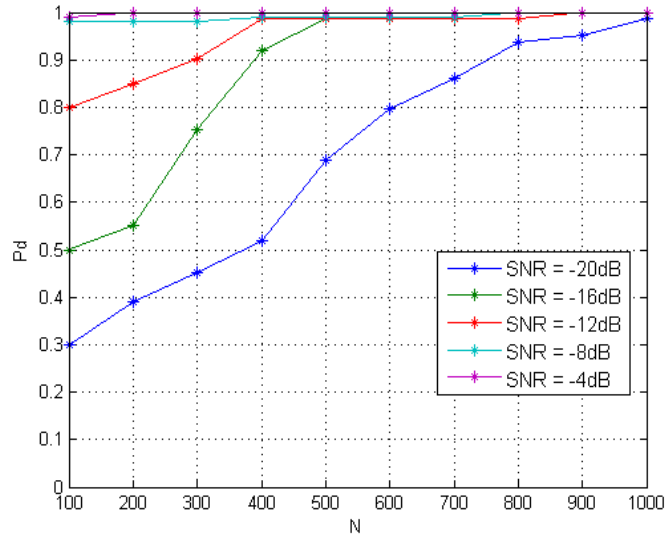

Figure 31: Pd as a function of the number of samples with variable SNR.

Figure 32 presents the $P_d$ as a function of *SNR* for several values of the threshold factor. $P_d$ and $P_{fd}$ have been simulated by varying the *SNR* from -20dB to +20dB for several values of the threshold factor with 1000 samples. As one can see, for high threshold factor (*Th_factor* = 4), $P_d$ decreases for a low *SNR* (-10dB). When the threshold factor decreases, $P_d$ increases. For a fixed threshold factor and low *SNR* values, $P_d$ decreases. Thus, $P_d$ increases with the increase of *SNR*. Also, the probability of detection decreases when the threshold increases.

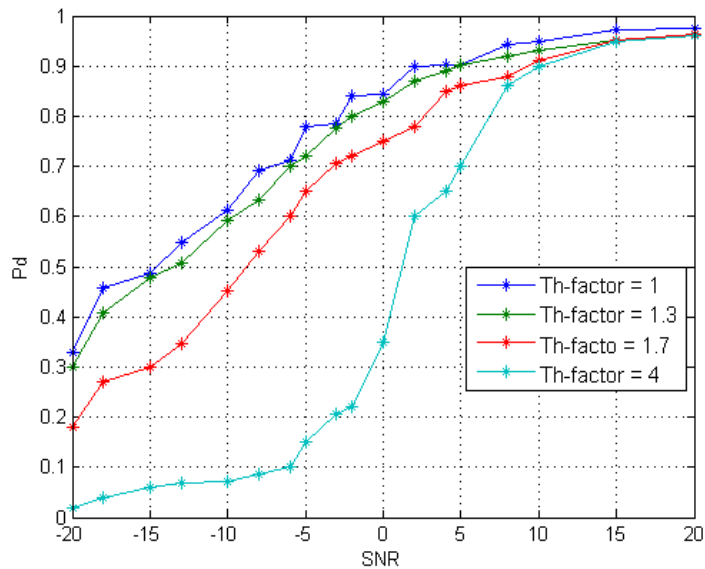

Figure 32: $P_d$ as a function of SNR with variable threshold.

Figure 33 represents $P_{fd}$ as a function of *SNR* for several values of the threshold factor. It can be seen that for a high threshold factor (*Th_factor* = 4), $P_{fd}$ decreases for low *SNR* (- 10dB). When the threshold factor decreases, $P_{fd}$ increases. For a fixed threshold factor and low *SNR*, $P_{fd}$ increases. Thus, $P_{fd}$ decreases with the increase of the threshold and *SNR*. The threshold impacts the performance of the matched filter detector.

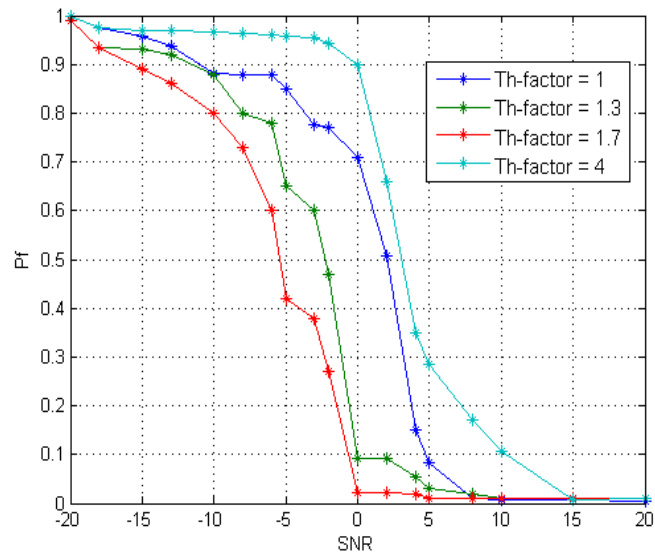

Figure 33: $P_{fd}$ Vs. SNR with variable threshold.

Through analyzing and comparing the simulation results, one can conclude that matched filter sensing technique with dynamic threshold can work under low SNR with a small number of samples. Table 3 compares the proposed technique with energy and autocorrelation based detection techniques to highlight their strengths and weaknesses.

Table 3: Spectrum sensing techniques comparison

| Energy detection | Autocorrelation based sensing | Matched filter detection based dynamic threshold |
|---|---|---|
| Low complexity | High data processing | High complexity |
| Prior knowledge of PU signal is not required | Knowledge of statistical distribution of the autocorrelation function is required | Prior knowledge of PU signal is required |
| Non-ability to distinguish between noise and signal after detection | Ability to distinguish between PU signal and noise after detection | High performance for low sample number |
| Good performance under high *SNR* | Robust against noise uncertainty | Ability to perform under low *SNR* |

Some researchers proposed new algorithms by combining the existing sensing techniques and make the most use of their advantages. For instance, in [162], the authors combined three different spectrum sensing techniques to form an intelligent spectrum sensing scheme, called I3S. I3S allows improving the efficiency of the spectrum utilization and decreasing the sensing time compared to a single sensing technique.

## IV.4 Conclusion

In this chapter, we presented one of our contributions related to the spectrum sensing techniques. we compared its results with the results of energy and autocorrelation based detection. As a conclusion, energy detection is easy to implement and does not require any information about the PU signal even though it is affected by the non-ability to distinguish between PU signal and interferences and non-reliability at low *SNR*. Matched filter sensing requires a perfect knowledge of the PU transmission, which may not be feasible in practice, but can work under low *SNR* through simple computation. Autocorrelation based sensing is mainly related to the autocorrelation features and can be adopted only when noise is Gaussian. Nevertheless, it is robust against noise uncertainty.

In the next chapter, we are going to present the proposed Bayesian compressive sensing based Circulant matrix technique and its performance evaluation.

# Chapter V

# BAYESIAN COMPRESSIVE SENSING WITH CIRCULANT MATRIX

In the previous chapter, we discussed a technique to improve the performance of the matched filter based sensing. In this chapter, we are discussing a Bayesian compressive sensing with Circulant matrix for signal sampling.

Compressive sensing has been proposed in spectrum sensing to speed up the radio spectrum scanning and reduce the hardware cost. Existing compressive sensing techniques suffer from uncertainty due to random measurements as well as low efficiency in recovering signals. Therefore, in this chapter, we propose two techniques, namely Bayesian compressive sensing with Circulant matrix sampling and Bayesian compressive sensing with Toeplitz matrix sampling for fast and efficient compressive sensing [163]. To the best of our knowledge, no paper has been published that investigates the efficiency of Bayesian recovery with Toeplitz matrix sampling [164].

The remaining of this chapter is organized as follows. In Section V.1, we present the existing techniques. In section V.2, we explain the simulation methodologies. The results of the simulation are discussed in section V.3. Finally, a conclusion is given at the end.

## V.1 Introduction

Compressive sensing involves three main processes: sparse representation, encoding, and decoding. In the first process, a signal, $x$, is projected in a sparse basis. In the second process, $x$ is multiplied by a sampling matrix, $M_c$, of $M$x$N$ elements to extract $M$ samples from $N$ of the signal where $M << N$. In the last process, $x$ is recovered from few $M$ measurements [6,7].

A number of sampling matrices have been proposed in the literature, including random matrix [165,166], Circulant matrix [167], Toeplitz matrix [168], and deterministic matrix [169]. Because of their simplicity, more interest has been paid to random matrices. These matrices are randomly generated with independent and identically distributed (i.i.d) elements such as Gaussian and Bernoulli distributions [166,167]. Random matrices satisfy the restrict isometry property for small RIC [165,166]. However, these matrices require a great deal of processing time and high memory capacity to store the matrix coefficients [168]. Because of the randomness, the results are uncertain, which makes the signal reconstruction inefficient.

Unlike random matrices, Circulant matrices are efficient, fast in terms of signal acquisition, require fewer measurements to perform, and less time to process [167]. A Circulant matrix is a structured matrix associated and determined using a predefined vector by cyclic permutation [168]. It satisfies the restrict isometry property for a small number of measurements [74]. Unlike random matrices, Circulant matrices are not universal. They have been used only with the $\mathcal{L}_1$ norm minimization technique [165-168]. Unlike random matrices, Circulant matrices are efficient, fast in terms of signal acquisition, require fewer measurements to perform, and require less time to process.

A number of algorithms that exploit the sparsity feature have been proposed in the literature for signal recovery [82,88,170]. A sparse signal can be estimated from a few measurements by solving the underdetermined system using different recovery algorithms. The iterative relaxation techniques solve the underdetermined system using linear programming. They are more accurate compared to Greedy algorithms, but they are complex, uncertain, require high measurements, and, thus high processing time [165,171,172]. Greedy algorithms consist of selecting a local optimal at each step in order to find the global optimum, which corresponds to the estimated signal coefficient. They are fast; however, they are inefficient, uncertain, and require more measurements for the recovery process [81,82,91]. Bayesian compressive sensing algorithms consist of using a Bayesian model to estimate the unknown parameters in order to deal with uncertainty in measurements. They are fast, accurate, require few measurements for a high recovery rate, and can deal with uncertainty [88,170]. They combine the strengths of the other two categories. Indeed, these techniques were used only with random matrices.

Therefore, we propose to combines the strengths of both Circulant matrices and Bayesian models to address the previously mentioned problems during the encoding and decoding processes for fast and efficient compressive sensing [163].

## V.2 Methodology

### V.2.1 Bayesian Compressive Sensing

Figure 34 represents the proposed approach. For sparsity representation process, we assume the signal to be sparse because of the underutilization of the spectrum. For the sensing matrix process, we adopt Circulant matrix. For the recovery process, we adopt the Bayesian recovery.

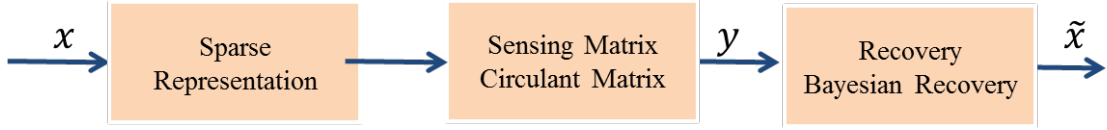

Figure 34: Block diagram of compressive sensing model [163].

A vector, $c$, is given as $(c_0, c_1, \ldots, c_{N-1})$. The Circulant matrix, $C$, is generated from $c$, where $C_{(i,j)} = C_{(j-i)\,mod(N)}$ for $i, j = 0 \ldots N$. It can be expressed as:

$$C = \begin{bmatrix} c_0 & c_1 & \cdots & \cdots & c_{N-1} \\ c_{N-1} & c_0 & c_1 & \cdots & c_{N-2} \\ \vdots & \vdots & \vdots & \vdots & \vdots \\ c_1 & c_2 & \cdots & c_{N-1} & c_0 \end{bmatrix} \qquad (42)$$

where c is a vector given as $(c_0, c_1, \ldots, c_{N-1})$. The values of $c$ are chosen randomly according to a suitable probability distribution to reduce the amount of randomness of the sensing matrix compared to random matrices [168]. During the encoding process, each column of the matrix $C$ is obtained by right cyclic shifting the previous column. Then, a *MxN* partial Circulant matrix, $M_c$, is defined as the sub-matrix of $C$ and considered for the signal sampling, where *M<<N*. The signal, *S,* is then multiplied by $M_c$ for signal compression. This multiplication is fast because of the reduced number of random coefficients in the Circulant matrix [167][168].

Under the Bayesian recovery, the *k*-sparse signal is acquired through a product with the Circulant matrix. A noise, *w,* is added to the signal measurements, which includes the noise measurements and the sparse representation error. Figure 35 illustrates the simulation methodology of our proposed model.

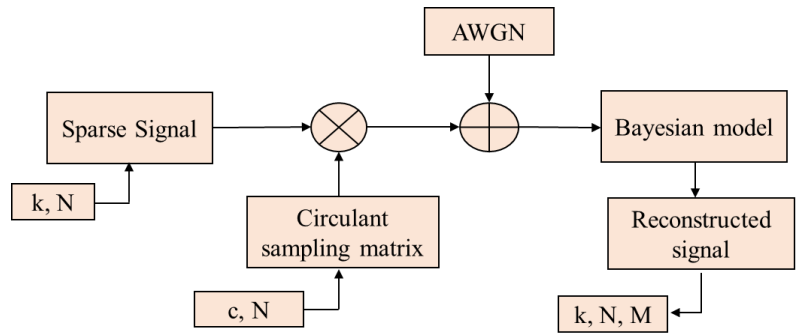

Figure 35: The model of the Bayesian compressive sensing with Circulant matrix [163].

The noisy measurements can be formulated as:

$$y = M_c\, x + w \qquad (43)$$

According to the theorem of central limit for N>>M, W can be approximated as a zero mean Gaussian noise with unknown variance $\delta_w$, which can be expressed as $N(0, \delta_w)$. The signal to be approximated can be considered as a Gaussian variable with $x = (x_1, x_2, ..., x_N)$. Therefore, the Bayesian model implies that the noisy measurements, $y$, is an i.i.d Gaussian and depends on the unknown S and $\delta_w$, which is given by

$$y|x, \delta_W \sim N(x, \delta_W) \tag{44}$$

Figure 36 illustrates the proposed Bayesian model in which the unknown signal, noise, and vector to generate the Circulant sampling matrix are parents of the noisy measurements. Noise variance $\delta_w$, signal mean $\mu_S$, and signal variance $\delta_s$ are the unknown parameters to estimate.

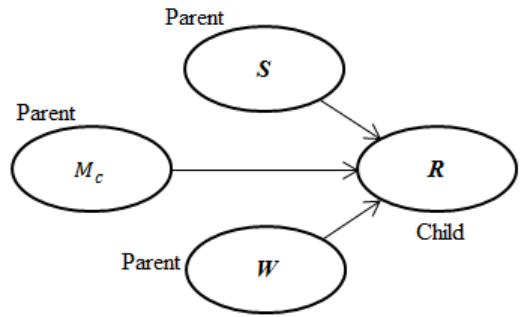

Figure 36: Graphical model of the Bayesian compressive sensing technique [163].

The Bayesian model specifies the conditional probabilities of the measurements $P(y/x)$ and the noise $P(y|\delta_w)$. $P(y/x)$ and $P(y|\delta_W)$ present the probability densities of $y$ given the values taken by the signal and the noise variance respectively[88]. The conditional probability of the signal to be estimated given the measurements can be expressed thought Bayes' rule as:

$$P(x/y) = \frac{P(y/x)P(x)}{\sum_{x\prime} P(y/x\prime)P(x\prime)} \tag{45}$$

where $x'$ represents the other alternative solution of the underdetermined system. Consequently, the Gaussian likelihood of the noisy measurements can be expressed from the previous conditional probability as:

$$P(y|x, \delta_w) = (2\pi\delta_w)^{-M/2} \exp\left(-\frac{1}{2\delta_w}\right) \|y - M_c x\|^2) \tag{46}$$

Given the vector $c$, the Circulant sampling matrix, $M_c$, coefficients are generated to be used for the recovery process. This matrix should be the same as the one adopted for the compression to reduce the randomness. With prior knowledge that $x$ is sparse and $M_c$ is known, $x$ and $\delta_w$ are the two quantities to be estimated. The prior density of the unknown signal $P(x/k)$ in terms of sparsity is expressed as:

$$P(x/k) = (k/2)^N e^{(-k \sum_i^N S_i)} \tag{47}$$

where $k$ is a parameter that represents the sparsity of $x = (x_1, x_2,…, x_N)$ [163]. Under the Bayesian estimation process, the underdetermined system becomes a linear problem with, $S$, sparse and can be reformulated as:

$$\tilde{x} = \underset{y = M_c x + w}{\mathrm{argmin}} \|y - M_c x\|_2^2 + z\|x\|_1 \tag{48}$$

where $z$ is a positive scalar. The Bayesian recovery aims to look for the posterior probabilistic distribution for $x$ and $\delta_w$ taking into account the known evidences. The maximum posterior probability corresponds to the sparsest solution of the underdetermined system. The algorithm computes the joint probability distribution of all unknown parameters and computes the prior distribution of each element of $x$ with the hyper parameters $a$ and $b$. The hyper parameter $a = (a_1, a_2,…, a_N)$ represents the initial posterior of the signal variance and the hyper parameter $b$ represents the initial posterior of the noise variance. The prior distribution of $S$ given the hyper parameters $a$ and $b$ can be expressed as the product of the conjugate prior of signal variance $\Gamma(\delta_{xi}|a, b)$ and the likelihood function of $S_i$, which is defined as a zero mean Gaussian prior for each signal coefficients $N(x_i|0, \delta_{wi})$. It is also called the marginal likelihood for Bayes estimation. This probability of the signal given $a$ and $b$ is given by

$$P(x|a,b) = \prod_i^N \int_0^\infty N(x_i|0, \delta_{xi}^{-1}) \Gamma(\delta_{xi}|a,b) d\delta_{xi} \tag{49}$$

In an iterative loop, the algorithm optimizes the hyper parameters $a$ and $b$, estimates their new values, and then maximizes the marginal likelihood using their new estimated values. The algorithm is based on the previous results for learning and searching for the new values of the hyper parameters $a$ and $b$. Taking into account the assumption about the knowledge of $a$ and $b$ in addition to $M_c$ and $y$, the posterior probabilistic distribution of $x$ can be then expressed as a Gaussian distribution $S \sim N(\mu_x, \delta_x)$ with mean $\mu_x$ and variance $\delta_x$ which are given by

$$\begin{cases} \mu_x = b\, \delta_x M_c^T y \\ \delta_x = (b\, M_c^T M_c + A)^{-1} \end{cases} \tag{50}$$

where $A = \mathrm{diag}(a_1, a_2, …, a_N)$. The last estimated value of $b$ will be the noise variance. At the end of the algorithm, the signal is approximated and the uncertainty is reduced [88].

## V.2.2 Performance Evaluation

In order to evaluate the efficiency of our model, its results have been compared to the results of the basis pursuit technique [169] using several metrics. These metrics are: mean square error, reconstruction error, correlation coefficients, processing time, recovery time, and sampling time. The reconstruction error, $R_e$, is a metric that calculates the norm of the difference between the expected signal, $\tilde{x}$, and the original signal, $x$, divided by the norm of the original signal. It is expressed as:

$$R_e = \frac{\|\tilde{x} - x\|}{\|x\|} \tag{51}$$

Mean square error (MSE) is a metric that measures the average magnitude of the squared difference between the reconstructed signal and the original signal. It corresponds to one of the loss functions used for error estimation. It is given by

$$MSE = \frac{1}{N}\sum_N (x - \tilde{x})^2 \tag{52}$$

It has the same measurement unit as the data being estimated. It is utilized for predictive modeling in order to analyze the variation in the error in the reconstruction algorithm for multiple times. Correlation measures the similarity between the original signal, $x$, and the reconstructed signal, $\tilde{x}$, to measure how similar they are. The measure of correlation is known as correlation coefficient, $C_c$, which is a scalar quantity. It can take values between -1 and 1. It is expressed as:

$$C_c = \frac{N \sum(x\tilde{x}) - (\sum x)(\sum \widetilde{x})}{\sqrt{N(\sum x^2) - (\sum x)^2}\sqrt{N(\sum x^2) - (\sum \tilde{x})^2}} \tag{53}$$

When $C_c$ is positive and less than 1, it implies that the two signals are positively correlated and the strength of the correlation is expressed as a percentage value. When $C_c$ is null, it implies there is no relationship between the two signals. When $C_c$ is negative and greater than -1, it implies that the two signals are negatively correlated and the strength of the correlation is expressed as a percentage value.

Recovery time, $t_r$, is the time required by the recovery process to reconstruct the signal. It allows defining the fastest reconstruction technique. Sampling time, $t_s$, is the time required by the sampling matrix process in order to compress the signal using a specific matrix. It allows defining the fastest sampling matrix technique. Finally, the processing time, $t_p$, is the time required to perform all compressive sensing processes.

## V.3 Results and Discussion

The two algorithms, Bayesian compressive sensing and basis pursuit, also called $\mathcal{L}1$ norm minimization, with the Circulant matrix, were implemented and extensively tested. Their efficiencies were analyzed and compared using the metrics previously mentioned ($R_e$, MSE, $C_c$, $t_p$, $t_r$ and $t_s$). In this performance evaluation, we investigated the efficiency of the Circulant matrix in sampling signals and compared its results with those of random matrix. we also investigated the performance of Bayesian recovery and compared its results with the basis pursuit technique.

Examples of the results are shown in Figure 37 to Figure 40. Figure 37(a) shows an example of the original signal with 15 spikes and a total of 200 samples. The noise was added to the original signal and fed to the two algorithms. Figure 37(b) shows the output signal after applying the Circulant sampling matrix and the Bayesian technique to the signal with added noise. Figure 37(c) represents the output signal after applying random sampling and basis pursuit technique to the original signal with added noise. As one can see in Figure 37(c), the output signal has more fluctuations than the output signal shown in Figure 37 (b). These fluctuations correspond to the null coefficients that are non-reconstructed as zero coefficients. Thus, the reconstruction with the Circulant matrix is more efficient compared to the reconstruction with random matrix.

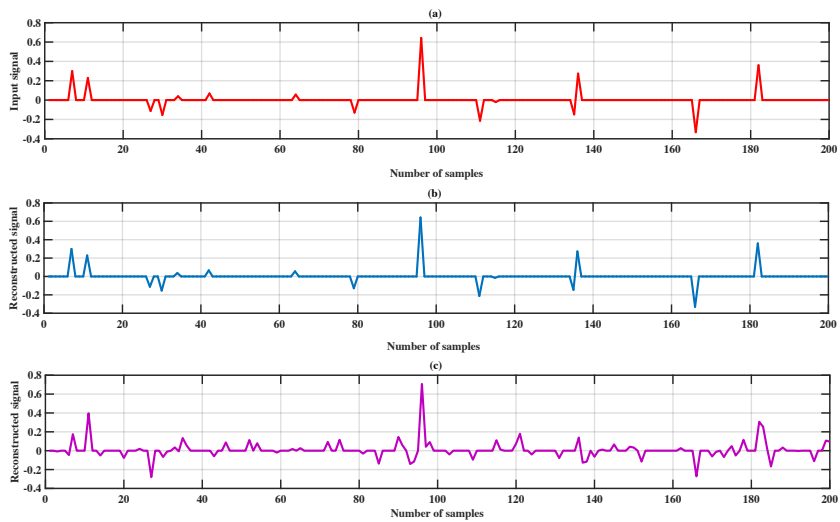

Figure 37: Example of input signal and the outputs after applying the recovery techniques (a) Input signal; (b) Output signal using Bayesian combined with Circulant matrix technique; (c) Output signal using Bayesian combined with random matrix technique.

Sampling time, $t_s$, was computed for each technique. The results show that random sampling matrix requires a great deal of time to process compared to the Circulant matrix. For example, for the signal seen in Figure 37(a), the $t_s$ of Circulant matrix is 0.06 ms while the $t_s$ of random matrix is 0.30 ms. This result shows how dense matrices are slow in terms of computation because of the high required number of measurements and the randomness in their coefficients.

Figure 38(b) and Figure 38(c) show the reconstructed signal using basis pursuit and Bayesian model with Circulant matrix technique, respectively. As one can see, for the Bayesian model, the output signal is similar to the original signal and the spikes are completely recovered. However, the output signal of basis pursuit presents more fluctuations than the output signal of the Bayesian model, as shown in Figure 38(b). Therefore, for high dimensional signal, the Bayesian reconstruction is more efficient than basis pursuit reconstruction. In addition, the sparsity level of the output signal is 14 for the Bayesian technique and 200 for the basis pursuit technique. The basis pursuit technique cannot estimate the exact value of each coefficient of the original signal, and it estimates the zero values as non-zero values with low magnitude. Moreover, the number of measurements to recover the signal needed by each technique is 15 for the Bayesian technique while it is 200 for the basis pursuit technique. Thus, the Bayesian technique is more efficient in reconstructing the original signal and also requires fewer measurements than the basis pursuit algorithm.

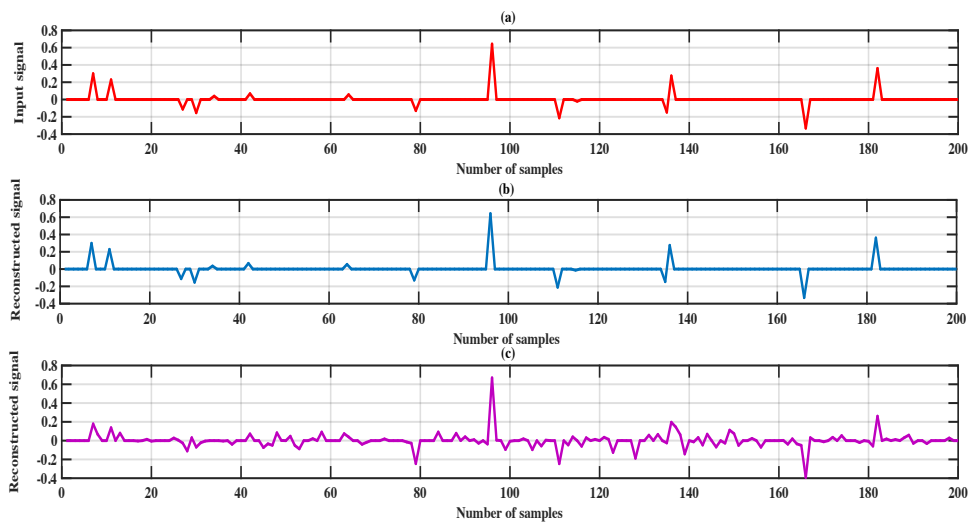

Figure 38: Example of input signal and the outputs after applying the recovery techniques (a) Input signal; (b) Output signal using Bayesian technique combined with Circulant matrix; (c) Output signal using basis pursuit technique combined with Circulant matrix.

Figure 39 shows *MSE* as a function of the number of samples for the two recovery techniques. As expected, for both techniques the *MSE* decreases with the increase of the number of samples. For *N* from 0 to 100, the Bayesian technique has lower *MSE* than the basis pursuit algorithm. For higher values of *N*, *MSE* of both techniques are slightly similar.

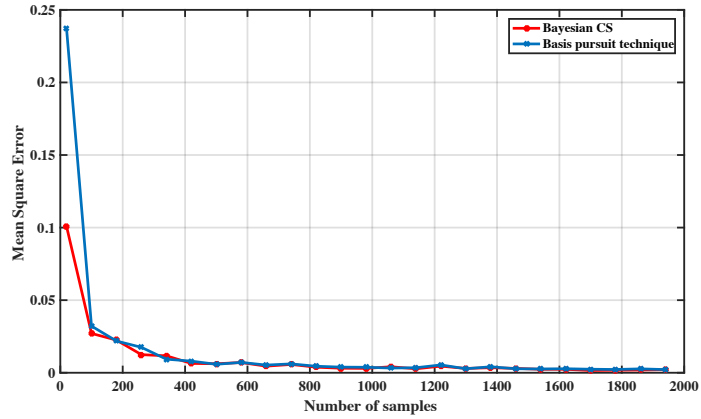

Figure 39: Mean Square Error as a function of number of samples N.

Figure 40 shows an example of results of MSE as a function of the sparsity level for the two techniques. As can be seen, the *MSE* values corresponding to the Bayesian technique and those corresponding to the basis pursuit technique are slightly similar and increase with the increase of the sparsity. This figure also shows that the more the number of non-zero elements of the signal increases, the more the reconstruction becomes inefficient. One can conclude that the two techniques minimize the *MSE* with the same way with the increase of sparsity level.

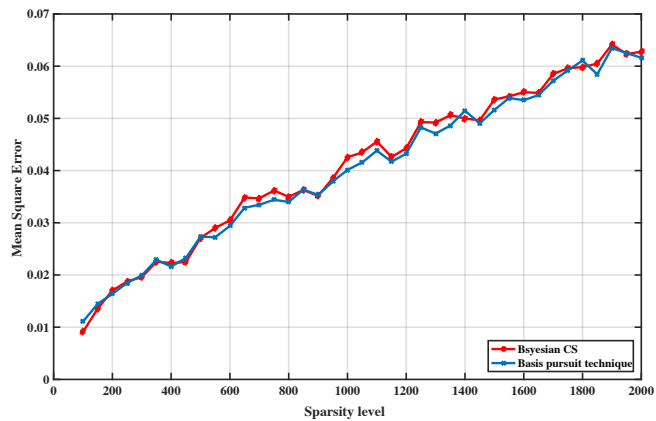

Figure 40: Mean square reconstruction error as a function of sparsity level *k*.

For the other metrics, Table 4 gives an example of results of the comparison performance. As shown in this table, for $R_e$, Bayesian technique has an average of 0.77% of recovery error. However, basis pursuit has 6.7% of recovery error. Thus, Bayesian is 80 times more precise than basis pursuit with the same matrix. Unlike basis pursuit, Bayesian technique permits to recover the signal with a very small errors number, which can be explained by the fact that it is able to deal with the uncertainty. Basis pursuit recovers the signal with high error level, which can be explained by the fact that it cannot handle uncertainty due to noisy measurements.

This table also shows that for the correlation metric, Bayesian technique presents an average of 100% of correlation while basis pursuit technique presents an average of 82.87%. Thus, both techniques present high correlation with values close to 100%, which indicates that the two signals are positively correlated. However, Bayesian technique presents better correlation.

For $t_r$, Bayesian technique requires an average of 0.90 ms to recover the original signal, but basis pursuit requires 7.20 ms, which represents 12 times higher, thus slower, than Bayesian technique. For $t_p$, Bayesian technique requires an average of 0.96 ms to process while basis pursuit requires an average of 7. 26 ms. Thus, Bayesian is 7 times faster than basis pursuit.

Table 4: Techniques comparison based on metrics[163]

|  | $R_e$ (%) | $C_c$ (%) | $t_r$ (ms) | $t_p$ (ms) |
| --- | --- | --- | --- | --- |
| Our technique [163] | 0.77 | 100.00 | 0.90 | 0.96 |
| Basis Pursuit with Circulant matrix [167] | 61.72 | 82.87 | 7.20 | 7.26 |

These examples of results show that the Bayesian technique with Circulant matrix is more accurate, faster, and deals with the uncertainty. In addition, our model requires less measurements, less sampling time, less recovery time, and less processing time. It also allows estimating the original signal with low sparsity level, high correlation, minimizes the mean square error, and handles the uncertainty during the encoding and decoding processes. Thus, our proposed approach includes the strengths of both Bayesian reconstruction and Circulant sampling matrix.

## V.4 Conclusion

In this chapter, we have proposed a compressive sensing approach that combines Circulant sampling matrix with Bayesian model. This approach reduces the randomness and deals with uncertainty during the compressive sensing processes. It also speeds up the scanning of the radio spectrum in cognitive radio networks. Its results are discussed and compared to those of basis pursuit with Circulant and random matrix techniques. The performance comparison involves several metrics including recovery success, speed, robustness, efficiency, memory, and certainty. The results show that Bayesian algorithm with Circulant matrix is more efficient and faster in compressing and recovering signals than Bayesian with random matrix as well as basis pursuit with either Circulant or random matrices.

In the next chapter, we are going to present the Bayesian recovery based Toeplitz matrix for wideband compressive sensing.

# Chapter VI

# BAYESIAN RECOVERY WITH TOEPLITZ MATRIX FOR COMPRESSIVE SENSING

In the previous chapter, we discussed the proposed Bayesian compressive sensing based Circulant matrix sampling. In this chapter, we are focusing on the Bayesian compressive sensing based Toeplitz matrix for signal sampling. To the best of our knowledge, no paper has been published that investigates the efficiency of Bayesian recovery with Toeplitz matrix sampling [164].

The remaining of this chapter is organized as follows. In Section VI.1, we analyze the related works. In section VI.2, we explain the simulation methodologies. The results of the simulation are discussed in section VI. 3. Finally, a conclusion is given at the end.

## VI.1 Introduction

Compressive sensing consists of directly acquiring the main information from a high dimension sparse signal using a sampling matrix technique. At the receiver, the original signal can be then recovered using recovery algorithm. For the sampling matrix, a few techniques have been proposed in the context of cognitive radio networks. These techniques are based on random measurements or random projections [22,103,173,174]. For instance, the authors of [103] proposed a random demodulator for wideband spectrum sensing that consists on multiplying the SU signal by pseudo-random numbers to compress it. This sampling technique can only be used for sparse signals with a finite set of pure sinusoids.

In [174], the authors proposed a random compressive sensing technique that allows capturing the SU signal by convolving it with generated random taps of a finite impulse response filter. The random taps, representing the sensing matrix coefficients, are generated as a normal Bernoulli distribution with zero mean and unit variance. In [98], the authors proposed another random compressive sensing technique for signal sampling based on the convolution of the SU signal with a random pulse. In [169], the authors proposed a sampling technique that multiplies the SU signal in its analog form by a random waveform at the Nyquist rate. In all these techniques, random matrices are used for sampling. However, these matrices are unstructured, dense, require high memory space, correspond to slow multiplication, and cannot deal with uncertainty [169]. Moreover, the implementation of such techniques using random measurements is costly.

For signal recovery, a number of recovery algorithms have been proposed for the spectrum sensing [174-178]. In [175], a cooperative compressive sensing technique was proposed for spectrum sensing using a distributed random sensing matrix. In [176], basis pursuit reconstruction technique was applied to reconstruct the original signal for spectrum sensing. Similarly, in [177][178], the authors developed a distributed compressive sensing in which a SU scans the radio spectrum by collaborating with other SUs. In [179], a time-frequency analysis approach was proposed to improve the compressive spectrum sensing performance and enable signal scanning under additive white Gaussian noise as well as Rayleigh and Rician fading channels. Basis pursuit was used for signal recovery for both channels.

However, all these recovery algorithms [174-178] use random matrices for sampling and do not deal with uncertainty in measurements. To address these problems, the authors of [130] proposed a Bayesian compressive sensing technique. Their method allows detecting the PUs location and transmission power with low sampling rate. The method is based on the error bar provided by the Bayesian model to measure the recovery efficiency. In another paper [179], the authors proposed a Bayesian compressive sensing technique that aims to estimate only the signal parameters instead of estimating the whole original signal using Bayesian estimation and discrete Fourier transform for sampling with random coefficients. Although these techniques [130][180] deal with uncertainty, they operate using random measurements, which limits their efficiency. As previously mentioned, random matrices are dense, not robust, and require high memory space, and processing time.

I propose a compressive sensing technique based on the Toeplitz matrix [180] for signal sampling and a Bayesian model for signal recovery. To the best of our knowledge, no paper has been published that investigates the efficiency of Bayesian recovery with Toeplitz matrix sampling. A very few papers performed the Bayesian recovery with a structured matrix [130,181]. Only in [178], a structured-based estimator has been developed using Bayesian recovery with Discrete Fourier Transform matrix [164].

Toeplitz matrices are structured and deterministic, which reduces the level of randomness in the matrix coefficients and also speeds up the signal acquisition process. The Bayesian model deals with uncertainty in the measurements.

## VI.2 Methodology

Figure 41 shows the block diagram of the proposed compressive sensing technique. This technique involves 3 processes: 1) sparsity representation, 2) signal acquisition using Toeplitz sensing matrix, and 3) signal recovery using Bayesian model.

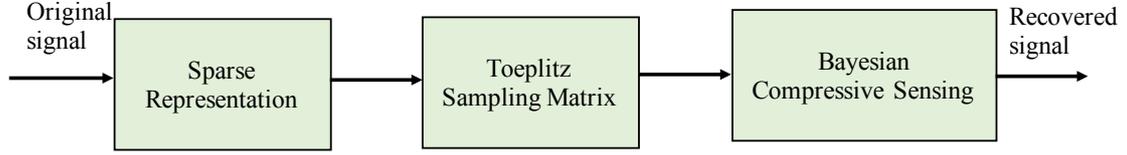

Figure 41: Block diagram for Bayesian compressive sensing model [164].

Figure 41 illustrates our methodology that includes the Toeplitz sensing matrix technique followed by the Bayesian recovery method. SU received signal is acquired through a multiplication with the Toeplitz sensing matrix, which is determined using a predefined vector, *T*, with *N*x*M* coefficients. AWGN is added to the product output to get the noisy measurements. The original sparse signal is then recovered from the noisy measurements by applying the Bayesian compressive sensing technique. The output of the model is the reconstructed signal with *N* samples and *k* sparsity.

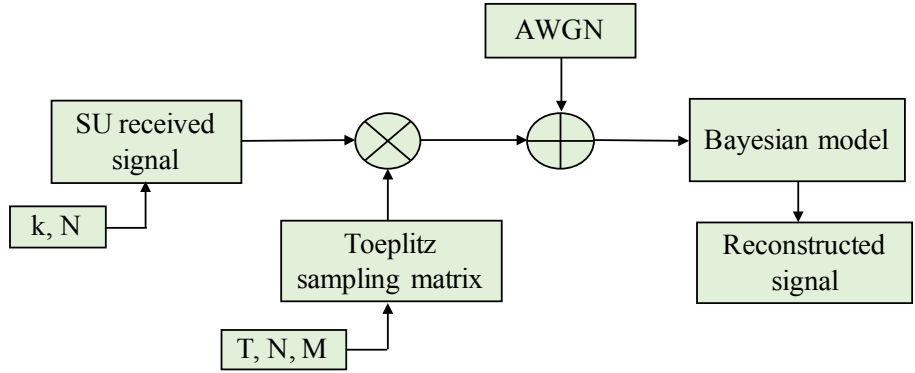

Figure 42: Bayesian compressive sensing with Toeplitz matrix [164].

Below are the Toeplitz sensing matrix and the Bayesian recovery techniques.

### VI.2.1 Toeplitz Sensing Matrix

Let consider a SU signal, $x \in R^N$, that is compressible and with *k* non-zero coefficients where $N \gg k$. Let *w* be a complex additive white Gaussian noise with zero mean and $\delta_w$ variance, and $y \in R^M$ be the noisy measurements with $M \ll N$, where *N* is the number of samples and *M* is the measurements number [164]. The noisy measurements can be expressed as:

$$y = M_T x + w \tag{54}$$

where $x = (x_1, x_2, \ldots, x_N)$ and $M_T$ is the sensing matrix, also known as sampling matrix or measurements matrix. Toeplitz matrix is an $M \times N$ structured partial Toeplitz matrix with a constant diagonal, $T$ [130]. It is a diagonal matrix whose rows are determined using a predefined vector and each left to right descending the diagonal is constant. The coefficients of $T$ respect the condition $t_{i,j} = t_{1+i,1+j}, \forall_{i,j}$, where $T = (t_1, \ldots, t_N)$ [180]. $T$ is determined with a probability distribution in order to generate structured values and minimize the stochasticity level in the sensing matrix compared to random matrices. Moreover, Toeplitz matrices satisfy the restrict isometry property with high probability, which guarantees the unicity of the recovered solution [182]. The restrict isometry property is given by

$$\exists \partial \epsilon \, (0,1) / \quad (1-\partial)\|x\|_2^2 \leq \|M_T x\|_2^2 \leq (1+\partial)\|x\|_2^2 \tag{55}$$

where $\partial$ is a positive scalar and $M_T$ is the partial Toeplitz sensing matrix. It is expressed as:

$$M_T = \begin{bmatrix} t_N & t_{N-1} & \cdots & t_1 \\ t_{N+1} & t_N & \cdots & t_2 \\ \vdots & \vdots & \vdots & \vdots \\ t_{N+M+1} & t_{N+M+2} & \cdots & t_M \end{bmatrix} \tag{56}$$

The signal compression consists on multiplying the original signal, $x$, with the Toeplitz sensing matrix. According to [180,182], the multiplication with the Toeplitz matrix is fast because of the small required number of measurements and the reduced random level in its coefficients. Subsequently, the compressed signal is recovered through the recovery algorithm. The number of measurements required to sample the signal with a conventional compressive sensing, i.e. using Gaussian random sampling matrix, is known to be equal to

$$M = O \, (\log \, (N \, / \, k)) \tag{57}$$

When using a Toeplitz matrix, the signal sampling requires less measurements number to acquire and recover all the signal coefficients, which makes the compressive sensing less complex than the conventional compressive sensing based on random matrices. For exact recovery, the required number is approximately proportional to the signal sparsity.

### VI.2.2 Recovery

For the recovery process, we used the Bayesian model to deal with uncertainty in measurements and enhance the compressive sensing efficiency. Bayesian recovery is based on the relevance vector machine. Estimating the unknown variables using this method consists of searching for the optimal solution of the underdetermined system using the available knowledge of the system. The optimal solution corresponds to the sparsest solution of the underdetermined system, which is a linear problem that can be expressed as:

$$\tilde{x} = argmin_{y=M_T x+n} \|y - M_T x\|_2^2 + z\|x\|_1 \tag{58}$$

where $\tilde{x}$ is the recovered signal and $z$ is a regulation parameter. To solve the equation (58), the Bayesian model uses the available knowledge about the system, which is signal sparsity assumption and the relationship between the noisy measurements and SU received signal.

Under the central limit theorem for $N>>M$, the original signal and the noise can be approximated as Gaussian variables with unknown parameters [183]. Thus, the noisy measurements can also be approximated as a Gaussian variable and associated to the unknown parameters, namely signal mean, $\mu_x$, signal variance, $\delta_x$, and noise variance, $\delta_n$. Bayesian model aims to estimate the unknown parameters by introducing new variables $\vartheta$ and $\tau$, called hyper parameters. Under the Bayesian analysis context, these parameters are called hyper to distinguish them from the other parameters that are not of prior distribution. The coefficients of the vector $\vartheta = (\vartheta_1, ..., \vartheta_N)$ represent the signal variance to be estimated. The scalar $\tau$ corresponds to the noise variance to be estimated. These two variables control the obtained solution in terms of accuracy and efficiency.

In our proposed model, noisy measurements, original signal, Toeplitz sensing matrix, and noise are the intervening variables. According to Bayes' rule, the Bayesian model aims to find the full posterior distribution for all the unknown variables, namely the original signal and the hyper parameters given the noisy measurements [184][185]. This probability is given by

$$P(x,\tau,\vartheta/y) = \frac{P(y/x,\tau,\vartheta)P(x,\tau,\vartheta)}{P(y)} \tag{59}$$

where $P(y/x, \tau,\vartheta)$ represents the conditional probability of the noisy measurements given the original signal and the hyper parameters, $P(x, \tau,\vartheta)$ represents the probability of the original signal and the hyper parameters, and $P(y)$ represents the prior probability of the noisy measurements. After simplifying the full posterior distribution formula, it can be expressed as:

$$P(x,\tau,\vartheta/y) = P(x/y,\tau,\vartheta)P(\tau,\vartheta/y) \tag{60}$$

The conditional probability density function of the noisy measurements given the original signal and the hyper parameter, $\tau$, can be formulated based on the $\mathcal{L}_2$ norm as:

$$P(y/x,\tau) = (2\pi\tau)^{-M/2} \exp(-\frac{1}{2\tau})\|y - M_T x\|_2^2 \tag{61}$$

This probability represents the Gaussian likelihood of the noisy measurements. It can be computed based on the prior information of the original signal and expressed as:

$$P(x/\vartheta) = \prod_1^M (2\pi\vartheta_i)^{-1/2} \exp(-\frac{\vartheta_i x_i^2}{2}) \tag{62}$$

Subsequently, the Toeplitz sensing matrix elements are generated based on the predefined vector $T$, which is similar to the one used for the SU received signal acquisition. In addition to the prior information about the original signal sparsity, the Toeplitz sensing matrix has to be considered known. Thus, the Bayesian recovery considers the prior density of the unknown signal $P(x/k_p)$ given the sparsity level and it can be expressed as:

$$P(x/k_p) = (k_p/2)^N \exp(-k_p \sum_I^N x_i) \tag{63}$$

where $k_p$ is the sparsity level of the signal $x$. This prior density expresses the available information about the signal sparsity that will be used for signal estimation. As previously mentioned, the Bayesian recovery computes the joint probability distribution of all the unknown parameters and the prior distribution of each coefficient of $x$ as well as the hyper parameters. In an iterative process, $\tau$ and $\vartheta$ are initialized as the initial posterior of the signal variance and the initial posterior of the noise variance, respectively.

Afterwards, the hyper parameters are estimated at each iteration and replaced by the new values. The marginal likelihood is them maximized based on the new values of $\tau$ and $\vartheta$. Based on the prior results from the previous iterations, the new values of $x_i$ are established considering the available information in terms of $\tau$, $\vartheta$, $M_T$, and $y$. Thus, the posterior probabilistic distribution of the original signal is defined as its conditional probability density given all the other parameters and it is given by

$$P(x/y,\tau,\vartheta) = (2\pi)^{-N/2} \exp(-\frac{1}{2}(x-\mu_x)^T \delta_x^{-1} (x-\mu_x)) \tag{64}$$

The algorithm is repeated until the convergence criterion of the posterior distribution is achieved and the algorithm converges to a global maximum with the respect of the values of the hyper parameters. At the end of the algorithm, the final estimated value of **τ** corresponds to the noise variance and the original signal is estimated. The estimated signal, $\tilde{x}$, can be expressed as a Gaussian distribution with mean $\mu_x$ and variance $\delta_x$:

$$\begin{cases} \mu_x = \tau \delta_x M_T' y \\ \delta_x = (V + \tau M_T' M_T)^{-1} \end{cases} \quad (65)$$

where $M_T'$ denotes the transpose of $M_T$ and $V = \text{diag}(\vartheta)$.

## VI.3 Results and Discussion

The proposed method as well as the two techniques classified under the iterative relaxation category and the Greedy category, namely the basis pursuit [22] and the orthogonal matching pursuit techniques [69], were implemented using Matlab as a platform. To evaluate and compare the performance of the proposed approach with those of the two other techniques, we generated over 100 different signals and used several metrics. These metrics are: sampling time, sparsity, required number of measurements, recovery time, processing time, recovery error, *SNR*, and MSE.

We first investigated the efficiency of the proposed sensing matrix and compared its results to the random sensing matrix using the sampling time, sparsity, and measurements number required for sampling. Then, we compared the performances of the three recovery techniques, namely Bayesian recovery, basis pursuit, and orthogonal matching pursuit in terms of recovery time, processing time, and Mean Square Error. The same results were found when using the 100 signals. Figure 43 to Figure 47 illustrate examples of the simulation results.

Figure 43 (a) presents an example of original signals with 20 spikes and 400 samples. Figure 43(b) represents the signal corresponding to the output of the Bayesian technique with Toeplitz matrix, while Figure 43(c) represents the signal corresponding to the output of the Bayesian recovery technique with the Gaussian random matrix. As can be seen, the output signal of Figure 43(b) is very similar to the original signal and has less fluctuations than the signal output of Figure 43(c). These fluctuations are the zero coefficients that are recovered as non-zero coefficients. Thus, sampling with Toeplitz matrix is more accurate than the sampling with the random matrix since both techniques used the Bayesian model for signal recovery.

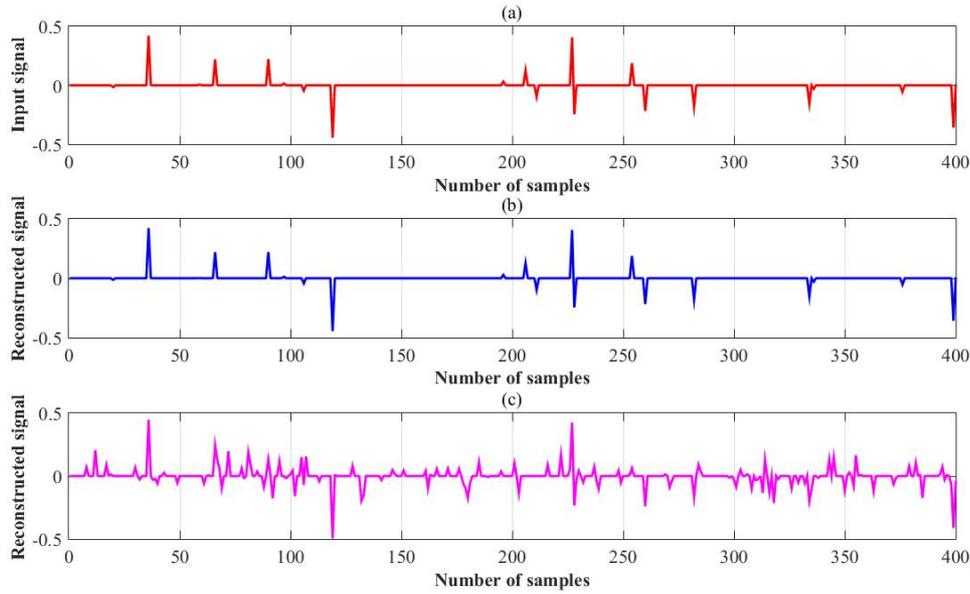

Figure 43: Example of input signal and the output signals after applying the Bayesian recovery technique (a) Input signal; (b) Output signal using Bayesian model with Toeplitz sensing matrix technique; (c) Output signal using Bayesian model with Gaussian random sensing matrix technique, SNR= 2dB, measurements ratio= 100/400.

Sampling time was calculated for each method, namely Toeplitz sensing matrix and random sensing matrix techniques with the Bayesian recovery. This metric corresponds to the period of time required to sample the signal. It evaluates the efficiency of the sampling method in terms of speed and how fast a method provides an output. The simulation results show that random sensing matrix requires a lot of time to acquire the signal than the Toeplitz sensing matrix. For instance, acquiring the signal shown is Figure 43(a) requires only 2.09 ms using Toeplitz matrix, while it requires 85.02 ms using random matrix. This result is due to the high measurements number required by the random matrix to compress the signal. The randomness of the matrix elements is another factor that increases the uncertainty level during the recovery process and results inefficient recovery. Thus, Toeplitz matrix is faster in terms of signal sampling speed and accurate in terms of randomness.

Figure 44(b), Figure 44(c), and Figure 44(d) show examples of output signals after applying the Bayesian method, basis pursuit, and orthogonal matching pursuit with Toeplitz sensing matrix method, respectively. As can be seen, for the Bayesian method, the spikes are completely recovered and the output signal is similar to the original signal shown in Figure 44 (a). However, the recovered signal using either the basis pursuit method or the orthogonal matching pursuit method represents more fluctuations, as illustrated in Figure 44(c) and Figure 44(d). Thus, the Bayesian method is more efficient than the basis pursuit method and orthogonal matching

pursuit with the same Toeplitz matrix, which demonstrates the superiority of our proposed approach.

The sparsity was computed for each technique. This metric corresponds to the number of spikes in the recovered signal after applying a recovery technique. For the signal example of Figure 44(a) with 20 spikes, the sparsity is 20 for the Bayesian method, while it is 400 and 209 for basis pursuit and orthogonal matching pursuit techniques, respectively. This result shows that the basis pursuit and orthogonal matching pursuit techniques are not able to estimate the accurate value of each spike of the output signal, which may be due to their inability to distinguish between spikes and null coefficients by associating non-null values to coefficients with very low amplitude. Moreover, since each simulated technique requires a specific number of measurements to process, the number of required measurements metric was calculated for each technique. The Bayesian method is able to achieve signal recovery with high recovery rate with only 123 measurements while the basis pursuit and the orthogonal matching pursuit require 397 and 400, respectively. Thus, compressive sensing with Bayesian method has better performance and it is more efficient in recovering high dimensional signals and it requires fewer measurements number than the other methods.

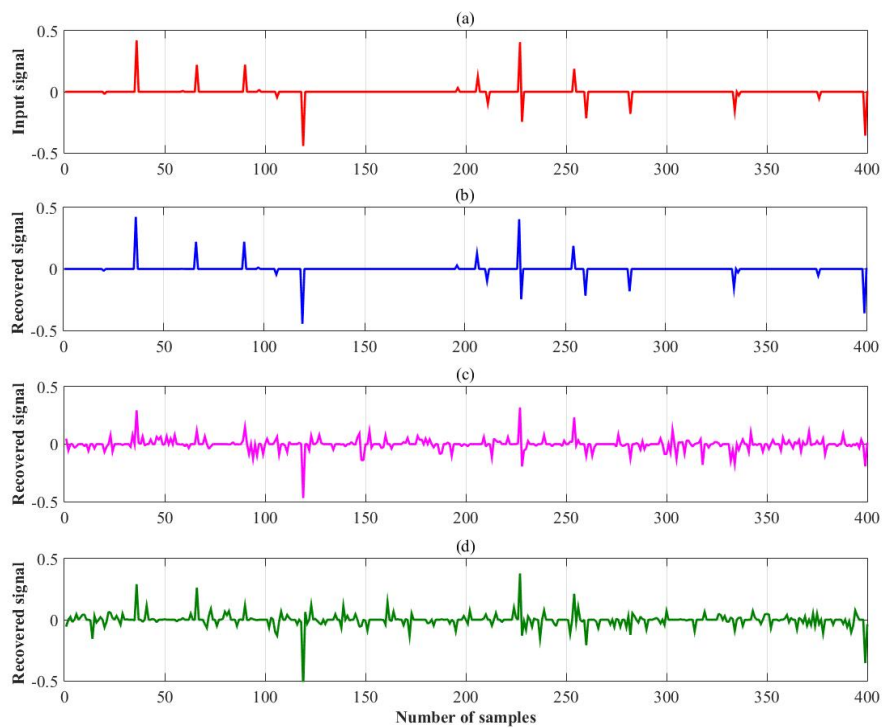

Figure 44: Example of input signal and the output signals after applying the recovery techniques (a) Input signal; (b) Output signal using Bayesian model with Toeplitz matrix; (c) Output signal using basis pursuit technique with Toeplitz matrix; (d) Output signal using orthogonal matching pursuit technique with Toeplitz matrix.

Recovery time is yet another metric used for the comparison. This metric corresponds to the period of time required to recover the compressed signal at the receiver. The Bayesian method requires 4.80 ms, while basis pursuit requires 70.11 ms and orthogonal matching pursuit requires 17.12 ms. The recovery time for the basis pursuit is 14 times higher than the one corresponds to the Bayesian method and the required recovery time for the orthogonal matching pursuit is 3 times higher than the one of the Bayesian method. Therefore, the Bayesian method is faster than the other methods.

In addition to the recovery time, the processing time was measured for each compressive sensing method. This metric represents the period of time required by the three compressive sensing processes for each technique. Using the Toeplitz sensing matrix method for signal sampling, the processing time required by the Bayesian method is around 7.01 ms which is 11 times faster than the basis pursuit (80.03 ms) and 5 times faster than the orthogonal matching pursuit (40.39 ms). Thus, the proposed approach is faster than the two other methods.

Recovery error metric was calculated to measure the error due to the reconstruction process. It corresponds to the norm of the difference between the original signal and the recovered signal divided by the norm of the original signal. It determines which technique represents a minimum error over the recovery process. For a signal with 400 samples, the recovery error for the proposed method is only 0.94 %, while it is 80.70 % for the basis pursuit and 33.42 % for the orthogonal matching pursuit. Thus, the recovery error is very small compared to the error of the two other techniques.

I also investigated the impact of the noise on the signal recovery. Figure 45 compares the recovery error of the recovery methods when the sampling matrix is a Toeplitz matrix as a function of the signal to noise ratio (*SNR*) from -10 dB to 10 dB. As expected, the recovery error is decreasing with the increase of *SNR* for all the three techniques. For low *SNR*, values under 4 dB, the recovery error of the Bayesian method is lower than those of the two other methods. For higher *SNR* (*SNR* > 4 dB), the recovery error of the three methods are very low and slightly similar. Thus, when SNR is under 4 dB, the Bayesian recovery is significantly more suitable and efficient than the other two techniques.

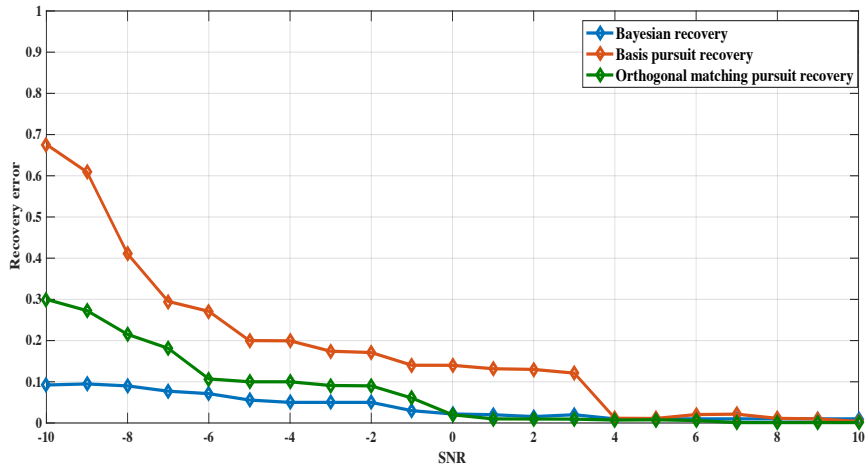

Figure 45: Recovery error as a function of *SNR* for Bayesian recovery, basis pursuit, and OMP with Toeplitz matrix, *N*=400.

Figure 46 shows $R_e$ as a function of the number of samples, *N*. As can be seen, the recovery error decreases with the increase of the number of samples for the three recovery methods. For a low number of samples, our proposed approach overcomes both recovery methods and minimizes the error. For a high number of samples, the recovery error of the Bayesian method is about 0.02, while it is about 0.2 for the other two recovery methods. Therefore, the Bayesian method presents the lowest recovery error even for a low number of samples.

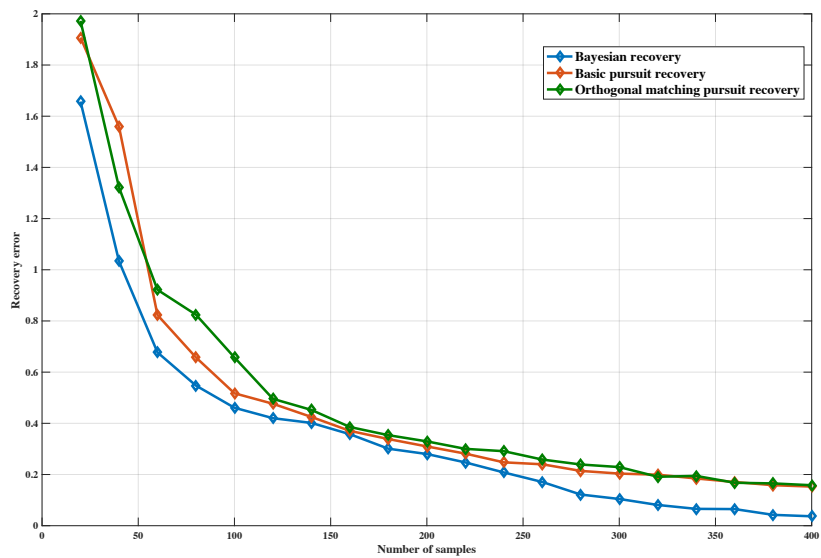

Figure 46: Recovery error as a function of *N* for Bayesian recovery, basis pursuit, and OMP with Toeplitz sensing matrix.

*MSE* was also used as a metric to assess the performance of the three recovery methods in terms of dealing with uncertainty due to the noise. This metric represents the average magnitude of the squared difference between the recovered signal and the original signal. Figure 47 shows *MSE* Vs. *SNR*. As expected, for all the methods, *MSE* decreases with the increase of *SNR*.

Bayesian recovery has lower *MSE* value than basis pursuit and OMP methods. At low *SNR*, uncertainty due to the high noise level increases and *MSE* of the Bayesian recovery is always lower than the other methods. Thus, Bayesian recovery is robust against noise and can deal with uncertainty in measurements leading to low $R_e$.

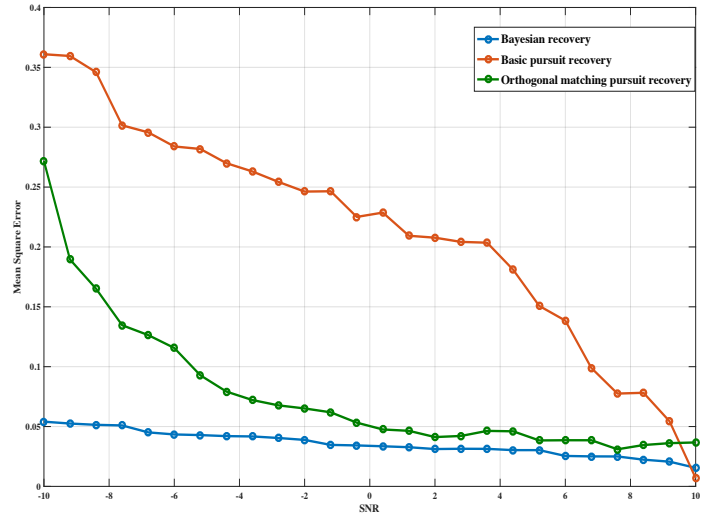

Figure 47: MSE as a function of *SNR* for Bayesian recovery, basis pursuit, and OMP with Toeplitz sensing matrix.

These results show that Toeplitz sensing matrix technique combined with Bayesian recovery is more efficient, robust, and handles uncertainty in measurements with low recovery error and *MSE* values. Bayesian recovery with Toeplitz matrix method represents less processing time, less recovery time, and less sampling time, which makes it faster than the existing methods. It also requires a smaller number of measurements to process and recover the signal.

Therefore, with our approach, the spectrum detection has been enhanced in terms of sensing time and accuracy. The time a SU needs to sense the spectrum becomes shorter compared to the existing techniques. In addition, fast sensing time allows enhancing the sensing security and protecting the spectrum from different attackers that can access to the spectrum while long cooperative sensing.

## VI.4 Conclusion

In this chapter, we have proposed an efficient and fast compressive sensing method for spectrum sensing, in which the Toeplitz sensing matrix and Bayesian model are combined to deal with uncertainty and reduce the randomness in measurements.

In the next chapter, we are going to present a real time spectrum scanning technique based compressive sensing for cognitive radio networks.

# Chapter VII

# REAL TIME SPECTRUM SCANNING BASED COMPRESSIVE SENSING FOR COGNITIVE RADIO NETWORKS

In the previous chapters, we discussed the Bayesian compressive sensing based Toeplitz sensing matrix to enhance the scanning process and reduce the proceeding time. In this chapter, we are presenting a real-time spectrum occupancy survey based on software defined radio units. we propose to adopt compressive sensing at the SU receiver before performing the spectrum sensing to speed up the processing time. The proposed approach aims to perform the wideband spectrum scanning on the sampling in real time using USRP equipment[186].

The remaining of this chapter is organized as follows. In Section VII.1, we analyze the related works. In section VII.2, we explain the simulation methodologies. The results of the simulation are discussed in section VII. 3. Finally, a conclusion is given at the end.

## VII. 1 Introduction

Spectrum sensing techniques allow measuring the spectrum occupancy over time and frequency. A number of spectrum occupancy surveys were conducted over the world whether for wideband or for licensed frequency bands. Examples of these surveys are presented and discussed in [187-191]. In [187], the authors conducted a spectrum occupancy study over a narrow range of frequency using a low noise amplifier followed by a spectrum analyzer. Energy detection was adopted as the simplest spectrum sensing technique with fixed threshold and 1% false alarm. In [188], a spectrum measurements survey was performed over a short frequency range to measure how much the spectrum is occupied and what are the free bands at different locations using energy detection with a predefined threshold. In [189], a campaign was conducted to measure the spectrum occupancy over space, time, and frequency simultaneously. The survey was performed using the power spectral density of the measured spectrum compared to a constant threshold. Similarly, in [190,191], energy detection technique was employed for spectrum scanning in a long frequency range with costly setup. In [45], the authors proposed a spectrum survey using Euclidian distance based detection using software defined radio (SDR) units, namely Universal Software Radio Peripheral (USRP) devices. The measurements were performed and compared to those of energy detection and autocorrelation based detection.

Most of these surveys [187-191] were performed in some frequency ranges using energy detection with a predefined threshold. This approach allows using only the signal power for spectrum usage measurements. Because of its simplicity, energy detection was the preferred choice for almost spectrum surveys and studies. Yet this technique has high false alarm rates and cannot perform well under random noise, which makes it inefficient for accurate spectrum measurements [6]. Only [45] performed the occupancy survey using autocorrelation and Euclidean distance based detection. In fact, these techniques require high sensing time to scan ranges of frequencies or several GHz. In addition, these techniques were performed by scanning channels one by one using several antennas over a limited frequency range, which results in high processing time and hardware cost. And, most of the surveys performed in specific and limited ranges of frequency, which does not include the wideband spectrum.

For wideband spectrum, these scanning techniques are not sufficient. Moreover, for measurements with a spectrum analyzer, it has been shown that these devices are costly, require filters and amplifier to get and analyze the measurements, and the results are not accurate compared to the measurements with USRP™ [192]. Hence, these techniques suffer from several limitations related to the scanning time and occupancy in the wideband spectrum. Thus, there is a great need for a spectrum measurements method that enables scanning the wideband spectrum in a very short time to allow SU find the available channels to use so quickly to balance between processing time, sensing capabilities, complexity, hardware cost, and measurements accuracy.

## VII.2 Methodology

Compressive sensing process is followed by the spectrum sensing process to detect the available channels through performing sensing on the compressed signals. At the same time, a SNR estimation technique is performed to help appropriately computing the probability of detection and the probability of false alarm. Indeed, two real-time experiments are performed simultaneously: (i) spectrum scanning with compressive sensing; (ii) spectrum sensing without compressive sensing [20]. Also, *SNR* estimation based particle swarm optimization (PSO) is performed in each experiment [192,193]. The experiments are performed simultaneously for several days to conclude about the spectrum occupancy in real time.

Figure 48 illustrates the general model of the experiments that include the processes: (i) compressive sensing process with Circulant matrix sampling [20,194]; (ii) spectrum sensing process with three detection techniques; (iii) SNR estimation based PSO, (iv) decision making process in which evaluation metrics are performed and analyzed for performance assessment.

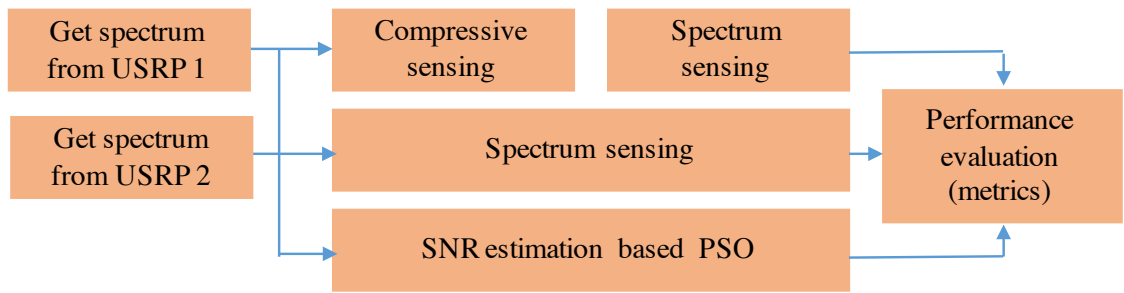

Figure 48: General model of the spectrum scanning experiments [186].

Compressive sensing consists on capturing the signal in its compressed form removing all the null coefficients to speed up the spectrum scanning and minimizes its complexity [194]. It involves three main processes, namely sparsity representation, sampling matrix, and recovery. For the sparsity representation, a signal is sparse when it has more null coefficients than non-null coefficients. SU received signals are assumed to be sparse based on the spectrum underutilization in cognitive radio context [20]. For the sampling matrix, the sparse signal is compressed by multiplying it with a measurements matrix. A number of compressive sensing techniques were proposed to enable acquiring signals and extracting only the main information instead of acquiring the whole signal [20]. Through comparing the different existing techniques [20][194], it has been demonstrated that Circulant matrix sampling technique can compress the signal with minimum processing time and measurements number, which justifies why adopting this technique for our experiments. For the recovery, the compressed signal can be recovered at the receiver using recovery algorithms.

In this work, the compressive sensing is performed only on the sampling without recovering the signal as it is not necessary to recover the compressed signal to fulfill our main objective, which is to perform the spectrum scanning on the sampled received signals and not the whole signal. The removed part of the signal after the matrix sampling includes only null coefficients based on the signal sparsity and the main information is kept for the spectrum sensing. As in [94], this compressive sensing approach is known as the compressive signal processing and allows solving the signal processing problems directly from the compressed signals.

### VIII.2.1 System Model

The first experiment consists of performing the spectrum scanning with compressive sensing. Let us consider the signal, $x$, as the received signals by the USRP unit to perform scanning over the wideband frequency. The signal of interest is a high dimension signal with N samples and it is assumed to be sparse. A partial Circulant matrix is an MxN matrix defined by a cyclic permutation of a predefined vector, where M<< N. This vector corresponds to the first row of the matrix and it is generated with a suitable probability reducing the randomness in the matrix.

Let us consider a vector, V, generated with low probability of randomness ($p = 0.1$) [20,194]. To generate the Circulant matrix, each of its elements has to respect the following condition:

$$M_{v(k,l)} = M_{v(l-k)} \tag{66}$$

where $V = (v_0, v_1, ..., v_{N-1})$, and $k, l \in [0, N]$. Circulant matrix is expressed as:

$$M_v = \begin{pmatrix} v_0 & v_1 & \cdots & v_{N-1} \\ v_{N-1} & v_0 & \cdots & \vdots \\ \vdots & \vdots & \vdots & v_1 \\ v_1 & \cdots & \cdots & v_0 \end{pmatrix} \tag{67}$$

Acquiring the signal consists on multiplying it with the partial Circulant matrix and extract only M samples from N of the signal, x. This multiplication is expressed as:

$$S_c = M_v\, x + w \tag{68}$$

where $x_c$ denotes the noisy measurements and w denotes the noise error. Hence, the compressive sensing system is given by

$$\begin{bmatrix} x_{c,1} \\ x_{c,2} \\ \vdots \\ x_{c,M} \end{bmatrix} = \begin{pmatrix} v_0 & v_1 & \cdots & v_{N-1} \\ v_{N-1} & v_0 & \cdots & \vdots \\ \vdots & \vdots & \vdots & v_1 \\ v_1 & \cdots & \cdots & v_0 \end{pmatrix} \begin{bmatrix} x_1 \\ x_2 \\ \vdots \\ \vdots \\ x_N \end{bmatrix} + \begin{bmatrix} w_1 \\ w_2 \\ \vdots \\ w_M \end{bmatrix} \tag{69}$$

After performing the compressive sensing and acquiring the SU received signals, spectrum sensing is then proceeded to detect the available channels among the scanned bands as:

$$y(n) = \begin{cases} w(n) & H_0: \text{PU is absent} \\ h * x_c(n) + w(n), & H_1: \text{PU is present} \end{cases} \tag{70}$$

where $n=1....N$, N is the number of samples, $y(n)$ is the SU received signal, $x_c(n)$ is the compressed signal, $w(n)$ is an AWGN noise, h is the complex channel gain of the sensing channel, and the hypothesis $H_o$ and $H_1$ denote respectively the absence and the presence of PU signal. The detection is performed by comparing the detector output with a threshold to decide between the two hypotheses $H_o$ and $H_1$[1][30]. The detection decision is performed as:

$$\begin{cases} \text{if } T \geq \gamma, & H_1 \\ \text{if } T < \gamma, & H_0 \end{cases} \quad (71)$$

where $T$ and $\gamma$ denote the detector output and threshold respectively, $H_0$ implies the PU signal is absent, and $H_1$ implies the PU signal is present. Three techniques were performed to enable spectrum sensing and allow getting knowledge about the spectrum occupancy. These techniques are energy, autocorrelation, and Euclidean distance based detection [30][31]. For each technique, the detector output is computed. The energy detector output is the signal energy as given by

$$T_{ED} = \sum_{n=0}^{N} y(n)^2 \quad n=1 \ldots N \quad (72)$$

where $T_{ED}$ denotes the received signal energy and $N$ denotes the number of samples [30]. The autocorrelation based detection output is based on the value of the autocorrelation coefficient of the received signal at lag $\tau$, which can be expressed as:

$$R_{y,y}(\tau) = \int_{-\infty}^{+\infty} y(t)\, y^*(t-\tau)\, dt \quad (73)$$

where $R_{y,y}(\tau)$ denotes the autocorrelation function of the signal $y(t)$, $t$ denotes time, and $\tau$ denotes the autocorrelation lag [30][31]. Euclidean distance based detection output is the distance between the signal autocorrelation and a reference line, which is expressed as:

$$D = \sqrt{\sum (R_{y,y}(\tau) - R)^2} \quad (74)$$

where $D$ denotes the Euclidean distance, $R$ denotes the reference line $R = (\frac{2}{M})i + 1$, $M$ denotes the number of lags of the autocorrelation, and $0 \leq e \leq \frac{M}{2}$ [4]. Therefore, the detector outputs are compared with suitable thresholds for each detection technique to decide about the PU existence at the scanned bands.

Spectrum scanning with compressive sensing experiment is performed in parallel to the *SNR* estimation based PSO in order to collect *SNR* values of the received signals by the USRP in real time. *SNR* plays a very important role in the wideband scanning process in the sense that it impacts the detection and false alarm detection rates. Knowing the *SNR* variable during the scanning process is required to help identify the available channels as well as taking the right decision about the spectrum occupancy. The SU received signals are noisy and the noise component needs to be distinguished from the signal component. SNR cannot be measured but only estimated because there are no methods yet that allow measuring the power of the signal or the noise from the received noisy signal. Thus, SNR can only be estimated with certain amount of accuracy using estimation techniques [17].

A number of SNR estimation techniques were proposed in the literature including maximum likelihood estimator [195], wavelet based SNR estimator [196], and Eigenvalue based SNR estimator [197]. Eigenvalue-based SNR estimation is based on the eigenvalues of the covariance matrix and it depends on the number of samples, the number of eigenvalues, and the Marchenko-Pastur distribution size [193]. To optimize these three key parameters, particle swarm optimization algorithm is adopted. The SNR estimation based PSO is performed by computing the eigenvalues of the covariance matrix that minimizes the mean square error of the received signals [198]. PSO being a global search optimization technique enables locating the optimal number of eigenvalues by minimizing the estimation error, which increases the accuracy of the SNR estimation. As *SNR* can be only estimated, we used the *SNR* estimation technique based on particle swarm optimization algorithm [192][193].

For the second experiment, it consists of performing spectrum scanning directly on the received signals without intervening the compressive sensing. It is the conventional spectrum scanning [20]. The three detection techniques are performed directly on the signals received from the USRP. Moreover, the *SNR* estimation based PSO is also performed to collect the SNR over the scanned bands in real time. The experiments were implemented and extensively tested for different values of the detection threshold and the number of samples in order to set up the optimal input parameters of the experiments. The performance evaluation is performed based on a number of metrics for performance assessment of the proposed method in terms of detection rate, accuracy, and complexity. These metrics include occupancy, probability of detection, probability of false alarm, *SNR*, processing time, and number of channels scanned.

### VIII.2.2 Experiments set up

The experiments, wideband spectrum scanning based compressive sensing and conventional spectrum scanning, were performed simultaneously in real time over a wide range of frequency based on SDR units. SDR is introduced to substitute the complex signal processing units for transistors and receivers with more flexibility, re-configurability, and low cost. SDR units were used for signal receiving over the wideband spectrum, namely USRP N200, which can operate from DC to 6 GHz. The USRP works using UBX 40 USRP daughterboard that covers 10 MHz to 6 GHz with 40 MHz of bandwidth. Thus, the materials list used in the experiments includes 2 USRP N200 equipment, 2 wideband antennas LP0965 Log Periodic PCB range from 850

MHz to 6.5 GHz to covers all the bands of interest and performs the scanning once instead of opting for multiple antennas, and 2 computers supporting Gigabit Ethernet where GNU Radio based Python is running to store the received data in real time as illustrated in Figure 49.

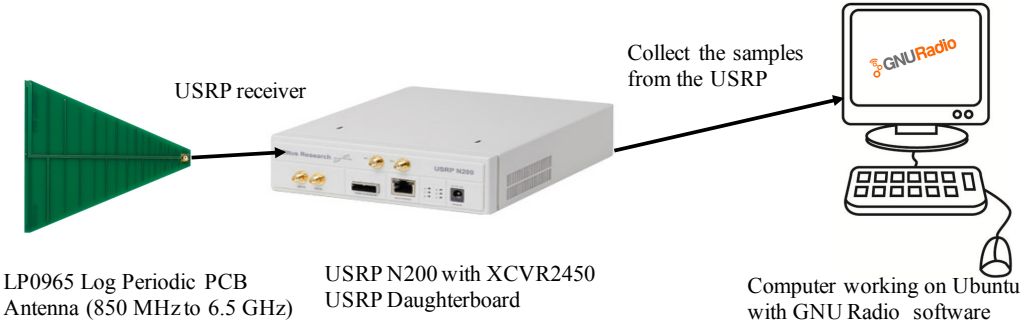

Figure 49: Experiments set up.

The spectrum survey aims to scan the wideband spectrum covering GSM 800 (D/L) and 1900 (D/L), Wi-Fi 2.5 and 5.4 GHz. Only ranges of frequencies for downlink are considered. Table 5 represents the list of frequency bands to be scanned with their ranges, channel spacing, and number of channels.

Table 5: Radio frequency bands for scanning.

| Radio frequency bands | Frequency ranges (MHz) | Channel spacing (MHz) | Number of channels |
|---|---|---|---|
| GSM-850 (D/L) | 869-894 | 3,2 | 11 |
| GSM-1900 (D/L) | 1930 -1990 | 3,2 | 25 |
| Wi-Fi 2.4 GHz | 2402-2497 | 5 | 20 |
| Wi-Fi 5.8 GHz | 5725 -5875 | 5 | 31 |

The experiments start by declaring and initializing all the input parameters with specific values. In order to get more data to analyze over time, the experiments are performed for several times, which can be equal to the number of times a channel is scanned. The second step consists on adjusting the USRP to the center frequency to identify which band and which channel to scan. This step will be repeated when moving to the next channel. Afterward, the real signals are acquired from the USRP using Circulant matrix for sampling. During this step, only the essential information is captured and stored for the followed processes. Subsequently, the three detection techniques are implemented and preceded sequentially. Energy, autocorrelation function, and Euclidean distance are computed as well as the thresholds to perform the sensing. Based on the sensing results, decisions are made and saved in a file with corresponding time and frequency. These steps are repeated until scanning all the channels. At the same time, the previous steps are performed without compressive sensing and the results are stored in a different file with corresponding time and frequency. On the other hand, *SNR* is dynamically estimated and stored for each channel. Figure 50 summarizes the different involved steps.

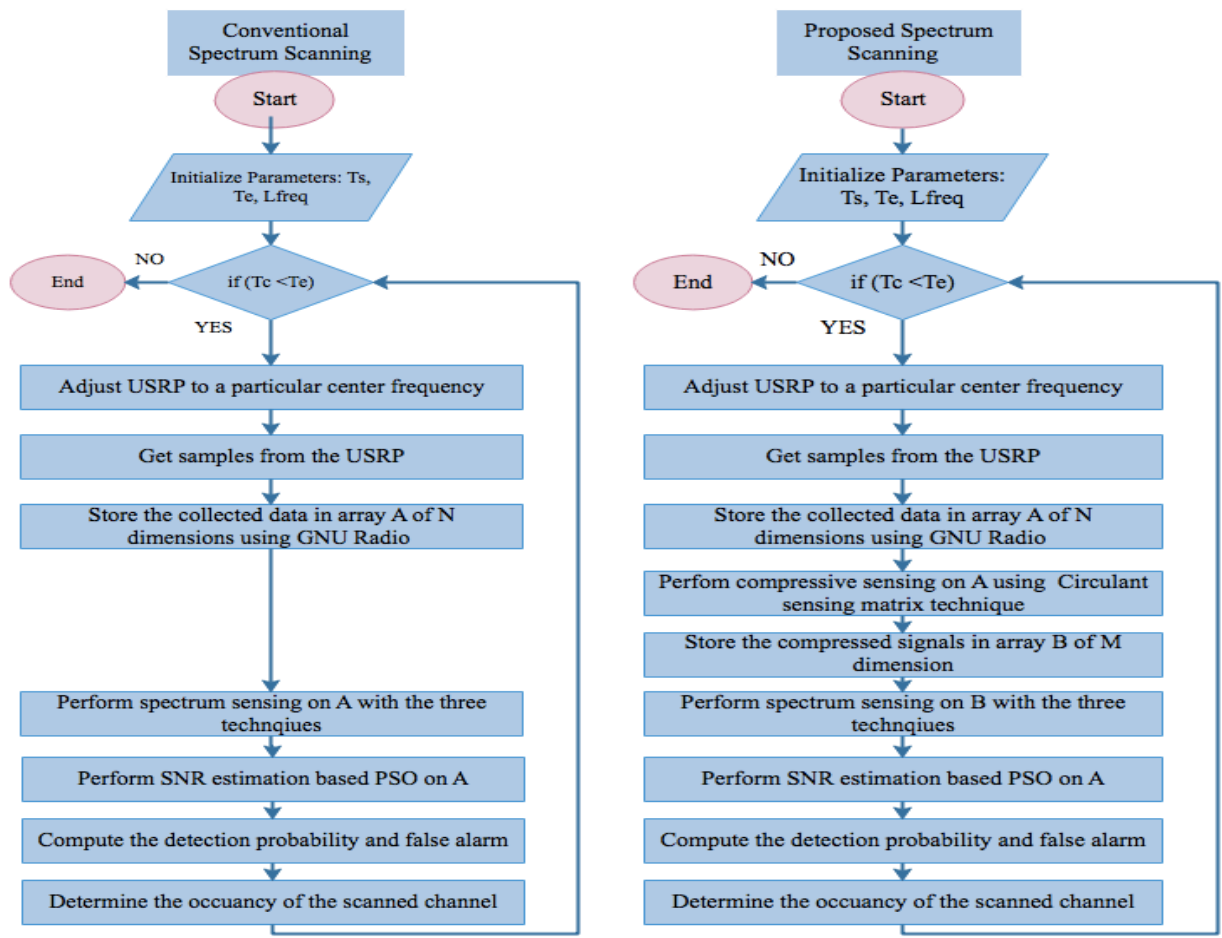

Figure 50: Flowchart of the spectrum scanning experiments.

## VII.3 Results and Discussion

The proposed method as well as the conventional scanning method were simultaneously implemented to analyze the wideband spectrum occupancy in real-time over the same location. We first investigated the efficiency of the proposed method in terms of occupancy level by comparing the three involved spectrum detection techniques, namely energy, autocorrelation, and Euclidean distance based detection. we also analyzed their probability of detection and false alarm taking into accounts the level of the estimated SNR. Then, we compared the performance of the spectrum scanning based compressive sensing and the conventional spectrum scanning in terms of processing time, and number of channels sensed per sec. The spectrum scanning surveys were conducted over the wideband frequency including all the channels for several weeks. Examples of these results are presented from Figure 51 to Figure 56.

Figure 51 to Figure 54 show the occupancy level of the spectrum scanning based compressive sensing during one week. The measurements were taken three times per day, namely 09 are to 12 pm, 12 pm to 3 pm, and 3 pm to 5 pm. Figure 51 represents the occupancy while scanning with energy detection using a fixed threshold ($1e^{-10}$). As expected, the channels are fully occupied all the period of times per day over the week. The high occupancy is due to the fact that energy detection suffers from high false detection and inability to distinguish between noise and signal as the energy includes also the noise. Thus, with energy detection, the wideband occupancy is very high and archives its maximum (100%) all the time.

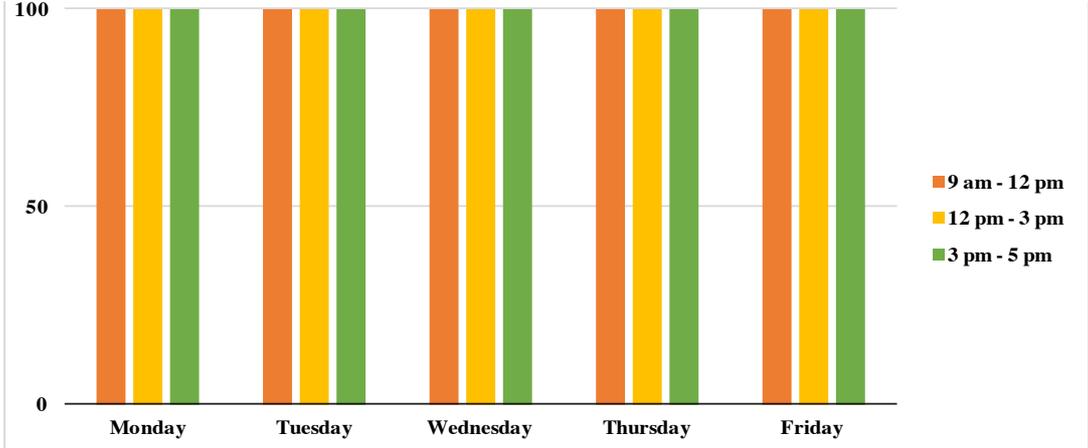

Figure 51: Occupancy of energy detection with compressive sensing, N= 3072, threshold= 1e-10.

Figure 52 illustrates the occupancy while scanning with autocorrelation based detection using a static threshold (0.90). As one can see, the occupancy varies over time during the day per week with ups and downs. Also, the occupancy is slightly higher during the rush time (12 pm to 3 pm) with peaks reaches 80%. Thus, the variable behavior of the occupancy can be explained by the fact that this technique is based only on the first lag of the autocorrelation, which may not have the entire information about the spectrum occupancy, or because of the internal thermal noise.

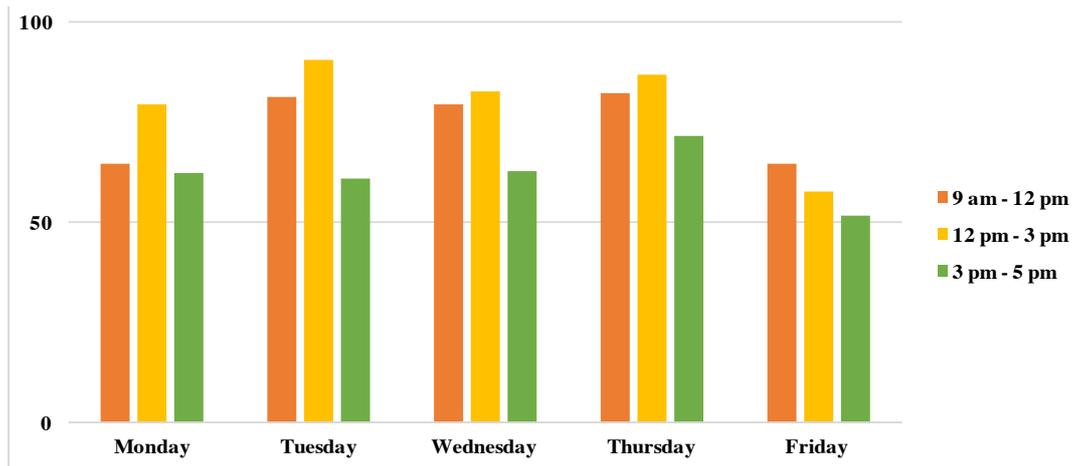

Figure 52: Occupancy of autocorrelation based detection with compressive sensing, N= 3072, threshold= 0.90.

Figure 53 shows the occupancy while scanning with Euclidean distance based detection with fixed threshold (0.95). As can be seen, the occupancy is low most of the time, which is close to be realistic. The survey represents some peaks during the rush time and up to 35% of the spectrum is occupied. Also, it represents some dips especially in the morning with 10% of occupancy. Thus, scanning using the Euclidean distance based detection is more reasonable as the peaks and dips correspond to the presence and absence of traffic during summer time in the location where the survey was conducted.

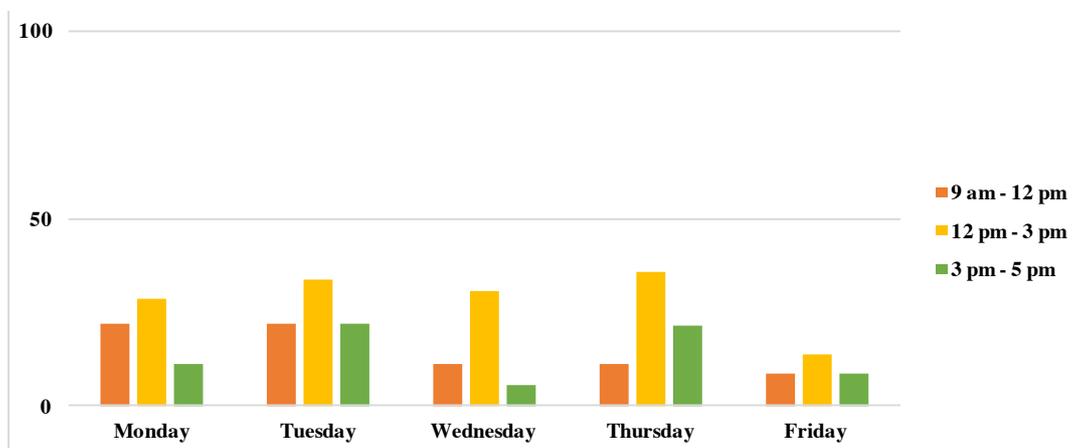

Figure 53: Occupancy of Euclidian distance based sensing with compressive sensing, N= 3072, Threhshold= 0.95.

Figure 54 summarizes the occupancies using the three sensing techniques with static thresholds over the day per week. The behavior of each sensing technique depends mainly on its performance. Through comparing the survey results, one can conclude that spectrum scanning with Euclidean distance based detection is close to be realistic as the survey was performed during summer time at the university campus where there is no much traffic during summer vacation. Also, the spectrum scanning was performed directly on the compressed signals and not the entire signals.

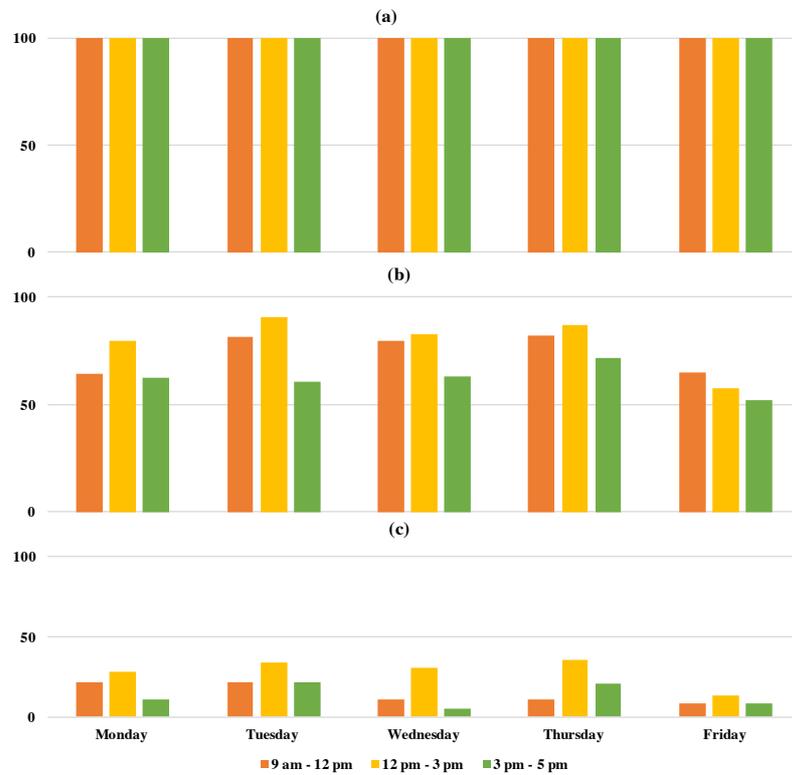

Figure 54: Spectrum occupancy (a) Energy; (b) Autocorrelation; (c) Euclidean distance based detection.

For the performance assessment of the sensing techniques, the probabilities of detection, $Pd$, and the probability of false alarm, $Pfa$, were measured taking into account the estimated *SNR*. Figure 55 shows the probability of detection against *SNR* for the three sensing techniques with selected thresholds for 3072 samples. As can be seen, the probability of detection increases with the increase of *SNR*, which was expected. Over *SNR*=-8dB, all the sensing techniques achieve the highest detection with 100% of $P_d$. Under *SNR*=-8dB, energy detection represents always high detection with 100% of $P_d$, which again confirms that this technique mixes up the signal with the noise leading to high detection rates. Autocorrelation and Euclidean distance based detection represent lower detection compared to energy detection. $P_d$ of the autocorrelation-based detection increases from 15% at *SNR*= -15dB to 100% at *SNR*=-8dB while $P_d$ of Euclidean distance based detection increases rapidly from 8% at *SNR*=-15dB to 100% at *SNR*=-10dB. Thus, Euclidean distance based detection represents better performance under low *SNR* over the scanning experiment on the sampling.

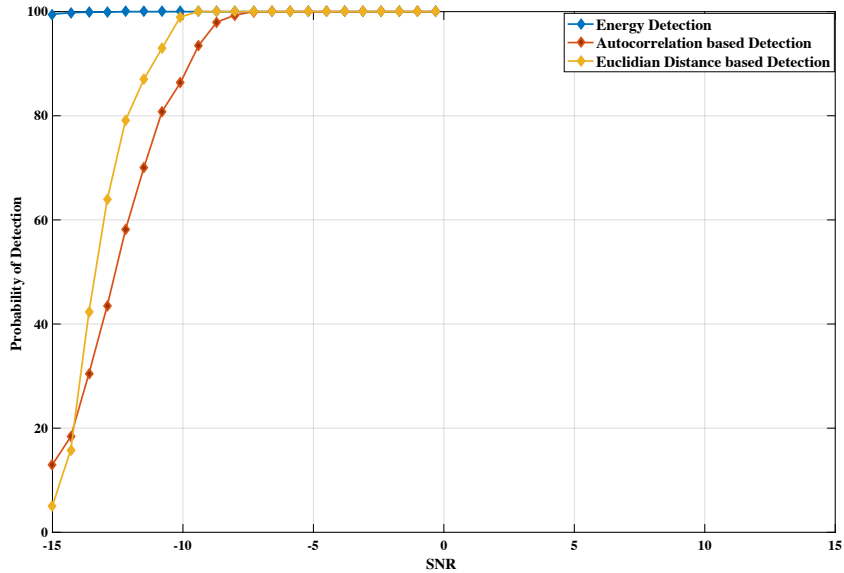

Figure 55: Probability of detection Vs. SNR.

Figure 56 represents the probability of false alarm against *SNR* of the three sensing techniques. As can be seen, for energy detection, the $P_{fd}$ decreases with the increase of *SNR* with high false alarm under low *SNR* (*SNR*<0dB), which is a normal behavior. It achieves its minimum value around *SNR*=3dB. For autocorrelation based detection, $P_{fd}$ varies very slightly with the increase of *SNR* with an average of 12% over the low and high *SNR*. For Euclidean distance based detection, $P_{fd}$ is almost null over the *SNR* range with very slight variation.

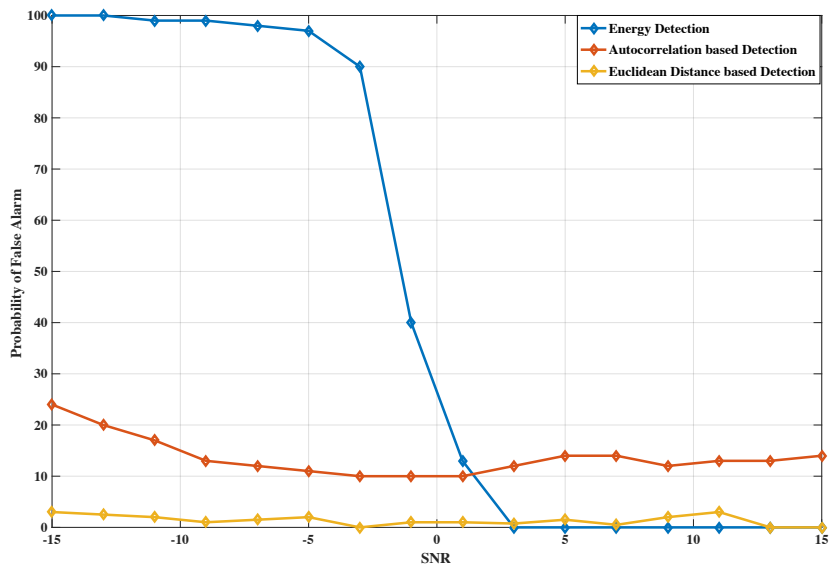

Figure 56: Probability of false alarm Vs. SNR.

After comparing and analyzing the performance of the spectrum scanning based compressive sensing using the three sensing techniques, we can conclude that Euclidean distance based detection performs well compared to energy and autocorrelation based detection with reasonable occupancy level over the week.

One the other hand, in order to evaluate the performance of our proposed method, it is necessary to compare it with the conventional scanning method by performing the scanning directly on the acquired signals from the USRP in real time. Metrics were introduced for this goal including, processing time and the number of channels sensed in a fixed period of time.

The measurements survey period, $t_p$, represents the total period of time to perform the occupancy survey for each scanning method. As the two surveys were performed for 8 hours per day per week, so the survey periods of the two methods are similar, but, the number of the acquired samples and the number of the sensed channels as well as the scanning time are different. In order to investigate how long each method takes to perform scanning, other metrics need to be considered and measured, namely, processing time and the number of channels sensed at a fixed period of time.

The processing time, also called scanning time, investigates how long does it take a scanning technique to sense one channel and it can be computed as the survey period divided by the total number of the sensed channels. It can be expressed as:

$$t_{sc} = \frac{T_s}{m * N_c} \tag{75}$$

where $t_{sc}$ denotes the scanning time, $T_s$ denotes the entire period of the survey, and *m* is equal to 3 and denotes the number of the performed sensing techniques assuming that they require the same amount to time to process, and *Nc* denotes the total number of the scanned channels during $T_s$. When performing the spectrum scanning on the compressed signals, the scanning time is around 2.75 secs per channel while it is around 9 secs when performing the spectrum scanning on the entire signals. More time is required for scanning without compressive sensing. Thus, the proposed method is faster than the conventional spectrum scanning method and does not require a great deal of time to process.

For the number of sensed channels metric, it provides how many channels can be scanned by each scanning method in a fixed period of time. For instance, during one hour of the survey, 5545 channels were scanned when using the proposed method while only 3236 channels were scanned when using the conventional scanning method. Less number of channels is scanned during the same amount of time without compressive sensing. Thus, the proposed method is faster and can sense a high number of channels in a fixed period of time compared to the conventional scanning method. Also, Compressive sensing allows speeding up the scanning process in spite of the high number of channels to be scanned or the high dimensional signals to be analyzed under the wideband spectrum scanning.

## VII.4 Conclusion

In this chapter, we have proposed an approach for wideband spectrum scanning based on compressive sensing. The proposed method as well as the conventional spectrum scanning method were implemented in real time using software defined radio units. These two experiments were implemented and simultaneously performed using USRP devices. Three sensing techniques were adopted, namely energy, autocorrelation, and Euclidean distance based detection. The two experiments were performed simultaneously to conduct a wideband spectrum occupancy survey over time.

The simulation results show that the proposed method requires less processing time for scanning and can sense a huge number of channels in a short amount of time while the conventional spectrum scanning can sense less channels in the same amount of time. Also, scanning with Euclidean distance based detection gives better results in terms of probability of detection and probability of false alarm as well as the occupancy level over time. The experiment may give better results if the survey was performed during a different time than summer with a recovery process.

In the next chapter, we are going to compare the one-bit compressive sensing and the conventional compressive sensing.

# Chapter VIII

# ONE-BIT COMPRESSIVE SENSING Vs. MULTI-BIT COMPRESSIVE SENSING

In the previous chapter, we presented a real time spectrum scanning technique based SDR unit. In this chapter, we are going to discuss and analyze one-bit compressive sensing and conventional compressive sensing.

To the best of our knowledge, a deep performance comparison between compressive sensing categories has not been investigated before [199-203]. Therefore, we reviewed the theory of both compressive sensing categories and analyzed their efficiency in a performance comparison to highlight our contribution. A number of metrics were used for the comparison: recovery SNR, recovery error, hamming distance, processing time, and complexity [204].

The remaining of this chapter is organized as follows. In Section VIII.1, we analyze the related works. In section VIII.2, we explain the simulation methodologies. The results of the simulation are discussed in section VIII.3. Finally, a conclusion is given at the end.

## VIII.1 Introduction

Compressive sensing approaches can be divided into two main categories: One-bit compressive sensing with one measurement and multi-bit compressive sensing with $M$ measurements, where $M$ is much less than the signal samples [20,205]. Multi-bit compressive sensing indicates the conventional compressive sensing [20,163]. It can sample high dimensional signals by only acquiring few measurements rather than acquiring the whole signal. It allows reducing the required number of measurements to acquire sparse signals. In fact, multi-bit compressive sensing is still face some problems in acquiring high dimensional signals under noise uncertainty [16,206]. Techniques with high recovery rate represent high processing time and complexity while fast techniques are not accurate, which resumes in a tradeoff between recovery rate and processing time [20,163,206]. Recently, one-bit compressive sensing has been proposed to overcome the multi-bit compressive sensing limitations and enhance the signal sampling efficiency. One-bit compressive sensing can recover sparse signals by using only the sign of each measurement with an extreme quantization. Preserving only the sign of the measurements is considered as the extreme case of sampling.

In one-bit compressive sensing, the extreme quantization adopted allows fast sampling and low hardware implementation cost, and complexity, which is due to the simplest way the quantizer works [200]. It compares the sign of the noisy measurements with zero, which requires very low cost and fast hardware devices. It is based on the most severe quantization that keeps only one bit per measurements. This bit represents its sign. The quantization corresponds to a simple comparator with zero with noiseless signals and exact sparsity. One-bit compressive sensing is easy to implement and can operate at very low power and high rates. It is also robust to noise [205].

For the sparsity requirement, sparsity is not always available in practical scenarios with multi-bit compressive sensing. In [205], the authors demonstrated the efficiency of one-bit compressive sensing to recover signals with a generalized sparsity model. This model considers the sparsity with the respect to a complete dictionary rather than a simple basis as in multi-bit compressive sensing. For the sensing matrix, one-bit compressive sensing can perform well with random matrices as well as the structured matrices including Circulant and Toeplitz matrices [207]. However, it cannot perform with non-Gaussian random measurements in contrast to multi-bit compressive sensing [208].

For the signal recovery, through analyzing the one-bit compressive sensing techniques for signal recovery, it has been shown that the binary iterative hard thresholding can be classified as the consistent algorithm among others because it performs with zero quantizer threshold and considers the signal sparsity while others do not [201,2013,209]. Therefore, there is a great need for deep performance analyze of the compressive sensing techniques to demonstrate the superiority of one-bit compressive sensing under certain conditions in cognitive radio networks. These conditions need to be identified by simulating a number of metrics that cover the most important aspects of signal sampling and recovery performance. In addition, very few papers performed the efficiency of one-bit compressive sensing for spectrum sensing [207,210].

A few papers that compare the efficiency of one-bit compressive sensing for spectrum sensing applications have been published [201,211]. These papers did not consider sufficient parameters for the performance analysis. The simulations were limited and the results were not compared to those of the conventional compressive sensing as an alternative solution. Moreover, to the best of our knowledge, a performance comparison between compressive sensing categories has not been investigated before. Thus, there is a great need for a comparison between the efficiencies of the compressive sensing categories under certain conditions for cognitive radio networks. These conditions need to be identified by simulating a number of parameters that cover the most important aspects in signal sampling and recovery performance.

## VIII.2 Methodology
### VIII.2.1 Compressive Sensing Approaches
#### VIII.2.1.1 One-bit compressive sensing

In one-bit compressive sensing, the noisy measurements, *y*, is quantized using a quantizer function, *f*. It is expressed as:

$$\begin{cases} y = f(Ax + w) & \text{Noisy model} \\ y = f(Ax) & \text{Noiseless model} \end{cases} \quad (76)$$

where $A \in R^{M*N}$ denotes the sensing matrix and e denotes AWGN noise added to the compressed signal during the acquisition process. Each coefficient of the measurements is quantized and only its sign is acquired. The one bit corresponds to one of the values $\pm 1$ for noiseless measurements model while it corresponds to a value that includes the noise error for noisy measurements model. One-bit compressive sensing shifts the concept of several measurements to one bit, which makes its implementation easy and fast [205,212].

One-bit compressive sensing has four main stages: sparsity, sensing matrix, measurements quantization, and recovery. It can handle only sparse signals and requires the RIP. A signal is considered sparse when most of its elements are null in some basis. RIP requires that the columns of the matrix have to be independent from each other with a size less than the signal sparsity [20,213].

In the sensing matrix stage, the signal is multiplied with a matrix to acquire the signal as in the conventional compressive sensing. Random matrix, Toeplitz, and Circulant matrix are examples of matrices used for signal sampling in one-bit compressive sensing [207]. In the measurements quantization stage, the compressed signal is forwarded to a quantizer block in which only the sign of each measurement is extracted forming the noisy measurements. The one-bit compressive sensing structure is illustrated in Figure 57.

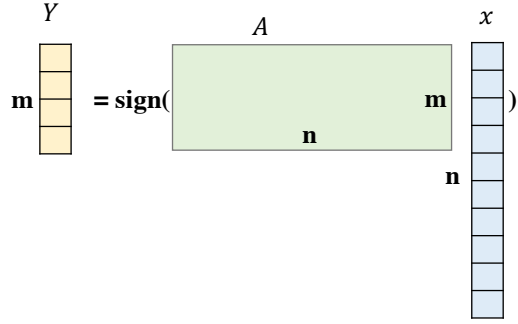

Figure 57: One-bit compressive sensing structure [204].

The noisy measurements can be expressed as:

$$Y = sign(Ax) \Leftrightarrow \begin{cases} Y_1 = sign(A_1 x) \\ \vdots \\ Y_m = sign(A_m x) \end{cases} \quad (77)$$

where $A \in \mathbb{R}^{n \times m}$ denotes the sensing matrix, $x \in \mathbb{R}^n$ denotes the original signal, and $y \in \mathbb{R}^m$ denotes the noisy measurements. The multiplication between each quantized measurement with the measurements must be always positive by respecting the following condition:

$$Ax_i \, sign(A_i x) \geq 0 \quad (78)$$

In the reconstruction stage, to recover the signal from only the measurements signs, a number of recovery algorithms have been proposed[201][203]. It is also possible to reconstruct the signal with algorithms dedicated to the conventional compressive sensing by treating the measurements as ±1 measurements when recovery error and noisy error are not considered. Recovering the original signal is performed by solving the following underdetermined system:

$$\tilde{x} = argmin_y \, \|x\|_1 \; s.t \; yAx \geq 0 \quad (79)$$

With only the sign of the compressed signal, it is clear that the signal magnitude is lost after the sampling. Recovering the signal can be only done by recovering the spikes directions (-1 or +1). However, it is possible to recover the whole signal with consistent reconstruction. Consistent reconstruction requires the sparse signal to be consistent with the measurements by adding another condition to the underdetermined system in Equation (75). This condition enforces the signal energy to be a unit with the use of $\mathcal{L}2$ norm [205]. The new underdetermined system can be reformulated:

$$\tilde{x} = arg_y min \, \|x\|_1 \; s.t \; y \, A \, x \geq 0 \; and \; \|x\|_2 = 1 \quad (80)$$

A number of reconstruction algorithms have been proposed to ensure a consistent reconstruction of the original signal. Examples of these techniques are fixed point continuation [202], renormalized fixed point iteration [214,215], gradient projection for sparse reconstruction [207], generalized approximate message passing [216], Bayesian for quantized compressed sensing [217], and binary iterative hard thresholding [201].

**VII.2.1.2 Multi-bit Compressive Sensing**

Multi-bit compressive sensing has three main stages: sparsity, sensing matrix, and reconstruction [20]. In the first stage, a high dimensional signal, $x \in R^n$, is presented in a specific basis in order to make it sparse. In the sensing matrix stage, the sparse signal is acquired by multiplying it with a sensing matrix to extract only m measurements from n. where *M* is smaller than *N*. This multiplication can be considered as the projection of the signal into a basis, also called dictionary, to only maintain the essential information of the signal. The multi-bit compressive sensing structure is illustrated in Figure 58.

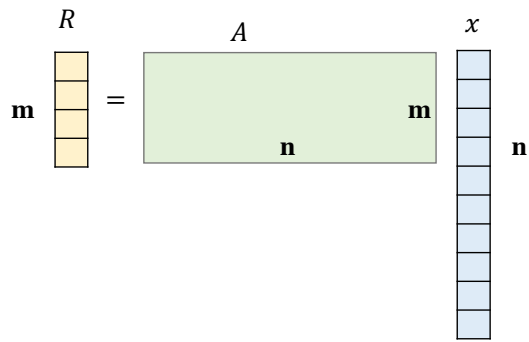

Figure 58: Multi-bit compressive sensing structure [20][195].

The compressed signal, *y*, can be expressed as the following in noiseless and noisy measurements models, respectively

$$y = A\,x \tag{81}$$

$$y = A\,x + w \tag{82}$$

where $A \in R^{m*n}$ denotes the sensing matrix and *w* denotes AWGN noise added to the compressed signal during the acquisition process.

A number of sensing matrix techniques have been proposed in order to achieve fast and efficient signal sampling. Examples of these techniques are random matrix, Toeplitz matrix, Circulant matrix, and deterministic matrix [20][163]. Multi-compressive sensing requires that these matrices meet a special requirement RIP [218].

In the reconstruction stage, the signal can be reconstructed from a few measurements at the receiver by estimating its coefficients using a reconstruction algorithm. This algorithm operates by solving the underdetermined system as an optimization problem [20][214]. In noisy model, the underdetermined system to be solved can be expressed as

$$\tilde{x} = argmin_{y=Ax+w}(\|x\|_1) \ s.t. \quad y = Ax + w \tag{83}$$

where $\tilde{x}$ denotes the reconstructed signal and it represents the sparsest solution among others. Examples of reconstruction algorithms include basis pursuit denoising (BPD), matching pursuit (MP), orthogonal matching pursuit (OMP), Bayesian recovery (BR), stage wise orthogonal matching pursuit (SOMP), compressive sampling matching pursuit (COSAMP), regularized orthogonal matching pursuit (ROMP), and generalized orthogonal matching pursuit (GOMP) [20,90,163,214].

In one-bit compressive sensing, the noisy measurements, $y$, is quantized using a quantizer function, $Q$. It is expressed as:

$$\begin{cases} y = Q(Ax + w) \text{Noisy model} \\ y = Q(Ax) \text{Noiseless model} \end{cases} \tag{84}$$

where each coefficient of the measurements is quantized and only its sign is acquired. The one bit corresponds to one of the values $\pm 1$ for noiseless measurements model while it corresponds to a value that includes the noise error for noisy measurements model. One-bit compressive sensing shifts the concept of several measurements to one bit, which makes its implementation easy and fast [211,202].

### VII.2.2 System Model
For the sparsity requirements, the signals are assumed to be sparse because of the underutilization of the spectrum in cognitive radio networks [20][6]. Let consider a sparse signal, $x$, with high dimension $n$. The simulation model of the compressive sensing is presented in Figure 59.

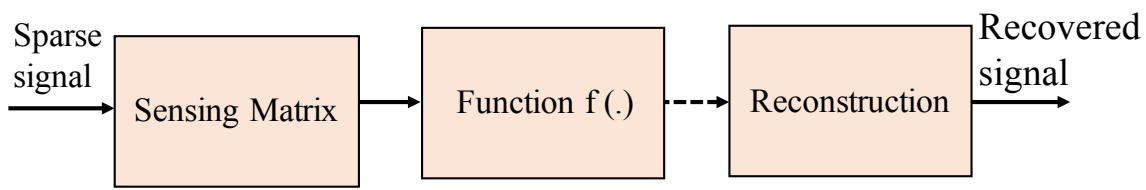

Figure 59: Compressive sensing system model [204].

For the sensing matrix, let *A* be a *m*x*n* partial Toeplitz matrix. The signal and the matrix are forwarded to the sensing matrix stage to perform the multiplication, *Ax*. When the noisy measurements model is considered, a noise is added to represent the sampling error and channel noise. The sparse signal, the partial Toeplitz matrix, and the noise are the inputs of our system model. This system model can be expressed as:

$$Y = f(Ax, w) \qquad (85)$$

where $f(.)$ represents the compressive sensing function. This function can be expressed as

$$f: \begin{cases} Ax + w & \text{Multi} - \text{bit compressive sensing} \\ sign\,(Ax + w) & \text{One} - \text{bit compressive sensing} \end{cases} \qquad (86)$$

$f(.)$ is a quantization function with one-bit compressive sensing.

For the recovery, each compressive sensing framework has a number of recovery algorithms with different performance. In our work, we selected binary iterative hard thresholding algorithm as an efficient recovery technique for one-bit compressive sensing, denoted as 1-Bit CS, while for the multi-bit compressive sensing three reconstruction algorithms were selected. Multi-bit compressive sensing recovery techniques can be classified into three categories: convex relaxation iterative, greedy, and Bayesian recovery algorithms [20]. In this work, we selected one technique from each category: basis pursuit denoising (BPD) from the convex relaxation iterative category, Bayesian recovery (BR) from the Bayesian category, and compressive sampling matching pursuit (COSAMP) from the Greedy category [20,214].

## VIII.3 Results and Discussion

The two techniques, one-bit compressive sensing and multi-bit compressive sensing with Toeplitz matrix, were implemented and extensively tested using a hundred of signals of different types to investigate their performances. To evaluate their performances, we used the metrics: recovered SNR, recovery error, hamming distance, processing time, and complexity.

The complexity of an algorithm corresponds to the efficiency of that algorithm to perform with high amount of data. The complexity of each of the two investigated techniques is given by:

$$M = O(log(N/k)) \qquad (77)$$
$$M_1 = O(k\,log(N/k)) \qquad (78)$$

where $M_1$ and $M$ denote the required measurements number by one-bit compressive sensing and multi-bit compressive sensing, respectively, $k$ is the signal sparsity, and $N$ is the sample number [2,21]. Processing time is also used as a metric. It corresponds to the time an algorithm needs to perform all the processes of the compressive sensing processes.

$R_e$ corresponds to the difference between the recovered signal and original one. It is defined as

$$R_e = \frac{\|x-\tilde{x}\|}{\|x\|} \tag{87}$$

Recovery signal to noise ratio metric, *RSNR*, represents the level of *SNR* at the receiver and it is computed by considering the original signal as the input and the reconstructed signal, $\tilde{x}$, as the output of the system [16,199]. It is defined as

$$\boldsymbol{RSNR} = \frac{\|x\|_2^2}{E(\|x-\tilde{x}\|_2^2)} \tag{88}$$

Hamming distance, $H_d$, between two signals is defined as the number of times two signals are different. It reflects the minimum number of times the original signal and the recovered signal are different. It is the number of non-zero elements of $H_d$ expressed as:

$$H_d = y - y_0 \tag{89}$$

where $y_0 = f(Ax, w)$ and $y = f(A\tilde{x}, w)$ [204].

In this work, we generated sparse signals with 2000 samples and 100 spikes at random positions. An additive white Gaussian noise is then added. While using similar sparse signals and sensing matrices, we performed different recovery techniques, namely BPD, BR, COSAMP, and 1-Bit CS. Examples of results are shown in Figure 60 through Figure 63.

Figure 60 shows the recovered *SNR* as a function of the number of samples of 1-Bit CS, BPD, BR, and COSAMP. It aims to compare the recovery performance of one-bit compressive sensing against multi-bit compressive sensing. As one can see, as the number of samples increases, the recovered *SNR* increases. Also, one-bit compressive sensing has higher SNR over the range of the number of samples than the multi-bit compressive sensing techniques, which have less *SNR* with slightly similar behavior. Thus, one can conclude that recovering signals with one-bit compressive sensing represents more efficiency and robustness to noise than recovering signals with multi-bit compressive sensing.

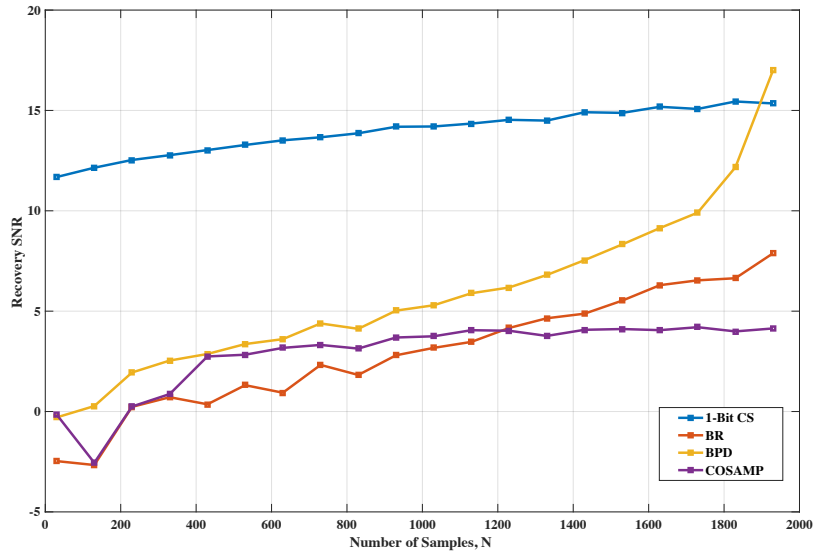

Figure 60: Recovery SNR as a function of the number of samples.

Figure 61 shows the recovery error as a function of *N*. It compares one-bit compressive sensing with BPD, BR, and COSAMP. As expected, with the increase of *N*, the recovery error decreases. In addition, one-bit compressive sensing represents less error when recovering high dimensional signals, followed by the COSAMP and BR. BPD represents high error, which decreases its efficiency in noisy measurements models. Thus, one-bit compressive sensing minimizes the recovery error and can reconstruct signals accurately.

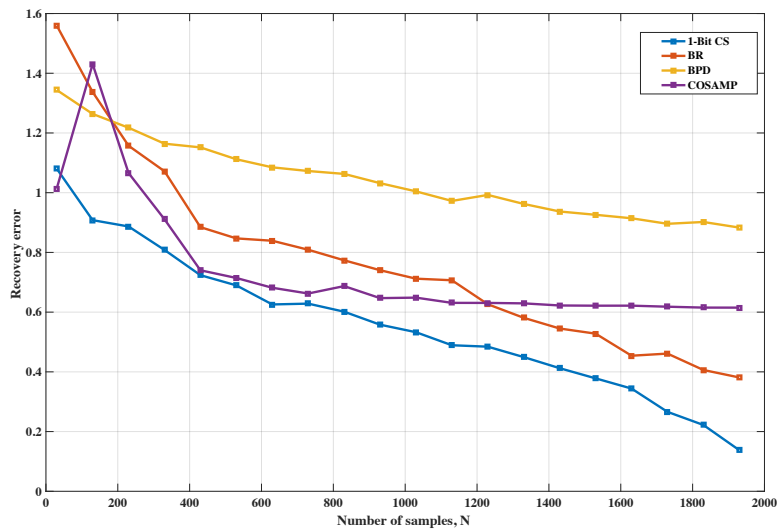

Figure 61: Recovery error as a function of the number of samples.

Figure 62 represents the hamming distance for different values of the number of samples. This metric shows how close is the recovered signal to the original one by counting the number of positions the original signal and the recovered signal are different. As it can be seen, the hamming distance of one-bit compressive sensing decreases with the increase of the number of samples. For *N* higher than 900, the hamming distance of one-bit compressive sensing is close to zero, which shows its efficiency in recovering high dimensional signals. COSAMP represents also less hamming distance and is slightly constant over high *N* starting at 800. BR represents more hamming distance compared to one-bit compressive sensing and COSAMP and it decreases with *N*. The hamming distance for BPD is constant and equal to 200 over *N*. This value corresponds to the number of measurements used in this simulation. Thus, one-bit compressive sensing represents less Hamming distance followed by COSAMP.

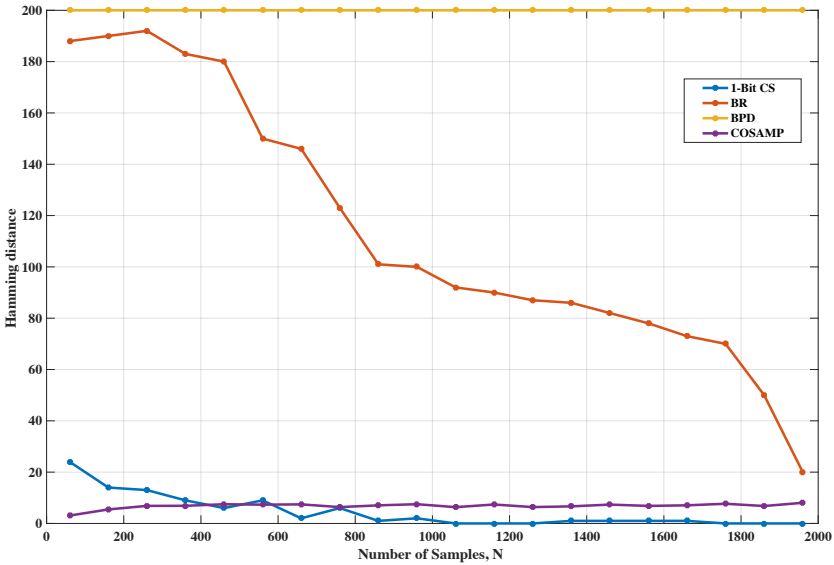

Figure 62: Hamming distance Vs. Number of samples.

By analyzing the number of measurements required by each technique, $M$ and $M_1$, represented in equation (77) and (78), one can conclude that multi-bit compressive sensing requires more measurements compared to one-bit compressive sensing. For instance, multi-bit compressive sensing requires 1073 measurements while one-bit compressive sensing requires only 500 when compressing a signal with 2000 samples and 50 non-zero coefficients.

Figure 63 shows an example of the results of the processing time as a function of the number of samples. As it can be observed, the processing time increases with the increase of the number of samples for all the reconstruction techniques. However, one-bit compressive sensing requires less processing time to perform high number of samples, while other techniques require high processing time to recover the original signal. One-bit compressive sensing requires less time followed by BR. BPD and COSAMP are very slow compared to the other techniques. Therefore, for high dimensional signals, one-bit compressive sensing is much better than the multi-bit compressive sensing including all its categories. It is fast and can perform well with high dimensional signals.

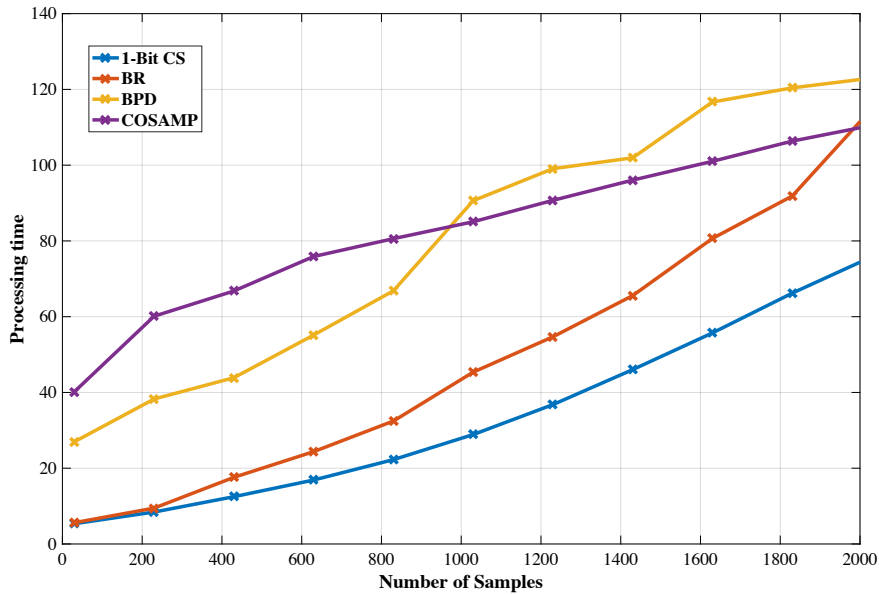

Figure 63: Processing time as a function of the number of samples.

Table 6 compares the two compressive sensing categories by summarizing the simulations results.

Table 6: Performance comparison

| Metrics | Multi-bit CS | One-bit CS |
|---|---|---|
| Recovered SNR | High | Low |
| Recovery error | High | Low |
| Hamming distance | High | Low |
| Processing time | Slow | Fast |
| Complexity | High | Low |
| Measurements number | High | Low |

These examples of results show that one-bit compressive sensing outperforms multi-bit compressive sensing in terms of recovered SNR, recovery error, and complexity. It is fast, less complex, and more efficient in recovering sparse signals.

## VIII.4 Conclusion

In this chapter, we have reviewed and analyzed both compressive sensing frameworks: one-bit compressive sensing and multi-bit compressive sensing. The simulation results shown that one-bit compressive sensing represents high recovered SNR, low recovery error, low processing time, reduced memory storage, and less complexity.

# CONCLUSIONS AND FUTURE DIRECTIONS

**Conclusions**

Cognitive radio network is a system able to learn from the environment and adjust its transmission parameters. With their awareness of the radio environment, SUs perform spectrum sensing to identify free channels in order to efficiency exploit these channels. As the sensing process requires a considerable amount of time, compressive sensing has been introduced as a low-cost solution to speed up the channels scanning process and improve the detection rate.

In this dissertation, we worked firstly on spectrum sensing techniques and we proposed to improve the matched filter detection technique by using a dynamic detection threshold. The proposed technique was compared to energy and autocorrelation based detection for performance evaluation. Simulation results show that a dynamic detection threshold enhances the efficiency of the matched filter detection technique compared to the same technique with a static detection threshold.

In Chapter III, we discussed the state-of-the-art of compressive sensing and classified the techniques in categories. Then, we proposed a compressive sensing technique exploiting the advantages of the Bayesian network to enhance the signal estimation and reduce the uncertainty impact on the measurements. we used Circulant matrix for signal sampling for its stable structure in terms of acquiring high dimensional signals in an uncertain environment. The results show that the proposed technique reduces the level of randomness and accelerates the spectrum scanning process.

In Chapter VI, we proposed a compressive sensing technique based on the Bayesian model and Toeplitz matrix in order to improve the efficiency of the compressive sensing processes. The simulation results show that the proposed technique is efficient and fast in terms of signal acquisition and signal reconstruction.

In Chapter VII, we described the implementation of a real time wideband spectrum scanning based on the compressive sensing using the software defined radio units (USRP). we measured the occupancies of channels to identify the white spaces in the spectrum in real time. The simulation results show that the proposed technique minimizes the scanning time and enhances the detection rate compared to the conventional spectrum scanning.

In addition, in chapter VIII, we performed a performance comparison between one-bit compressive sensing and conventional compressive sensing. Simulation results show that one-bit compressive sensing is faster and perform well compared to multi-bit compressive sensing.

I can conclude that with compressive sensing, the spectrum scanning is faster and its efficiency is improved. The performance of the compressive sensing is mainly related to the selected sensing matrix and the recovery algorithm, which limits its efficiency.

**Challenges**

The implementation of the spectrum scanning based compressive sensing requires designing specific and practical sensing matrices, particularly in real scenarios. The signal acquisition with a sensing matrix can result in the information loss of the original signal. The cost of implementing such fast and structured matrices is also challenging in terms of design complexity and memory storage. For the signal sparsity, it is an assumption that is not always correct, particularly in real environments where the signal sparsity is unknown. Moreover, the performance of the spectrum scanning is degraded when considering the interferences and the high level of uncertainty. Acquiring non-linear signals is another challenge to the spectrum scanning, which involves real signals with non-linear measurements. Advanced ADCs devises are required to address these challenges and limitations to make the research going.

**Future directions**

For future directions, there are several interesting research directions related to the compressive sensing. For instance, advanced ADCs can be considered for future works. They are highly needed to support the high sampling rate presented in the cognitive radio networks with the high increase of wireless network services and mobile users. Moreover, hardware implementation is yet another future direction in terms of designing fast and inexpensive ADCs devises for signal sampling and integrating the compressive sensing algorithms on these devises.

In addition, implementing in hardware the compressive sensing techniques is another direction to overcome the problems of synchronization, calibration, and uncertainty in measurements. In addition, developing new and efficient signal acquisition models based compressive sensing is an interesting direction of interest to cover all the signal models presented in real radio environments. Also, handling the uncertainty and the imperfections of real radio networks by designing practical compressive sensing techniques is an open door for researchers in this field.

# Appendix A

# TECHNIQUES FOR DEALING WITH UNCERTAINTY IN COGNITIVE RADIO NETWORKS

In this appendix, we are going to present the different techniques to deal with uncertainty in cognitive radio networks. The remaining of this appendix is organized as follows. In Section A.1 and A.2, we define the uncertainty and its categories. In Section A.3, we describe the different techniques to handle uncertainty and their utilization in cognitive radio networks. Finally, a conclusion is given at the end.

## A.1 Introduction

Cognitive radio system can perform the following processes: (1) sensing, which is the comprehension and awareness of the environment; (2) deciding, which is the analysis of results and reliable decision-making based on what is sensed from the environment; and (3) acting intelligently by adapting, changing, and adjusting radio parameters to enhance the performance and overcome the spectrum scarcity issues.

The first process is critical since it is the stage where the measurements are taken and the spectrum sensing is performed. However, due to multipath fading, shadowing, or varying channel conditions [18][219], uncertainty affects this first process. In the observation stage, measurements taken by the SUs are also uncertain. In the next stage, the SUs make a decision based on what has already been observed using their knowledge basis, which may have impacted by uncertainty in the sensed measurements, leading to wrong decisions. In the last process, uncertainty spreads in the cognitive radio cycle, and sometimes the wrong actions are taken. Thus, uncertainty propagation impacts all the radio spectrum processes which degrades the cognitive radio performance.

Thus, it is necessary to address these problems in cognitive radio cycle by sensing the spectrum correctly, making the correct decision, and taking the right action. Existing spectrum sensing techniques, such as energy [49], matched filter [6], wavelet [220], and autocorrelation based detection [45], do not consider uncertainty when the measurements are missing or uncertain. Mitigating uncertainty techniques can be classified into four main categories: probabilistic, fuzzy set theory, possibility theory, and evidence theory methods.

## A.2 Uncertainty Classifications

According to its origin, uncertainty is classified into two main classes [13][175][176], aleatoric and epistemic. In general, aleatoric uncertainty is a statistical uncertainty that reflects the inherent randomness in nature. It represents unknowns that differ each time the same experiment is done. It cannot be eliminated or predicted by collecting more information or knowledge. The studied system can eventually behave differently depending on this uncertainty. In simple terms, it is simply random [221]. Epistemic uncertainty is a systematic uncertainty that is due to a lack of knowledge and subsequent ability to model the studied system. When data are available, epistemic uncertainty can be presented using probabilities and it can be decreased by collecting more information about the studied system. Both categories exist in real applications. Aleatoric uncertainty arises from stochastic behavior and epistemic uncertainty arises from parameter estimation. Uncertainty type should be first identified in order to mitigate its spread in a specific system. In [222], the two classes were combined in one as a hybrid framework when both are propagated in a dynamic system.

In order to handle uncertainty and data deficiency and avoid imprecise decisions, several methods have been proposed under the epistemic type [177][224][225]. These methods are classified into 4 categories: probabilistic, fuzzy, possibility, and evidence based theories. Figure 64 illustrates the classification of the epistemic uncertainty mitigation techniques.

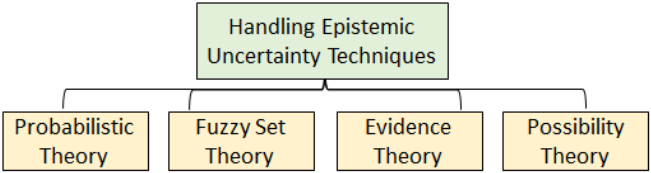

Figure 64: Representation of methods to handle uncertainty.

Probability theory is the main tool used to estimate all the uncertain measurements. It is a mathematical approach aiming to analyze random phenomena based on random variables, stochastic processes, and events [16]. Fuzzy set theory is an alternative way to handle uncertain and imprecise information to make reliable decisions [177]. Evidence theory is an alternative approach to probabilistic approach for modeling the epistemic uncertainty [224]. Possibility theory is a method to mitigate with uncertainty and incomplete or imprecise data in multisource information [225].

## A.3 Handling uncertainty techniques
### A.3.1 Probabilistic Theory Based Techniques

Probabilistic methods can handle both epistemic through experiments and subjective aleatory uncertainty. Under this category, the degree of belief replaces the knowledge about a system state. The degree of belief is attached to all possible events for the studied system and it is expressed using probabilities since knowledge provides a degree of belief and not certain information. Probabilities relate statements to a state of knowledge. They are expressed as P(*A/B*), changing with new evidence C to be expressed as P(*A/B, C*) [16]. Using probabilistic theory allows the studied system to choose the best action. Graphical models are examples of techniques classified under this category. Bayesian network [226] and Markov network [181,182] are examples of graphical models.

### A.3.1.1 Bayesian Network

Bayesian network is used to present knowledge about an uncertain domain and model how intervening variables influence one another. It allows a system to handle uncertain contexts and express conditional probabilities of events where data are missing [229]. It is a graphical representation of probabilistic relationships between a set of random variables and their conditional dependencies related via a directed acyclic graph, which reflects the conditional relations between variables based on direct links between them [230][231]. These links indicate direct influence from one variable to another, which is the direct dependence between them. The lack of connection identifies the conditional independence between variables.

Bayesian models are based on joint and conditional probabilities to operate and provide the probability of how an event is true [17]. The chain rule is used to write any joint probability distribution as an incremental product of conditional distributions.

$$P(x_1, x_2, \ldots, x_n) = \prod_i P(x_i / x_1, x_2, \ldots, x_{i-1}) \tag{90}$$

where $x_i$ is an event, $i=1\ldots n$, $P(x_i)$ is the prior probability of $x_i$, and $P(x_i/x_j)$ is the conditional probability of $x_i$ given $x_j$. Joint probability distribution is expressed as a function of conditional probabilities associated with each node $x_i$ under the conditional independence hypothesis.

$$P(X=x) = \prod P(x_i / P_a(x_i)) \tag{91}$$

where $P_a(x_i)$ is the probability of the parent $x_i$ of the child $x_j$ if $x_j$ depends on $x_i$. Bayes' rule expresses the relation between events using conditional probabilities and it has the form

$$P(x_i / x_j) = P(x_j / x_i) P(x_i) / P(x_j) \tag{92}$$

This rule computes probabilities when there is not direct information about an event[190]. It allows the representation of causal dependencies between various contextual events and the obtainment of probability distributions. Bayesian models have been used in cognitive radio networks to overcome the uncertainty issues in spectrum sensing [225-229]. SUs need to sense and find the available channels for transmission. Spectrum sensing problem is reformulated as

$$y(k) = \begin{cases} n(k), & H0 \\ \sqrt{E_x}e^{j\varphi(k)} + n(k), & H1 \end{cases} \quad (93)$$

where $y(k)$ is the SU received signal, $E_x$ is the PU signal energy, $\varphi(k)=(0, \pi)$, and $n(k)$ is an Additive White Gaussian Noise (AWGN) with zero mean and variance $N_o/2$. Based on Bayesian criterion, decision tests can be written as

$$\left[\frac{P(y/H_1)}{P(y/H_0)}\right] \gtreqless \left[\frac{P(H_0)(C_{pq}-C_{pp})}{P(H_1)(C_{qp}-C_{qq})}\right] \quad (94)$$

where $P(y/H_1)$ and $P(y/H_0)$ are the probability density functions (PDF) of $H_1$ and $H_0$ respectively; $P(H_1)$ and $P(H_0)$ are the prior probabilities of $H_1$ and $H_0$ respectively, which are assumed to be known. The PDF of the SU received signal $y(n)$ over an $N$ symbol duration under the hypotheses $H_0$ and $H_1$, respectively, can be written as [23]:

$$P(y/H_0) = \prod_{k=0}^{N-1} \frac{e^{-y^2(k)/N_0}}{\sqrt{\pi N_0}} \quad (95)$$

$$P(y/H_1) = \prod_{k=0}^{N-1} \frac{e^{-\frac{\left(y(k)-\sqrt{E_s}e^{j\varphi(k)}\right)^2}{N_0}}}{\sqrt{\pi N_0}} P(\varphi(k)) \quad (96)$$

Then, from equations (6) and (7), the likelihood ratio test (LRT) of $H_1$ and $H_0$ can be defined as

$$LRT(y) = \frac{P(y/H_1)}{P(y/H_0)} = \prod_{k=0}^{N-1} e^{-E_x/N_0} \cosh(x(k)) \quad (97)$$

where $x(k) = (2\sqrt{(E_x)})/N_0 y(k)$. From (7) and (8), the test decision can be deduced by comparing LRT to the threshold $\delta$. The Bayesian detector is based on the Bayesian decision rule, which aims to minimize the expected posterior cost (EPC) [230]. The EPC can be formulated as a function of the cost $C_{ij}$ associated to the decision $H_i$ if the state is $H_j$ for i, j=0, 1

$$EPC = \sum_{i=0}^{1}\sum_{j=0}^{1} C_{ij} P(H_j) P(H_i/H_j) \quad (98)$$

From (97), the optimal Bayesian detector can be derived as

$$\sum_{k=0}^{N-1} \ln\left(\cosh(x(k))\right) > N\gamma + \ln(\delta) \quad (99)$$

The threshold is determined as

$$\delta = \frac{P(H_0)(C_{10}-C_{00})}{(PH_1)(C_{01}-C_{11})} \tag{100}$$

The suboptimal Bayesian detector is obtained using approximations of the low and high SNR. The Bayesian rule is also used to derive expressions for $P_d$ and $P_f$ [230].

Consider a cognitive radio network with $N$ SUs and $M$ licensed channels, the network is modeled by a graph in which each node represents a SU and each edge represents the communication link between the two corresponding nodes (SUs). In [231], Bayesian model is used for each SU for decision making and channel selection. In [232], spectrum sensing is reviewed under noise uncertainty caused by the channel conditions, which are realistic time-varying multipath fading channels. The authors proposed a spectrum sensing scheme based on energy detection technique. This scheme deals with uncertainty by estimating both the PU state and time-variant multipath gains.

In [233], the authors proposed a new approach based on Bayesian models and game theory, also called Bayesian games, which allows the analysis of the influence of incomplete data and uncertain information in cognitive radio networks. They addressed the uncertainty about several "players" decision (SUs) in a cooperative system. In [234], a generalized likelihood ratio test (GLRT) is used to address the incomplete knowledge about the sensing measurements by considering some known and unknown parameters and studying the influence and the interactions between them. In [183], in order to learn about the environment conditions, the authors proposed a Bayesian network that represents the relationships between variables indicating how the system is performing and identifying which variables affecting bit error rate (BER). Those variables are bit energy to noise spectral density ratio, carrier to interference ratio, modulation scheme, Doppler phase shift, and BER.

In [235], a probabilistic approach based Bayesian models were used to estimate the interferences in wireless networks by modeling the signal SINR and dynamically handling the uncertainty affecting it. This uncertainty resides in the frequencies, transmission power, the number of nodes and their locations. In [236], The authors proposed to apply Bayesian models to mitigate uncertainty behind the interference power in order to minimize interference to the PU when the SU is using the shared channel. The authors probabilistically modeled the parameters affecting the interference power such as path loss and shadowing. In [17], in the context of compressive spectrum sensing [190], the authors used a Bayesian-based recovery to recover the compressed signal at the received by exploiting the advantages of Bayesian models to deal with uncertainty in measurements during the compressive sensing processes.

### A.3.1.2 Markov Models

Markov models are undirected graphical models that can be categorized in four divisions used in different situations, according to the system observed (autonomous or controlled) and the system state. These models are Markov chain, Hidden Markov model, Markov decision process, and Partially Observable Markov decision process [228]. The Hidden Markov model is a statistical Markov model in which the modeled system is assumed to be a Markovian process with unobserved states. It starts with a Markov chain and adds a noisy observation about the state at each time [182].

Markov approach is used in spectrum sensing to model the interactions between PU and SU as continuous-time Markov chains. Hidden Markov models are used to identify the signal features processing in which spectrum activities are Markovian. In [237], the authors proposed a Markov model to obtain the channel holding time, which is the period that SU uses the free band without PU signal interruption. In [238], a new scheme is proposed to reduce the spectrum utilization: channels are grouped in clusters, only one channel must be sensed, and the other channels are estimated using the historical states and their correlation with the sensed channel in the same cluster. Markov model is used to model the influence of historical states on the current state in which the channel correlation and Markov model reduces the number of channels, which must be sensed to improve the sensing performance [228].

### A.3.2 Fuzzy Set Theory

Fuzzy set theory is an alternative approach for handling uncertain and imprecise data to make decisions and allows expressing real situations with a mathematical model [229]. Fuzzy set theory aims to provide a mathematical model for reasoning under uncertainty in detection and decision making. It describes to what degree a decision is certain and reliable using uncertain and imprecise knowledge based on subjective estimation and expert opinions and experiences. Under fuzzy logic, statement truth is not always clear, but depends on a degree of membership to a set or a class; statement truth takes values between 0 and 1, in contrast to classical logic with two possible values: true (1) or false (0). The degree of membership to this set (i.e., the closeness to 0 or 1) indicates the degree of truth of the statement. Fuzzy logic indicates to what degree $X$ is in various sets. The degree of membership in the interval [0, 1] is associated with each element in the fuzzy set [239].

In [241], a fuzzy inference system is implemented following a four-stage procedure: fuzzification, fuzzy inference engine, fuzzy rule base, and defuzzification. The system inputs are the measurements affected by uncertainty. The inputs are forwarded to determine the membership function in the fuzzification stage in which the fuzzy rule IF-THEN is applied to determine the new outputs set. The fuzzified measurements are used in the next stage and all rules are aggregating using a fuzzy logic operation (e.g., OR, AND, Union, Intersection). In the final stage, fuzzy set is converted to a defuzzified set which corresponds to the chance value. Fuzzy probability is an extension to handle mixed probabilistic/non-probabilistic uncertainty.

Fuzzy logic is applied to cognitive radio networks in order to make decisions under uncertain environments. SUs make decisions based on available information from the spectrum sensing stage, which includes measurements, incomplete data, and their own knowledge. SUs decide to whether transmit their data depending on what is sensed and then select which channel frequency to use without creating interference to the PUs. The fuzzy logic-based approach can be used to make decisions under incomplete and doubtful data. It can deal with uncertainty in data by transforming imprecise data into precise data based on fuzzy set inference [240]. SUs need specific information about the channel state; this information depends on the spectrum sensing method used in the sensing stage. A fuzzy-based spectrum handoff is proposed in [241] in order to avoid interference to the PUs. The fuzzy inference system estimates the distance between the SU and the PU and the required SU power for avoiding interference. The results show that fuzzy logic outperforms classical spectrum sensing. In [242], a fuzzy power control scheme is proposed in which a transmit power control system is designed using a fuzzy logic system to permit dynamic control of the transmitted signal power depending on the interference level experienced by the PUs. In [243], fuzzy conditional entropy maximization is used to design energy detection for cooperative spectrum sensing in order to minimize uncertainty in the threshold selection. In [244], the authors proposed cooperative spectrum sensing based on fuzzy integral theory in the context of cognitive radio. The proposed scheme seeks to handle uncertainty in the information provided by the local SUs.

### A.2.3 Evidence Theory

Evidence theory, also called Dempster-Shafer theory (DST), was developed as an alternative to probability theory [245]. It is a mathematical theory for reasoning and modeling the epistemic uncertainty that reasons with belief and plausibility. Belief provides all the available evidence for a specific hypothesis; plausibility offers all the evidence that is consistent with this hypothesis. The two concepts give an interval of probabilities, including true probability with some certainty. In other words, it combines distinct evidence from several sources to compute the probability of the hypothesis. This evidence is represented by a mathematical function called belief function. This function considers all evidence available for the hypothesis; a hypothesis and its negation do not have any causal relationship between them, which proves that an ignorance of belief does not involve disbelief, but reflects a state of uncertainty.

In order to represent uncertainty in a hypothesis, uncertainty is replaced with belief or disbelief as evidence, which is the concept behind DST. This method suffers from high mathematical complexity and needs all possible states to compute the exact probability while representing good certainty with additional information about the degree of belief. DST rule for combination can be defined as the procedure of combining distinct states of evidence. DST rule for combination consists of a space of mass $W$ and belief mass as a function denoted $m$.

$$m: 2_x \rightarrow [0,1] \qquad (101)$$

where $X$ is the set that includes all possible states and $2_x$ is the set of all the subsets of $X$. DST is able to handle uncertainty, imprecision, ignorance, and lack of data because it is based on the estimation of imprecision and conflict from several sources. In [246], evidence theory was used to deal with uncertainty in a mixed aleatory-epistemic model.

DST is applied in cognitive radio context to handle uncertainty and imperfect knowledge in radio cognition cycle. In [247], the authors proposed a distributed spectrum sensing model to ensure reliable and credible spectrum sensing decision. The proposed scheme combines the local decisions from several SUs to make the final decision. Each local decision $m_i$ is associated with a credibility parameter $m_i(c)$, which quantifies the channel condition and sends the information to the access point to make the final decision about the state of the observed band. It was shown that the proposed scheme performed well in terms of $P_d$ and $P_f$ based on the credibility for $H_0$ and $H_1$ [248]. In [247], a cooperative model was proposed to enhance the reliability of spectrum detection based on energy detection with a double threshold.

### A.2.4 Possibility theory

Possibility theory is a technique to handle certain types of uncertainty, based on fuzzy set concepts. It is an alternative to the probability theory and an extension to fuzzy logic theory. It permits handling uncertainty in multisource information [225]. It is characterized by the possibility, *Poss*, measures, which assigns numbers to each subset, $W$, based on fuzzy logic [249]. Possibility theory rules are different from probability theory rules. It is defined as:

$$Poss: _{W \to [0, 1]}, \max (A \in W, Poss (A) = 1) \tag{102}$$

Conditional independence in possibility theory allows modeling the dependence between uncertain variables as in probability theory, and can be defined as:

$$Poss (A, B/C) = Poss (A/C) \otimes Poss (B/C) \tag{103}$$

In order to model the available information, possibility and necessity measures are considered based on possibility distribution. Possibility theory is used to handle only epistemic uncertainty, and it is a special case or a subset of the evidence theory. Possibility logic was applied in many fields including signal processing to handle noise uncertainty in signal propagation. Noise nature cannot be assumed, estimating its level in each sample based on some signal characteristics was treated by applying possibility theory, which considered the lack of knowledge about noise. Noise uncertainty is due to random perturbation propagated through the transmission channel [249]. Theory of possibility was applied on cognitive radio as a special case of the evidence theory already reviewed, especially for optimization problems.

## A.4 Techniques Comparison

Several methods to handle uncertainty propagation were proposed in the literature. Selecting a solution depends on several criteria, including requirements, conditions, and especially the tradeoff between precision and complexity. Through analyzing the different methods, we can conclude that methods with good precision represent high complexity. All methods had their weaknesses and strengths. The methods choice depends on the problem to be solved.

Probabilistic theory is a powerful tool for handling uncertainty if precise uncertainty is required. Bayesian network as a probabilistic method is more efficient and it is the most used to present, reason, and model uncertainty. However, it needs prior probability distributions because parameters are not always Gaussian in practice measurements. Fuzzy theory is easy to implement; it is more suitable for problems that did not require precise knowledge of uncertainty, but it handles a degree of truth and not uncertainty. For evidence theory, it provides additional information about the degree to which information is available. DST is more general than possibility and probability theory. Table 7 compares the listed techniques.

Table 7: Techniques comparison

| Probability theory | Fuzzy logic theory | Evidence theory | Possibility theory |
|---|---|---|---|
| -Deals with randomness and subjective uncertainty<br>-Handles epistemic and aleatory uncertainty through experiments | -Deals with imprecise vague information<br>-Describes of what degree a decision is certain and reliable using uncertain and imprecise knowledge | -Deals with epistemic uncertainty<br>-Used where there is some degree of ignorance (incomplete model) | Deals with incomplete or imprecise data in multisource information |
| -Powerful for handling uncertainty in case precise uncertainty is required<br>-Based on mathematical probability to predict an event | -Precision and stability not guaranteed<br>-Performance measured a Posterior<br>-Based on user experience and interpretation | More Flexible | Improves the precision of evidence |
| -Applies to all systems<br>-Complex data handling<br>-Inexact / incorrect | -Applies to systems that are difficult to model<br>- Easy to implement and interpret | High complexity | -Computationally simple<br>-Complexity close to classical logic |
| One valued approach (Truth is one-valued) | Set valued approaches (Truth is many-valued) | Two valued approaches (Truth is 2-valued) | Two valued approaches (Truth is 2-valued) |
| Degree of belief | Degree of membership to set | Belief and Plausibility | Possibility measures |

Understanding uncertainty in the context of the cognitive radio network is an important task in modeling this uncertainty and mitigating it, identifying its type, its source, and its influence on the performance cognitive radio system. As shown in this survey, Bayesian models and fuzzy logic are the most used methods in spectrum sensing for reasoning under uncertainty and making decisions.

## A.5 Conclusion

Due to the randomness features of communication channels, uncertainty impacts all the processes of the cognitive radio cycle, including the sensing results at the SUs. Making decisions under uncertainty is a challenge that cognitive radio systems must face. This work evaluated different models to deal with uncertainty and enhance the fidelity of sensing outcomes and decisions.

# Appendix B
# SNR ESTIMATION BASED PSO

In this appendix, we describe the estimation based PSO technique to estimate SNR in real time for the wideband spectrum scanning experiments presented in chapter VII. The remaining of this appendix is organized as follows. In Section B.1 and B2, we introduce the SNR estimation and highlight some of its proposed techniques. In section B.3, we describe the SNR estimation based PSO technique. Finally, a conclusion is given at the end.

## B.1 Introduction

SNR plays a very important role in the wideband spectrum scanning in the sense that it impacts the detection and false alarm detection rates [136]. Estimation of the SNR has become an integral part of wireless communication systems, particularly in cognitive radio systems. Approximating this parameter can help in the spectrum sensing process to identify the available channels and in the decision-making phase to estimate the usage level of different channels of the radio spectrum [193].

## B.2 SNR estimation techniques

A number of SNR estimation techniques have been proposed in the literature. These techniques can be classified into two main categories: data aided and non-data aided techniques. Data aided categories require knowing the characteristics of the transmitted data sequences. Examples of techniques in this category include split symbol moment estimator [236,250,251], maximum likelihood estimator [251], squared signal to noise variance estimator [252], second and fourth order moment estimator [45,253] and low bias algorithm negative SNR estimator [254]. Non-data aided category does not require any information about the features of the transmitted data sequences to perform. Non-data aided techniques are based on extracting and analyzing the inherent characteristics of the received signal to estimate the signal and the noise powers. Examples of these techniques include wavelets based SNR estimation [255], sixth-order statistics based non-data aided SNR estimator [256], and eigenvalue-based SNR estimation [257-259].

## B.3 SNR estimation based PSO

Eigenvalue-based SNR estimation is based on the eigenvalues of the covariance matrix and it depends on the number of samples, the number of eigenvalues, and the Marchenko-Pastur distribution size [193]. To optimize these three key parameters, particle swarm optimization (PSO) algorithm is adopted. The SNR estimation based PSO is performed by computing the eigenvalues of the covariance matrix that minimizes the mean square error of the received signals [198]. PSO being a global search optimization technique enables locating the optimal number of eigenvalues by minimizing the estimation error, which increases the accuracy of the SNR estimation.

The authors of [258] proposed a method based on the eigenvalues of the covariance matrix formed from the received samples. This method initially detects the eigenvalues as in [256-269]. Then, the minimum descriptive length criterion is used to split the signal and noise corresponding to these eigenvalues. This technique is highly dependent on the number of received samples, $N$, the number of eigenvalues, $L$, and the Marchenko-Pastur distribution size, $K$. To implement the technique, the received samples are comprised of both noise components and signal components. Thus, it is necessary to determine the noise variance to estimate SNR. The received signal, $x(n)$, is acquired and stored in an array as

$$[x(0), x(1), x(2), \ldots, x(N-1)] \tag{104}$$

A value known as the smoothing factor is chosen and denoted as $L$. An $L$x$M$ dimension matrix is formed, where each row of the matrix is comprised of L time shifted versions of the received signal samples as

$$\boldsymbol{X} = \begin{pmatrix} x_{1,1} & \cdots & x_{1,N} \\ \vdots & \ddots & \vdots \\ x_{L,1} & \cdots & x_{L,N} \end{pmatrix} \tag{105}$$

where $x_{i,j}$ is the received signal vector sample, $L$ is the number of eigenvalues and $N$ is the length of the received signal vector. Similar to the approach in [158], the sample covariance matrix is computed as the product of matrix, $\boldsymbol{X}$ and its Hermitian transpose averaged over $N$ samples which is given by

$$\widehat{\boldsymbol{R}}_x = \frac{1}{N} \boldsymbol{X}\boldsymbol{X}^H \tag{106}$$

The eigenvalues of the resultant $L$ x $L$ matrix are computed and sorted in descending order to form an $L$-element array. The descending order sort is performed based on the MDL criterion which implies that the first $M$ eigenvalues represent the transmitted signal component and the remaining ($L$-$M$) eigenvalues represent the noise component. The array of eigenvalues is represented as:

$$[\lambda_1, \lambda_2, \ldots, \lambda_M, \lambda_{M+1}, \ldots, \lambda_L] \tag{107}$$

The value of $M$ is estimated using the MDL criterion as

$$\hat{M} = argmin_M \left(-(L-M)N \log\left(\frac{\theta(M)}{\phi(M)}\right) + \frac{1}{2}M(2L-M)\log N\right), 0 \leq M \leq L-1 \qquad (108)$$

where $\theta(M) = \prod_{i=M+1}^{L} \lambda_i^{\frac{1}{L-M}}$ and $\phi(M) = \frac{1}{L-M}\sum_{i=M+1}^{L} \lambda_i$, and $\hat{M}$ is the estimated value of $M$. Then, the array of eigenvalues is split up based on the noise group and transmitted signal group as $\lambda_{Signal} = [\lambda_1, \lambda_2, \ldots, \lambda_{\hat{M}}]$ and $\lambda_{Noise} = [\lambda_{\hat{M}+1}, \lambda_{\hat{M}+2}, \ldots, \lambda_L]$.

To estimate the noise power using the array $\lambda_{Noise}$, $\sigma_{z1}^2$ and $\sigma_{z2}^2$ are computed as

$$\sigma_{z1}^2 = \frac{\lambda_L}{\left(1-\sqrt{c}\right)^2} \qquad (109)$$

$$\sigma_{z2}^2 = \frac{\lambda_{\hat{M}+1}}{\left(1+\sqrt{c}\right)^2} \qquad (110)$$

where, $c = L/N$. In random matrix theory, the Marchenko-Pastur law provides the probability density function of singular values of large rectangular random matrices [193]. In this case, the matrix $X$ is the rectangular random matrix whose entries $x_{i,j}$ are independent and identically distributed random variables with mean zero and variance $\sigma^2$. A set of $K$ linearly spaced values in the range $[\sigma_{z1}^2, \sigma_{z2}^2]$ is generated and denoted as $\pi_k$, where $1 \leq k \leq K$. The Marchenko-Pastur density of the parameters $(1-\hat{\beta})c$ and $\pi_k$ where $\hat{\beta} = \frac{\hat{M}}{L}$ is given by

$$MP_d = MP\left((1-\hat{\beta})c, \pi_k\right) = \frac{\sqrt{\left(v-\left(\pi_k\left(1-\sqrt{(1-\hat{\beta})c}\right)\right)^2\right)*\left(\left(\pi_k\left(1+\sqrt{(1-\hat{\beta})c}\right)\right)^2-v\right)}}{2*\pi*\pi_k^2*(1-\hat{\beta})c*v} \qquad (111)$$

where $\left(\pi_k\left(1-\sqrt{(1-\hat{\beta})c}\right)\right)^2 \leq v \leq \left(\pi_k\left(1+\sqrt{(1-\hat{\beta})c}\right)\right)^2$

The empirical distribution function of the noise group eigenvalues, $\lambda_{Noise}$, is computed by

$$E_d = F_n(t) = \frac{N \leq t}{n} = \frac{1}{n}\sum_{i=1}^{n} 1_{\lambda_{Noise}(i) \leq t} \qquad (112)$$

where $n = L - \hat{M} + 1$ and N is the number of sample values. Both the arrays ($MP_d$ and $E_d$) are compared and a goodness of fitting ($D(\pi_k)$) is used to find the best estimate of $\pi_k$, thereby estimating the value of the noise power $\sigma_z^2$ [193]. The goodness of fitting is given by

$$D(\pi_k) = \|E_d - MP_d\|_2 = \sqrt{\sum(E_d - MP_d)^2} \qquad (113)$$

From the array of values of $D(\pi_k)$, the index of the minimum value of $D(\pi_k)$ is obtained and the corresponding value of the array $\pi_k$ for the obtained index is the estimate of noise variance $\widehat{\sigma_z^2}$ which is given by

$$\widehat{\sigma_Z^2} = argmin_{\pi_k}\big(D(\pi_k)\big)$$

Once the noise power has been estimated, the signal power can be calculated as the difference between the total received signal power and the estimated noise power. The total received signal power is given by

$$\widehat{P_t} = \left(\frac{1}{NL}\sum_{j=1}^{L}\sum_{i=1}^{N}|x_{i,j}|^2\right) \tag{114}$$

Therefore, the SNR $\hat{\gamma}$ is given by

$$\hat{\gamma} = \frac{\widehat{P_s}}{\widehat{\sigma_Z^2}} = \frac{\widehat{P_t}-\widehat{\sigma_Z^2}}{\widehat{\sigma_Z^2}} = \left(\frac{1}{NL\widehat{\sigma_Z^2}}\sum_{j=1}^{L}\sum_{i=1}^{N}|x_{i,j}|^2\right) - 1 \tag{115}$$

The received samples are stored in an array, and based on the value of the smoothing factor variable $L$, an $L$x$N$ received sample matrix is generated. The sample covariance matrix of the newly generated matrix is computed and its eigenvalues are extracted. From this set of eigenvalues, the value of $M$ is estimated, which aids in the splitting of the noise group and signal group eigenvalues. From the noise group eigenvalues, the Marchenko-Pastur density values and the empirical distribution are computed, from which the noise power is computed. With the availability of the noise power and the total received power, the signal power can be calculated as their difference. The ratio of the signal power to the noise power yields the estimated value of the SNR [192,193].

## B.4 Conclusion

In this appendix, the SNR estimation technique based on particle swarm optimization algorithm that was used in some of our experiments is presented. It is based on the computation of eigenvalues of the covariance matrix of the received signal samples.

# Appendix C
# SOLVING THE UNDERDETERMINED SYSTEM

In this appendix, we describe how the underdetermined system can be solved using optimization algorithms. The remaining of this appendix is organized as follows. In Section C.1, we define an underdetermined system and its mathematical model. In section C.2, we describe the different algorithms used to solve underdetermined systems for sparse signals. Finally, a conclusion is given at the end.

## C.1 Introduction

In the estimation theory, unknown parameters can be estimated by solving a system of polynomial equations. This system can be determined, underdetermined, or overdetermined. A system of equations is called determined when the number of equations is equal to the number of the unknown variables. A system is called underdetermined system if there are less equations than unknowns [233]. A system of equations is called overdetermined system if there are more equations than the unknown variables. An unknown variable is considered as a degree of freedom while an equation is considered as a constraint restricting one unknown variable [173].

Let's consider the variables, $x \in \mathcal{R}^N$ and $y \in \mathcal{R}^M$, where $x = (x_1, x_2, \ldots, x_N)$ and $M \ll N$, and the following system of equation:

$$y = \Phi x \qquad (116)$$

where $\Phi$ is a matrix. The system in this equation is an underdetermined system as we have $M$ equations and $N$ unknown variables where $M \ll N$. To solve this underdetermined system and approximate $x$ coefficients, the problem is considered as an optimization problem. Various recovery algorithms have been developed to solve the following optimization problem.

$$\tilde{x} = \underset{y=\Phi x}{\mathrm{argmin}} \|x\|_1 \qquad \text{Subject to} \quad y = \Phi x \qquad (117)$$

where the estimated variable $\tilde{x}$ is the sparsest solution from many possible solutions of the optimization problem. $\|x\|_1 = \sum_i |x_i|$ is the $\mathcal{L}_1$ norm of $x$ and represents the sum of the absolute values of $x$ coefficients [61][62].

In order to solve the underdetermined system, a regularization parameter is introduced as an additional information to solve the optimization problem using machine learning algorithms. The underdetermined system presented in Equation (117) can be reformulated as:

$$\tilde{x} = \operatorname*{argmin}_{y = M_c x + w} \|y - M_c x\|_2^2 + z\|x\|_1 \qquad (118)$$

where $z$ is the regularization parameter, also called regularization term. The regularization term ca be computed by

$$min_x \sum_1^N A(x_i, y_i) + Bz(x) \qquad (119)$$

where $A$ is the loss function that describes the cost of estimating the unknown quantity $x$ given $y$ and $B$ is a parameter controlling how important is the regularization term in solving the underdetermined system. The regularization term, $z$, controls the penalty on the complexity of the variable $x$.

## C.2 Estimation algorithm

The underdetermined system can have infinite number of solutions, but only the sparsest solution is considered as the unique solution that respects the RIP and the coherence requirements described in chapter V. The sparest solution is the solution $\tilde{x}$ that minimizes the $\|x\|_1$ of the system $y = \Phi x$.

Let's consider a matrix $\Phi$ of $M$x$N$ with linearly independent rows, where M<<N. Any solution of the underdetermined system can be formulated as

$$x = x_0 - b \qquad (120)$$

where $x_0$ is a particular solution of $y = \Phi x$ and $b \in \mathcal{N}(\Phi)$ and $\Phi b=0$. Thus, the problem of finding a solution $\tilde{x}$ of the optimization problem $y = \Phi x$ can be transformed into a problem of finding a vector $\tilde{b} \in \mathcal{N}(\Phi)$ that minimizes the $\|x_0 - b\|$. Then, the solution $\tilde{b}$ satisfies the system of equation $\tilde{x} = x_0 - \tilde{b}$.

The underdetermined system can be solved in different ways according to the used solver. A number of algorithms have been proposed to solve the underdetermined system using as a linear programming problem, such as $\mathcal{L}_1$ norm minimization, gradient descent, iterative thresholding, matching pursuit, and orthogonal matching pursuit [22,61-68].

$\mathcal{L}_1$ norm minimization, also called basis pursuit, performs by minimizing the cost function

$$\tilde{x} = \min\|y - \Phi x\|_2 + z\|x\|_1 \tag{121}$$

where z is the regularization term. The algorithm works step by step as follows:

- The first guest is initialized ($x = x_0 = \Phi'y$), which represents the minimal energy of the signal and the cost function ($\min\|y - \Phi x_0\|_2$) is set.
- For *k* iterations, the new sensing matrix is computed by selecting the required measurement number and the signal coefficients are updated to reach the minimized form.
- The algorithm operates iteratively until obtaining enough signal coefficients less than the signal sparsity level.

Matching pursuit algorithm performs by selecting one position of a non-zero element of the signal, *x*, which corresponds to selecting one column from the matrix. It permits to estimate the signal by decomposing it into a linear expansion of waveforms selected from a dictionary. It consists in selecting a column from A that maximizes the inner product of the current residual.

- The vector that corresponds to the longest projection of *x* is selected from the dictionary.
- The signal *x* is orthogonalized by removing any element of the selected vector from *x* to get the residual of *x*, that has the lowest energy.
- The two previous steps are repeated to the remaining of the dictionary in an iterative process until the residual norm is low than a threshold *c*.

Orthogonal matching pursuit algorithm is inspired from matching pursuit algorithm by removing not only elements of a selected vector from *x*, but also from the basis before repeating the process. It performs by adding an orthogonal projection to the residual computing.

## C.3 Conclusion

In this appendix, we described how the underdetermined system is solved using different solvers and algorithms. Most of these algorithms were described with more details in the previous chapters (Chapter III, Chapter IV, Chapter V, and Chapter VI)